\pdfoutput=1
\documentclass[12pt,preprint]{aastex}
\usepackage{graphicx}
\usepackage{natbib}
\usepackage{aas_macros}
\usepackage[section] {placeins}
\usepackage{subfigure}
\usepackage{float}

\input mydefs.sty

\def\picturesize{7.05}
\def\pictureskip{-1.9}
\def\picturecapskip{-1.0}

\begin{document}

\title{The Nature of Damped Lyman Alpha Systems and Their Hosts in the Standard Cold Dark Matter Universe}
 
\author{
Renyue Cen$^{1}$
} 
 
\footnotetext[1]{Princeton University Observatory, Princeton, NJ 08544;
 cen@astro.princeton.edu}

\begin{abstract} 

Using adaptive mesh-refinement cosmological hydrodynamic simulations 
with a physically motivated supernova feedback prescription
we show that the standard cold dark matter model 
can account for extant observed properties of damped Lyman alpha systems (DLAs).
We then examine the properties of DLA host galaxies.
We find:
(1) While DLA hosts roughly trace the overall population of galaxies at all redshifts,
they are always gas rich and have tendencies of being slightly smaller and bluer.
(2) The history of DLA evolution is cosmological in nature and 
reflects primarily the evolution of the underlying cosmic density, galaxy size and galaxy interactions. 
With higher density and more interactions at high redshift
DLAs are larger in both absolute terms and in relative terms with respect to virial radii of halos.
(3) The variety of DLAs at high redshift is richer with 
a large contribution coming from galactic filaments, created through close galaxy interactions.
The portion of gaseous disks of galaxies where most stars reside makes relatively small contribution
to DLA incidence at $z=3-4$.
(4) The vast majority of DLAs arise in halos of mass $M_h=10^{10}-10^{12}\msun$
at $z=1.6-4$, as these galaxies dominate the overall population of galaxies then.
At $z=3-4$, 20-30\% of DLA hosts are Lyman Break Galaxies (LBGs), 
10-20\% are due to galaxies more massive than LBGs and 50-70\% are from smaller galaxies.
(5) Galactic winds play an indispensable role in shaping the kinematic properties of DLAs. 
Specifically, the high velocity width DLAs are a mixture of those arising 
in high mass, high velocity dispersion halos and those arising
in smaller mass systems where cold gas clouds are entrained to high velocities by galactic winds.
(6) In agreement with observations, we see a weak but noticeable evolution in DLA metallicity.
The metallicity distribution centers at $[Z/H]=-1.5$ to $-1$ 
and spans more than three decades at $z=3-4$,
with the peak moving to $[Z/H]=-0.75$ at $z=1.6$ and $[Z/H]=-0.5$ by $z=0$.
(7) The star formation rate of DLA hosts is concentrated in the range $0.3-30\msun$/yr at $z=3-4$,
gradually shifting lower to peak at $\sim 0.5-1\msun$/yr by $z=0$.

\end{abstract}
 
\keywords{Methods: numerical, ISM: kinematics and dynamics,
Galaxies: interactions,
Galaxies: evolution,
intergalactic medium}
 
\section{Introduction}

Damped $\lya$ systems (DLAs) are fundamentally important,
because they contain most of the neutral gas in the universe at all times 
since cosmological reionization
\citep[e.g.,][]{2000StorrieLombardi,2003Peroux,2009Prochaska}.
Molecular clouds, within which
star formation takes place, likely condense out of cold dense neutral atomic gas contained in DLAs,
evidenced by the fact that the neutral hydrogen (surface) density in DLAs and molecular hydrogen (surface) 
density in molecular clouds form a continuous sequence \citep[e.g.,][]{1998bKennicutt, 2006Zwaan}.
Therefore, DLAs likely hold key to understanding the fuel for and ultimately galaxy formation.
A substantial amount of theoretical work has been devoted to studying the nature of DLAs
\citep[e.g.,][]{1997Gardner,1997bGardner,1998Haehnelt,2001Gardner,2001Maller,
2003dCen,2004fNagamine,2004gNagamine,2007aNagamine,
2006Razoumov,2008Razoumov,2008Pontzen, 2009Tescari, 2010Hong},
since the pioneering investigation of \citet[][]{1996bKatz} in the context of the cold dark matter (CDM) cosmogony.
A very interesting contrast is drawn between the observationally based inference
of simple large disk galaxies possibly giving rise to DLAs 
\citep[][]{1986Wolfe,1997Prochaska}
and the more naturally expected hierarchical buildup of structures in the CDM cosmogony
where galactic subunits may produce some of the observed kinematics of DLAs
\citep[][]{1998Haehnelt}.
Clearly, the implications on the evolution of galaxies in the two scenarios are very different.

We have carried out a set of Eulerian adaptive mesh refinement (AMR) simulations 
with a resolution of $0.65~$kpc proper and a sample size of several thousand galaxies
with mass $\ge 10^{10}\msun$ to statistically address the physical nature of DLAs in the current
standard cosmological constant-dominated CDM model (LCDM)
\citep[][]{2010Komatsu}.
Mechanical feedback from star formation driven by supernova explosions and stellar winds
is modeled by a one-parameter prescription that is physically and energetically sound.
Part of the motivation was to complement and cross-check studies to date that are largely based on 
smooth-particle-hydrodynamics (SPH) simulations.
With the simulation set and detailed analysis performed here 
this study represents a significant extension of previous works to simultaneously
subject the LCDM model to a wider and more complete range of comparisons with observations.
We examine in detail the following properties of DLAs in a self-consistent 
fashion within the same model:
DLA column density distribution evolution, line density evolution, 
metallicity distribution evolution, size distribution evolution,
velocity width distribution evolution,
kinematic structural parameters evolution, neutral mass content evolution and others.
A gallery of DLAs is presented to obtain a visual understanding of the physical 
richness of DLA systems, especially the effects of galactic winds and large-scale gaseous structures.
In comparison to the recent work of 
\citet[][]{2010Hong} we track the metallicity distribution and evolution explicitly 
and show that the metallicity distribution of DLAs is, 
in good agreement with observations, very wide,
which itself calls for a self-consistent treatment of metals transport.
In agreement with the conclusions of \citet[][]{2010Hong}, although not in the detailed process,
we show that galactic winds are {\it directly} responsible for a large fraction of wide DLAs at high redshift,
by entraining cold clouds to large velocities and causing large kinematic velocity widths.
We find that the simulated \ion{Si}{2} $\lambda$1808 line velocity width, kinematic shape measures and 
DLA metallicity distributions that are all in excellent agreement with observations. 
Taking all together, we conclude that 
the standard LCDM model gives a satisfactory account of all properties of DLAs.
Finally, we examine the properties of DLA hosts,
including their mass, 
star formation rate, HI content, gas to stellar mass ratio and colors, 
and show that DLAs arise in a variety
of galaxies and roughly trace the entire population of galaxies at any redshift.
This may reconcile many apparently conflicting observational evidence of identifying DLAs with
different galaxy populations.
Specifically, at $z\sim 3$ we show that $20-30\%$ of DLAs 
are Lyman Break Galaxies (LBGs) \citep[][]{1996bSteidel},
while the majority arise in smaller galaxies.

The outline of this paper is as follows.
In \S 2 we detail our simulations, method of making galaxy catalogs, method of making DLA catalogs
and procedure of defining \siii line profile shape measures. 
Results are presented in \S 3.
In \S 3.1 we present a gallery of twelve DLAs.
We give \siii line velocity width distribution functions in \S 3.2,
demonstrating excellent agreement between simulations and observations,
particularly at high velocity width end.
\S 3.3 is devoted to the three kinematic measures of the \siii absorption line
and we show the model produces results that are consistent with observations.
Column density distribution and evolution, line density and neutral gas density evolution 
are described in \S 3.4,
where, while simulations are consistent with observations,
we emphasize large cosmic variance from region to region with different large-scale overdensities.
The focus is shifted to metallicity distribution and evolution in \S 3.5
and simulations are found to be in excellent agreement with observations where comparisons can be made.
The next subsection \S 3.6 performs a detailed analysis of the size of DLAs
and finds that the available observed QSO pairs with DLAs are in accord with
the expectation of our model.
Having found agreement between simulations and observations in all aspects pertinent to DLAs,
we turn our attention to the properties of DLA hosts in \S 3.7,
where we show that DLAs, while slightly favoring galaxies that are more gas rich,
less massive and bluer in color, and have higher HI mass and higher gas to stellar mass ratio,
roughly trace the entire population of galaxies at all redshifts.
Conclusions are given in \S 4.

\section{Simulations}\label{sec: sims}

\subsection{Hydrocode and Simulation Parameters}

We perform cosmological simulations with the adaptive mesh refinement (AMR) 
Eulerian hydro code, Enzo 
\citep[][]{1999aBryan, 1999bBryan, 2004OShea, 2009Joung}.  
First we ran a low resolution simulation with a periodic box of $120~h^{-1}$Mpc on a side.
We identified two regions separately, one centered on
a cluster of mass of $\sim 2\times 10^{14}\msun$
and the other centered on a void region at $z=0$.
We then resimulate each of the two regions separately with high resolution, but embedded
in the outer $120h^{-1}$Mpc box to properly take into account large-scale tidal field
and appropriate boundary conditions at the surface of the refined region.
We name the simulation centered on the cluster ``C" run
and the one centered on the void  ``V" run.
The refined region for ``C" run has a size of $21\times 24\times 20h^{-3}$Mpc$^3$
and that for ``V" run is $31\times 31\times 35h^{-3}$Mpc$^3$.
At their respective volumes, they represent $1.8\sigma$ and $-1.0\sigma$ fluctuations.
The initial condition in the refined region has a mean interparticle-separation of 
$117h^{-1}$kpc comoving, dark matter particle mass of $1.07\times 10^8h^{-1}\msun$.
The refined region is surrounded by two layers (each of $\sim 1h^{-1}$Mpc) 
of buffer zones with 
particle masses successively larger by a factor of $8$ for each layer, 
which then connects with
the outer root grid that has a dark matter particle mass $8^3$ times that in the refined region.
Because we still can not run a very large volume simulation with adequate resolution and physics,
we choose these two runs to represent two opposite environments that possibly bracket the average.
At redshift $z>1.6$, as we will show, the average properties of most quantities concerning DLAs
in ``C" and ``V" runs are not very different,
although the abundances of DLAs in the two runs are already very different.
It is only at lower redshift where we see significant divergence of some quantities of DLAs
between the two runs, presumably due to different dynamic evolutions in the two runs.

We choose the mesh refinement criterion such that the resolution is 
always better than $460h^{-1}$pc physical, corresponding to a maximum
mesh refinement level of $11$ at $z=0$.
We also ran an additional simulation for ``C" run with a factor of two lower resolution
to assess the convergence of the results which we name ``C/2" run and, as we will show in the Appendix,
the convergence is excellent for all quantities examined here.
The simulations include
a metagalactic UV background
\citep[][]{1996Haardt},  
and a model for shielding of UV radiation by neutral hydrogen 
\citep[][]{2005Cen}.
They also include metallicity-dependent radiative cooling 
\citep[][]{1995Cen}.
Star particles are created in cells that satisfy a set of criteria for 
star formation proposed by \citet[][]{1992CenOstriker}.
Each star particle is tagged with its initial mass, creation time, and metallicity; 
star particles typically have masses of $\sim$$10^6\msun$.

Supernova feedback from star formation is modeled following \citet[][]{2005Cen}.
Feedback energy and ejected metal-enriched mass are distributed into 
27 local gas cells centered at the star particle in question, 
weighted by the specific volume of each cell, which is to mimic the physical process of supernova
blastwave propagation that tends to channel energy, momentum and mass into the least dense regions
(with the least resistance and cooling).
We allow the whole feedback processes to be hydrodynamically coupled to surroundings
and subject to relevant physical processes, such as cooling and heating, as in nature.
As we will show later, the extremely inhomogeneous metal enrichment process
demands that both metals and energy (and momentum) are correctly modeled so that they
are transported into right directions in a physically sound (albeit still approximate 
at the current resolution) way.
The primary advantages of this supernova energy based feedback mechanism are three-fold.
First, nature does drive winds in this way and energy input is realistic.
Second, it has only one free parameter $e_{SN}$, 
namely, the fraction of the rest mass energy of stars formed
that is deposited as thermal energy on the cell scale at the location of supernovae.
Third, the processes are treated physically, obeying their respective conservation laws (where they apply),
allowing transport of metals, mass, energy and momentum to be treated self-consistently 
and taking into account relevant heating/cooling processes at all times.
We use $e_{SN}=1\times 10^{-5}$ in these simulations.
The total amount of explosion kinetic energy from Type II supernovae
with a Chabrier IMF translates to $e_{GSW}=6.6\times 10^{-6}$.
Observations of local starburst galaxies indicate
that nearly all of the star formation produced kinetic energy (due to Type II supernovae)
is used to power GSW \citep[e.g.,][]{2001Heckman}.
Given the uncertainties on the evolution of IMF with redshift (i.e., possibly more top heavy at higher redshift)
and the fact that newly discovered prompt Type I supernovae contribute a comparable
amount of energy compared to Type II supernovae, it seems that our adopted value for
$e_{SN}$ is consistent with observations and within physical plausibility.

We use the following cosmological parameters that are consistent with 
the WMAP7-normalized \citep[][]{2010Komatsu} LCDM model:
$\Omega_M=0.28$, $\Omega_b=0.046$, $\Omega_{\Lambda}=0.72$, $\sigma_8=0.82$,
$H_0=100 h {\rm km s}^{-1} {\rm Mpc}^{-1} = 70 {\rm km} s^{-1} {\rm Mpc}^{-1}$ and $n=0.96$.

Convergence test of results are presented separately in Appendix A in order not to 
disrupt the flow of the presentation in the results section.
The tests show that our results are quite converged and should be robust 
at the accuracies concerned here,
suggesting that our resolution has reached an adequate level
for the present study.
The reader may go to the Appendix any time to gauge the convergence of 
relevant computed quantities.
The fact that most of the contributions to DLA incidence come from galaxies of mass $\sim 10^{11}\msun$
that are well above our resolution, the results of our convergence tests are self-consistent.

\subsection{Simulated Galaxy Catalogs}

We identify galaxies in our high resolution simulations using the HOP algorithm 
\citep[][]{1999Eisenstein}, operated on the stellar particles, which is tested to be robust
and insensitive to specific choices of concerned parameters within reasonable ranges.
Satellites within a galaxy are clearly identified separately.
The luminosity of each stellar particle at each of the Sloan Digital Sky Survey (SDSS) five bands 
is computed using the GISSEL stellar synthesis code \citep[][]{Bruzual03}, 
by supplying the formation time, metallicity and stellar mass.
Collecting luminosity and other quantities of member stellar particles, gas cells and dark matter 
particles yields
the following physical parameters for each galaxy:
position, velocity, total mass, stellar mass, gas mass, 
mean formation time, 
mean stellar metallicity, mean gas metallicity,
star formation rate,
luminosities in five SDSS bands (and various colors) and others.

\subsection{Simulated Damped Lyman Alpha System Samples}

While our simulations also solve relevant gas chemistry 
chains for molecular hydrogen formation \citep[][]{1997Abel},
molecular formation on dust grains \citep[][]{2009Joung}
and metal cooling extended down to $10~$K \citep[][]{1972Dalgarno},
at the resolution of the simulations,
molecular clouds are not properly modeled.
To correct for that, we use the \citet{2002Hidaka} observation that 
at $n_c=5$HI/cm$^3$ \h2 fraction is about 50\% 
and then implement the following prescription to 
remove neutral gas in extrapolated high density regions and put it in \h2 phase.
In detail, we assume that the density profile is isothermal below
our resolution, which would translate 
the fraction of mass in \h2 is $\min(1,0.5(n_c/n_{\rm res})^{-1/2})$.
Thus, we post-process the neutral neutral density in the simulation
by the following transformation: $n_{HI}({\rm after}) = n_{HI}({\rm before}) (1-\min(1,0.5(n_c/n_{\rm HI}({\rm before}))^{-1/2}))$,
where $n_{HI}({\rm before})$ is the HI density directly from the simulation,
and $n_{HI}({\rm after})$ is that after this processing step.
A very precise choice of the parameter in the above equation is unimportant;
changing $0.5$ to $1.0$ makes marginally noticeable differences in the results.
The primary effect of doing this is to remove very high HI column DLAs
and causes the HI column density distribution function to steepen at $N_{HI}\ge 22.5$,
in agreement with observations.
In addition, because of that, the total amount of neutral gas in DLAs also become convergent and more stable.

After the above post-process step,
we shoot rays through the entire refined region of each simulation  along all three orthogonal directions 
using a cell size of $0.915h^{-1}$kpc comoving.
In practice, this is done piece-wise, one small volume of the simulation box at a time,
due to limited computer memory.  The spectral bin size is $3$km/s.
All physical effects are taken into account, including temperature broadening and peculiar velocities.
Both intrinsic Lorentzian line profile and Doppler broadening are taken into account
for both $\lya$ and \ion{Si}{2} $\lambda$1808 line, although, in practice, for DLAs, Doppler broadening is important for 
\ion{Si}{2} $\lambda$1808 line 
and Lorentzian profile for $\lya$ line.
All relevant atomic data are taken from \citet[][]{2003Morton}.
A DLA is defined, as usual, a system with HI column larger than $10^{20.3}$cm$^{-2}$.
We assume that the fractional abundance of \siii
is equal to fractional abundance of HI.
Since, as we will see later, the HI regions of DLAs are ``peaky" with well-defined line-of-sight boundaries 
and since DLAs are very optically opaque to ionizing photons,
any refined treatment of radiative self-shielding etc is unlikely to have any significant effect.
Note that we have already included a crude self-shielding method during the simulation,
which should work well for optically opaque regions.
As a side, one numerical point to note is that, because of the very large dynamic range of both line 
cross sections as a function
of frequency shift from the line center and the delta function like cross section shapes in the line core regions,
the convolution operations involved in the detailed calculations of optical depths require at least 
$64-$bits precision for floating point numbers.

For each DLA, we compute the HI column weighted metallicity,
register its position relative to the center of the primary galaxy (i.e, the impact parameter),
and for DLAs that are physically connected by at least one cell side in projection
we merge them and in the end compute projected area $A$
of each connected region to define it size $r_{\rm DLA}=(A/\pi)^{1/2}$.
For each galaxy we also register the maximum velocity width $v_{\rm 90,max}$ among its associated DLAs.

We are able to identify more than one million DLAs through ray tracing at each redshift examined 
in each of the runs.
So the statistical errors are very small for each specific run at any redshift.
But that does not speak to cosmic variance and as we shall show later,
cosmic variance is indeed quite large concerning quantities that directly or indirectly pertain
to the number density of DLAs.
Other quantities, such as size, metallicity, kinematic properties, etc., however,
appear to depend weakly on environments and their variances are small.
A DLA ``belongs" to the largest galaxy in the region,
within whose virial radius the DLA lies.
For example, a DLA that is physically more closely located to a satellite galaxy
that in turn is within the virial radius of a larger galaxy
is said to belong to that larger galaxy.

\subsection{Kinematic Measures for \siii Line}

We do not add instrumental noise to the simulated spectra,
but we adopt the same observational procedure to 
compute the kinematic measures for the \siii absorption lines.
For all relevant measures for the \siii line,
we follow identically the procedures and definitions in \citet[][]{1997Prochaska}.
We generate synthetic spectra for both $\lya$ and \siii line with $3$km/s pixels
and then smooth it with a 9-pixel boxcar averaging procedure.
We define the velocity width of a \siii absorption line 
associated with a DLA to be the velocity interval of 90\% of the total optical depth, $v_{90}$.
For the three kinematic shape measures for \siii line
we use all intensity troughs (optical depth peaks) 
without the $0.1 \le I(v_{pk})/\bar I \le 0.6$ constraint,
where $\bar I$ is the continuum flux, as re-emphasized by \citet[][]{2010bProchaska}.
The kinematic shape measures, $f_{\rm mm}$, $f_{\rm edg}$
and $f_{\rm 2pk}$, are defined exactly the same way as in \citet[][]{1997Prochaska}.

\section{Results}

\subsection{A Garden Variety of DLAs}

\begin{figure}[ht]
\centering
\vskip \pictureskip cm
\resizebox{\picturesize in}{!}{\includegraphics[angle=0]{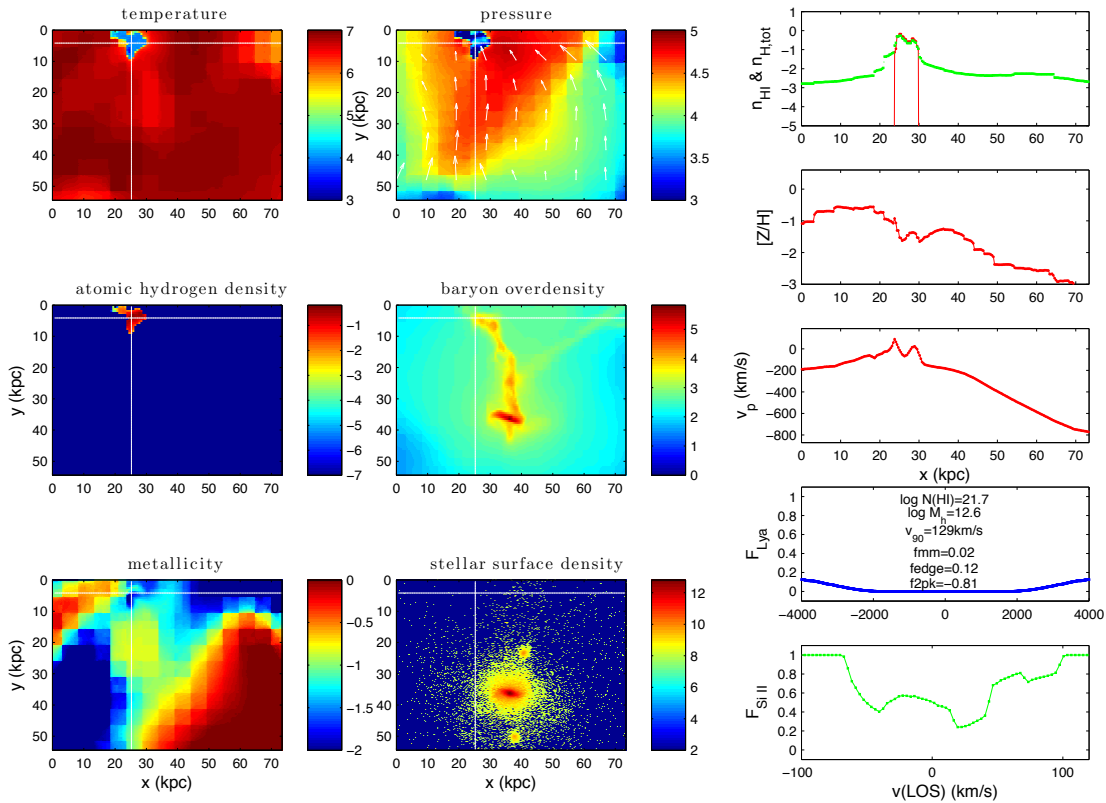}}
\vskip \picturecapskip cm
\caption{\footnotesize 
Top left: temperature (K); 
middle left: atomic hydrogen density (cm$^{-3}$); 
bottom left: metallicity (solar units); 
top middle: pressure (Kelvin cm$^{-3}$);
the above maps have a thickness of $1.3$kpc.
Middle middle: baryonic overdensity;
bottom middle: SDSS U band luminosity surface density ($\lsun$/kpc$^2$);
these two maps are projected over the virial diameter of the galaxy. 
Included in pressure map  
is peculiar velocity field with $5$kpc corresponding to $500$km/s.
The five panels on the right column, from top to bottom, are:
atomic hydrogen density (cm$^{-3}$; red solid curve)
with total hydrogen density (dotted green curve), 
gas metallicity (solar units), 
LOS proper peculiar velocity,
$\lya$ flux and finally \ion{Si}{2} $\lambda$1808 flux.
The top three panels are plotted against physical distance,
whereas the bottom two versus LOS velocity.
Indicated in the second from bottom panel
are properties of the DLA: $\log$ N(HI),
$\log {\rm M}_h$, $v_{90}$, $f_{\rm mm}$, $f_{\rm edg}$, $f_{\rm 2pk}$.
}
\label{fig:pic2}
\end{figure}

\begin{figure}[ht]
\centering
\vskip \pictureskip cm
\resizebox{\picturesize in}{!}{\includegraphics[angle=0]{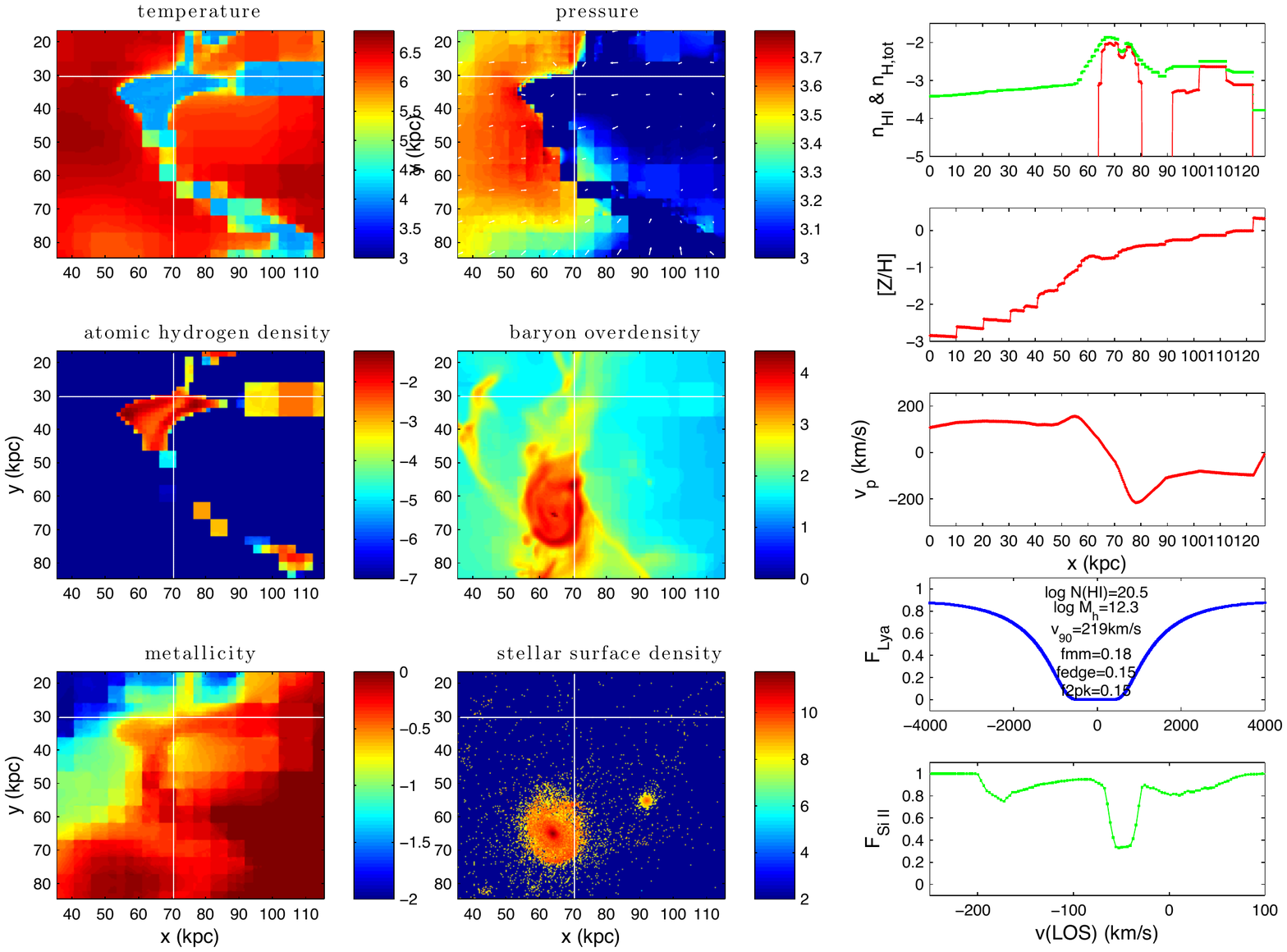}}
\vskip \picturecapskip cm
\caption{\footnotesize 
Top left: temperature (K); 
middle left: atomic hydrogen density (cm$^{-3}$); 
bottom left: metallicity (solar units); 
top middle: pressure (Kelvin cm$^{-3}$);
the above maps have a thickness of $1.3$kpc.
Middle middle: baryonic overdensity;
bottom middle: SDSS U band luminosity surface density ($\lsun$/kpc$^2$);
these two maps are projected over the virial diameter of the galaxy. 
Included in pressure map  
is peculiar velocity field with $5$kpc corresponding to $500$km/s.
The five panels on the right column, from top to bottom, are:
atomic hydrogen density (cm$^{-3}$; red solid curve)
with total hydrogen density (dotted green curve), 
gas metallicity (solar units), 
LOS proper peculiar velocity,
$\lya$ flux and \ion{Si}{2} $\lambda$1808 flux.
The top three panels are plotted against physical distance,
whereas the bottom two versus LOS velocity.
Indicated in the second from bottom panel
are properties of the DLA: $\log$ N(HI),
$\log {\rm M}_h$, $v_{90}$, $f_{\rm mm}$, $f_{\rm edg}$, $f_{\rm 2pk}$.
}
\label{fig:pic3}
\end{figure}

\begin{figure}[ht]
\centering
\vskip \pictureskip cm
\resizebox{\picturesize in}{!}{\includegraphics[angle=0]{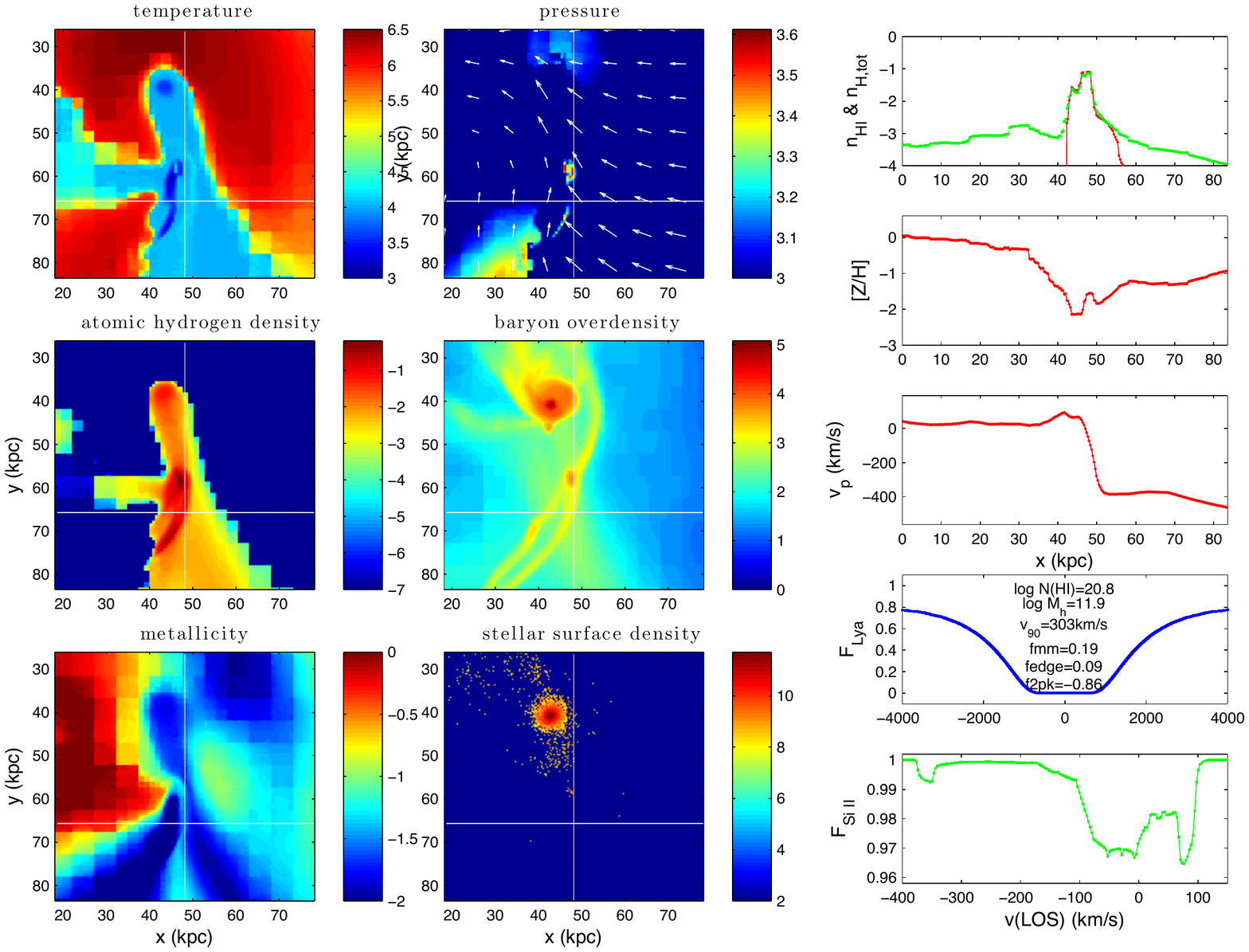}}
\vskip \picturecapskip cm
\caption{\footnotesize 
Top left: temperature (K); 
middle left: atomic hydrogen density (cm$^{-3}$); 
bottom left: metallicity (solar units); 
top middle: pressure (Kelvin cm$^{-3}$);
the above maps have a thickness of $1.3$kpc.
Middle middle: baryonic overdensity;
bottom middle: SDSS U band luminosity surface density ($\lsun$/kpc$^2$);
these two maps are projected over the virial diameter of the galaxy. 
Included in pressure map  
is peculiar velocity field with $5$kpc corresponding to $500$km/s.
The five panels on the right column, from top to bottom, are:
atomic hydrogen density (cm$^{-3}$; red solid curve)
with total hydrogen density (dotted green curve), 
gas metallicity (solar units), 
LOS proper peculiar velocity,
$\lya$ flux and \ion{Si}{2} $\lambda$1808 flux.
The top three panels are plotted against physical distance,
whereas the bottom two versus LOS velocity.
Indicated in the second from bottom panel
are properties of the DLA: $\log$ N(HI),
$\log {\rm M}_h$, $v_{90}$, $f_{\rm mm}$, $f_{\rm edg}$, $f_{\rm 2pk}$.
}
\label{fig:pic5}
\end{figure}

\begin{figure}[ht]
\centering
\vskip \pictureskip cm
\resizebox{\picturesize in}{!}{\includegraphics[angle=0]{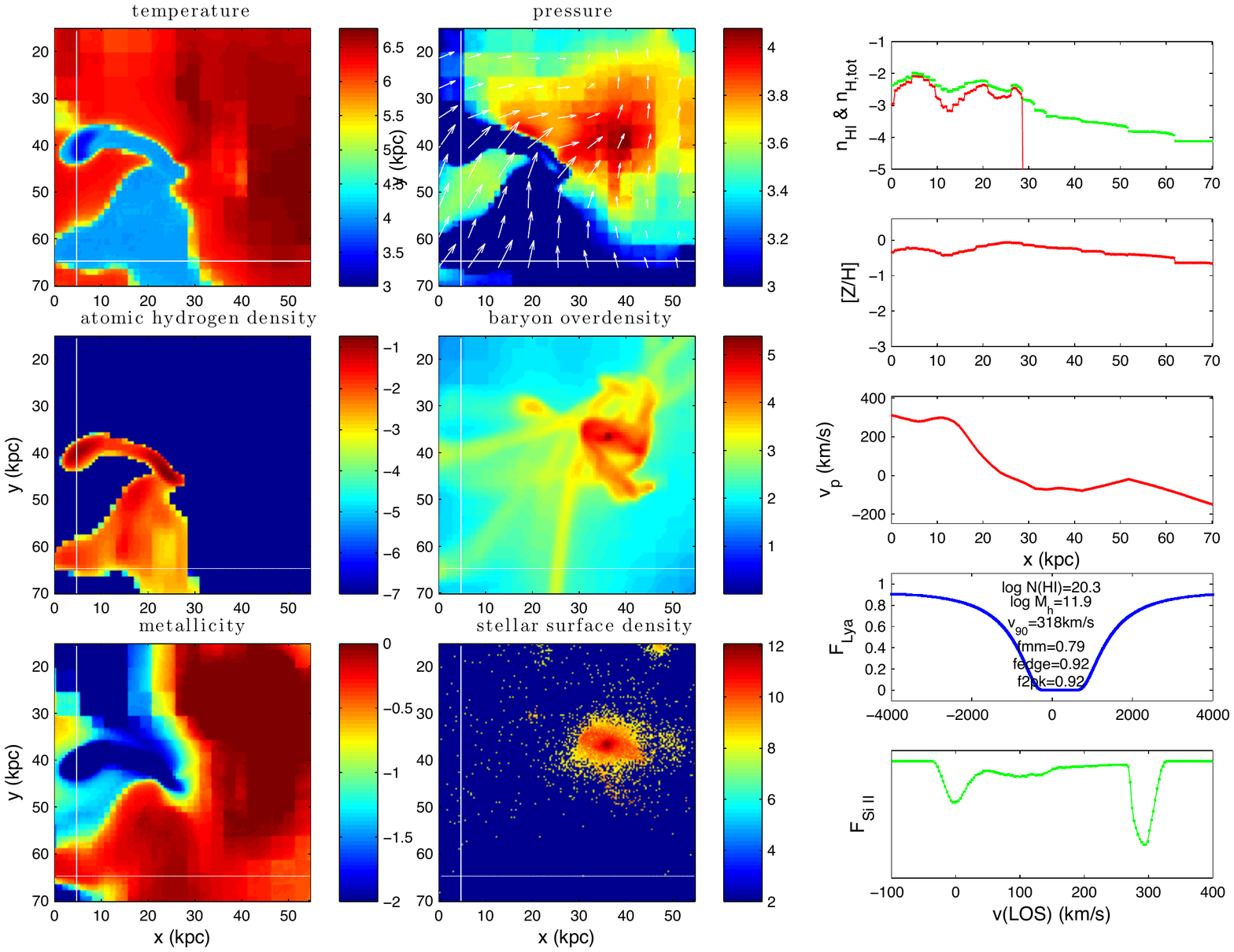}}
\vskip \picturecapskip cm
\caption{\footnotesize 
Top left: temperature (K); 
middle left: atomic hydrogen density (cm$^{-3}$); 
bottom left: metallicity (solar units); 
top middle: pressure (Kelvin cm$^{-3}$);
the above maps have a thickness of $1.3$kpc.
Middle middle: baryonic overdensity;
bottom middle: SDSS U band luminosity surface density ($\lsun$/kpc$^2$);
these two maps are projected over the virial diameter of the galaxy. 
Included in pressure map  
is peculiar velocity field with $5$kpc corresponding to $500$km/s.
The five panels on the right column, from top to bottom, are:
atomic hydrogen density (cm$^{-3}$; red solid curve)
with total hydrogen density (dotted green curve), 
gas metallicity (solar units), 
LOS proper peculiar velocity,
$\lya$ flux and \ion{Si}{2} $\lambda$1808 flux.
The top three panels are plotted against physical distance,
whereas the bottom two versus LOS velocity.
Indicated in the second from bottom panel
are properties of the DLA: $\log$ N(HI),
$\log {\rm M}_h$, $v_{90}$, $f_{\rm mm}$, $f_{\rm edg}$, $f_{\rm 2pk}$.
}
\label{fig:pic7}
\end{figure}

\begin{figure}[ht]
\centering
\vskip \pictureskip cm
\resizebox{\picturesize in}{!}{\includegraphics[angle=0]{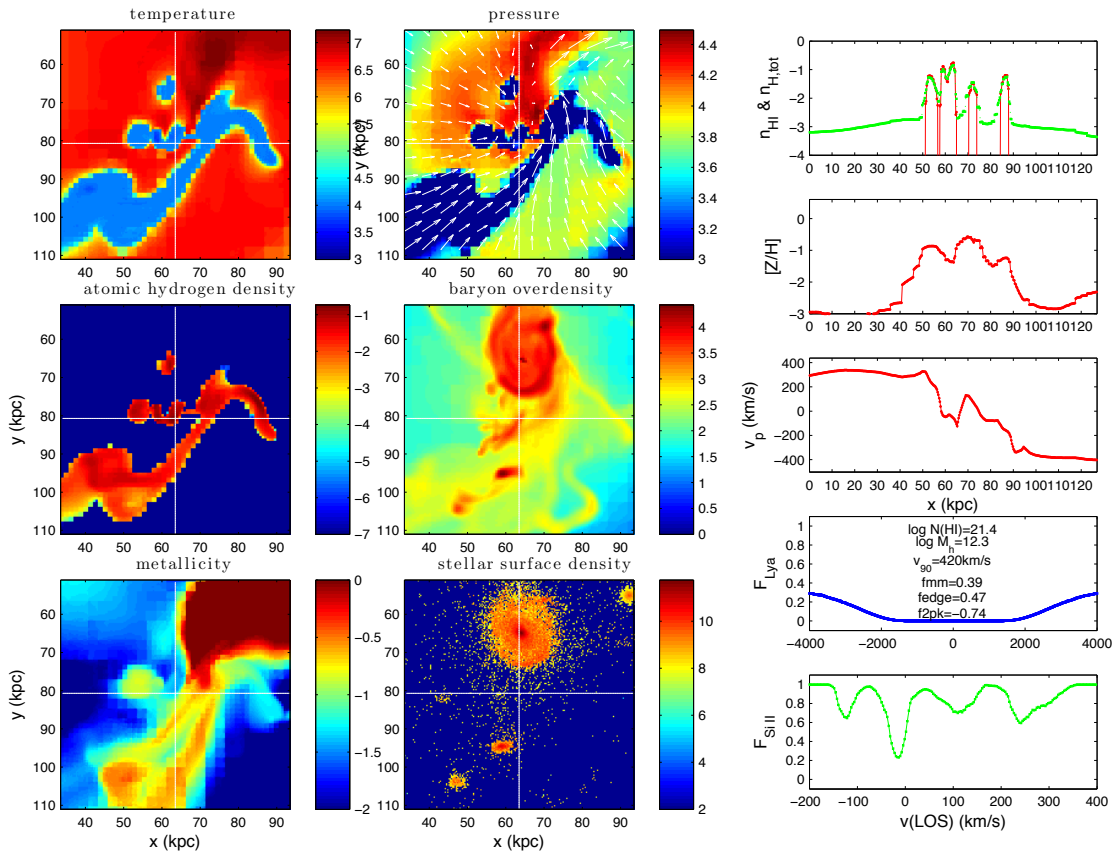}}
\vskip \picturecapskip cm
\caption{\footnotesize 
Top left: temperature (K); 
middle left: atomic hydrogen density (cm$^{-3}$); 
bottom left: metallicity (solar units); 
top middle: pressure (Kelvin cm$^{-3}$);
the above maps have a thickness of $1.3$kpc.
Middle middle: baryonic overdensity;
bottom middle: SDSS U band luminosity surface density ($\lsun$/kpc$^2$);
these two maps are projected over the virial diameter of the galaxy. 
Included in pressure map  
is peculiar velocity field with $5$kpc corresponding to $500$km/s.
The five panels on the right column, from top to bottom, are:
atomic hydrogen density (cm$^{-3}$; red solid curve)
with total hydrogen density (dotted green curve), 
gas metallicity (solar units), 
LOS proper peculiar velocity,
$\lya$ flux and \ion{Si}{2} $\lambda$1808 flux.
The top three panels are plotted against physical distance,
whereas the bottom two versus LOS velocity.
Indicated in the second from bottom panel
are properties of the DLA: $\log$ N(HI),
$\log {\rm M}_h$, $v_{90}$, $f_{\rm mm}$, $f_{\rm edg}$, $f_{\rm 2pk}$.
}
\label{fig:pic8}
\end{figure}

\begin{figure}[ht]
\centering
\vskip \pictureskip cm
\resizebox{\picturesize in}{!}{\includegraphics[angle=0]{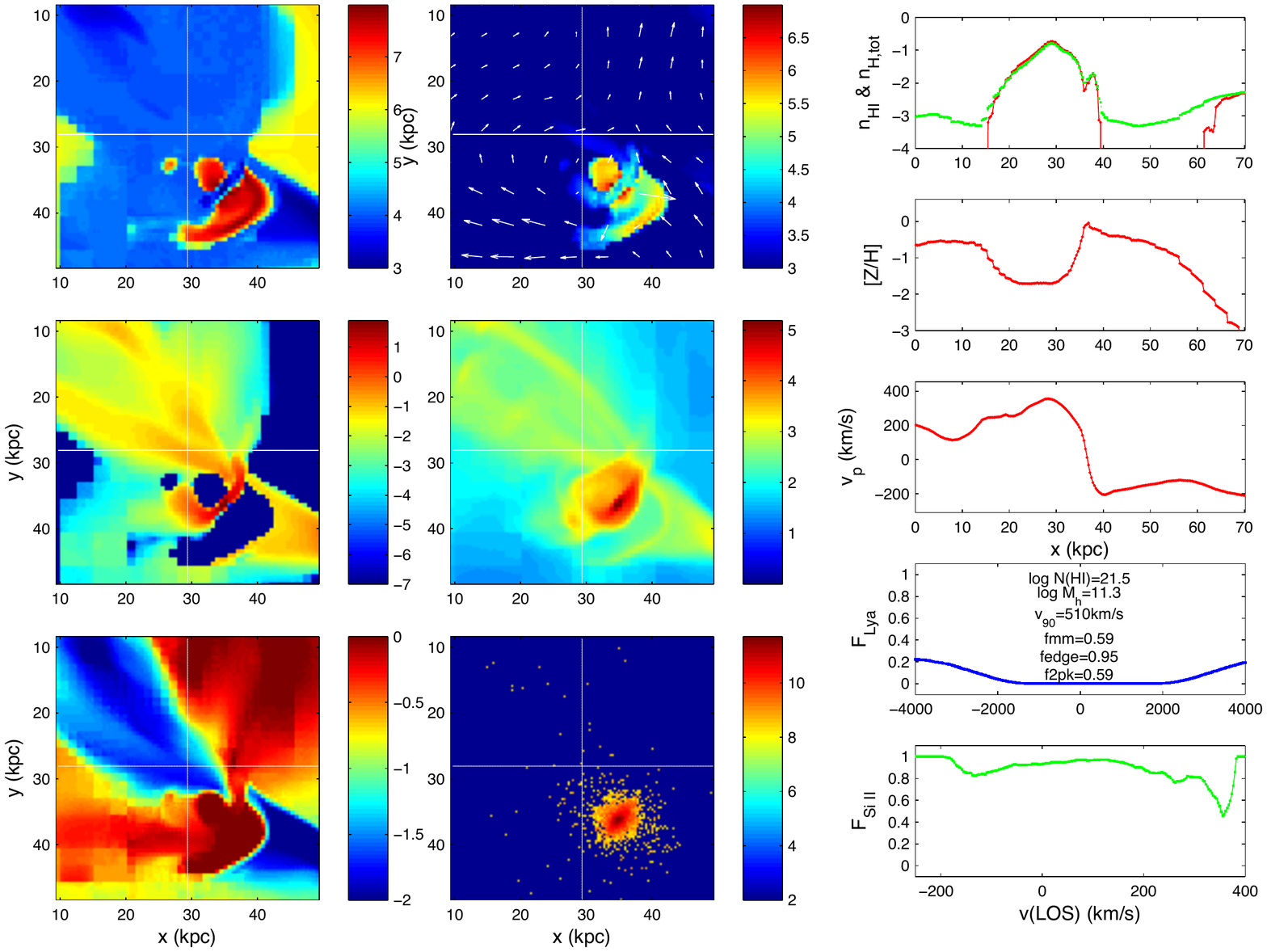}}
\vskip \picturecapskip cm
\caption{\footnotesize 
Top left: temperature (K); 
middle left: atomic hydrogen density (cm$^{-3}$); 
bottom left: metallicity (solar units); 
top middle: pressure (Kelvin cm$^{-3}$);
the above maps have a thickness of $1.3$kpc.
Middle middle: baryonic overdensity;
bottom middle: SDSS U band luminosity surface density ($\lsun$/kpc$^2$);
these two maps are projected over the virial diameter of the galaxy. 
Included in pressure map  
is peculiar velocity field with $5$kpc corresponding to $500$km/s.
The five panels on the right column, from top to bottom, are:
atomic hydrogen density (cm$^{-3}$; red solid curve)
with total hydrogen density (dotted green curve), 
gas metallicity (solar units), 
LOS proper peculiar velocity,
$\lya$ flux and \ion{Si}{2} $\lambda$1808 flux.
The top three panels are plotted against physical distance,
whereas the bottom two versus LOS velocity.
Indicated in the second from bottom panel
are properties of the DLA: $\log$ N(HI),
$\log {\rm M}_h$, $v_{90}$, $f_{\rm mm}$, $f_{\rm edg}$, $f_{\rm 2pk}$.
}
\label{fig:pic9}
\end{figure}

\begin{figure}[ht]
\centering
\vskip \pictureskip cm
\resizebox{\picturesize in}{!}{\includegraphics[angle=0]{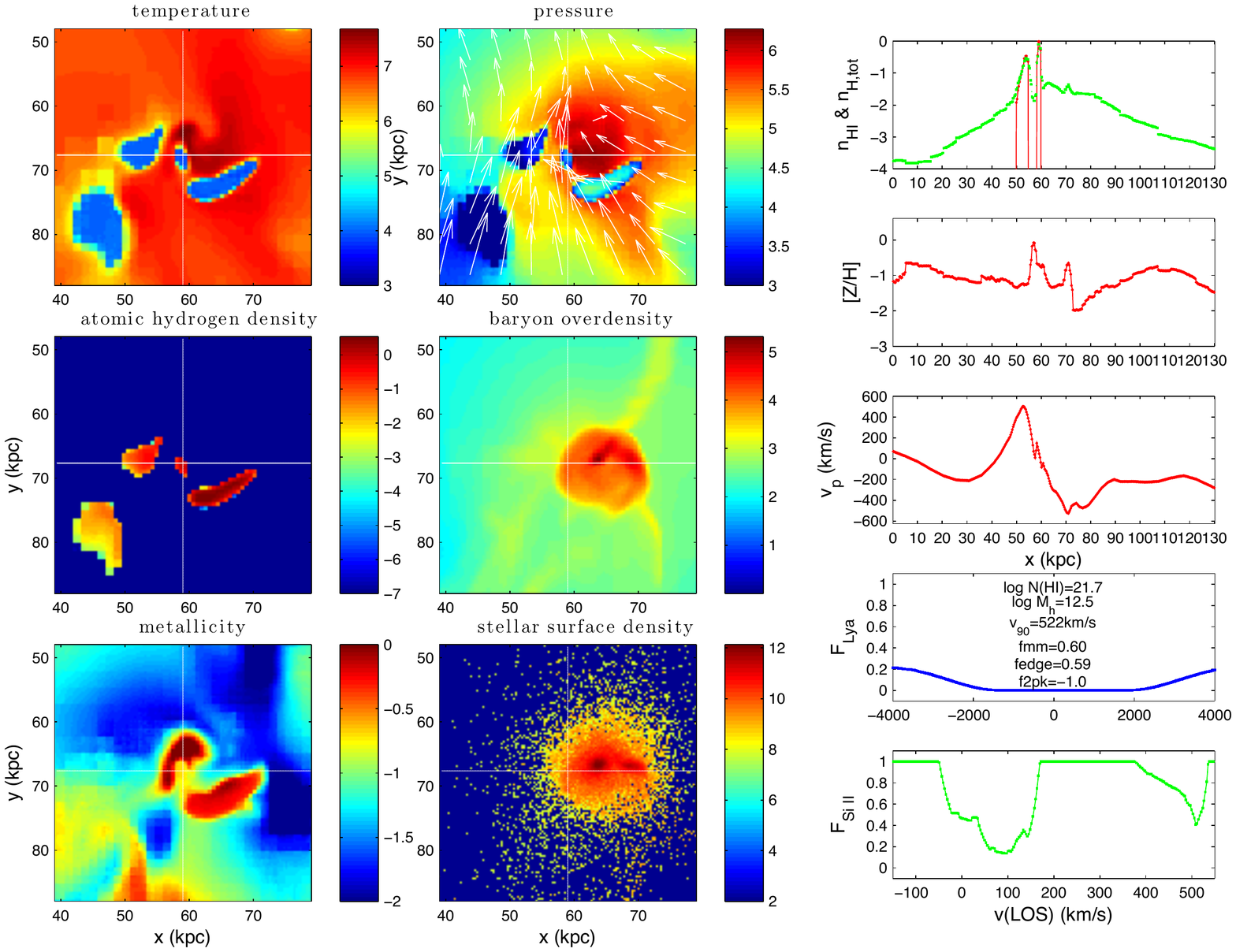}}
\vskip \picturecapskip cm
\caption{\footnotesize 
Top left: temperature (K); 
middle left: atomic hydrogen density (cm$^{-3}$); 
bottom left: metallicity (solar units); 
top middle: pressure (Kelvin cm$^{-3}$);
the above maps have a thickness of $1.3$kpc.
Middle middle: baryonic overdensity;
bottom middle: SDSS U band luminosity surface density ($\lsun$/kpc$^2$);
these two maps are projected over the virial diameter of the galaxy. 
Included in pressure map  
is peculiar velocity field with $5$kpc corresponding to $500$km/s.
The five panels on the right column, from top to bottom, are:
atomic hydrogen density (cm$^{-3}$; red solid curve)
with total hydrogen density (dotted green curve), 
gas metallicity (solar units), 
LOS proper peculiar velocity,
$\lya$ flux and \ion{Si}{2} $\lambda$1808 flux.
The top three panels are plotted against physical distance,
whereas the bottom two versus LOS velocity.
Indicated in the second from bottom panel
are properties of the DLA: $\log$ N(HI),
$\log {\rm M}_h$, $v_{90}$, $f_{\rm mm}$, $f_{\rm edg}$, $f_{\rm 2pk}$.
}
\label{fig:pic10}
\end{figure}

\begin{figure}[ht]
\centering
\vskip \pictureskip cm
\resizebox{\picturesize in}{!}{\includegraphics[angle=0]{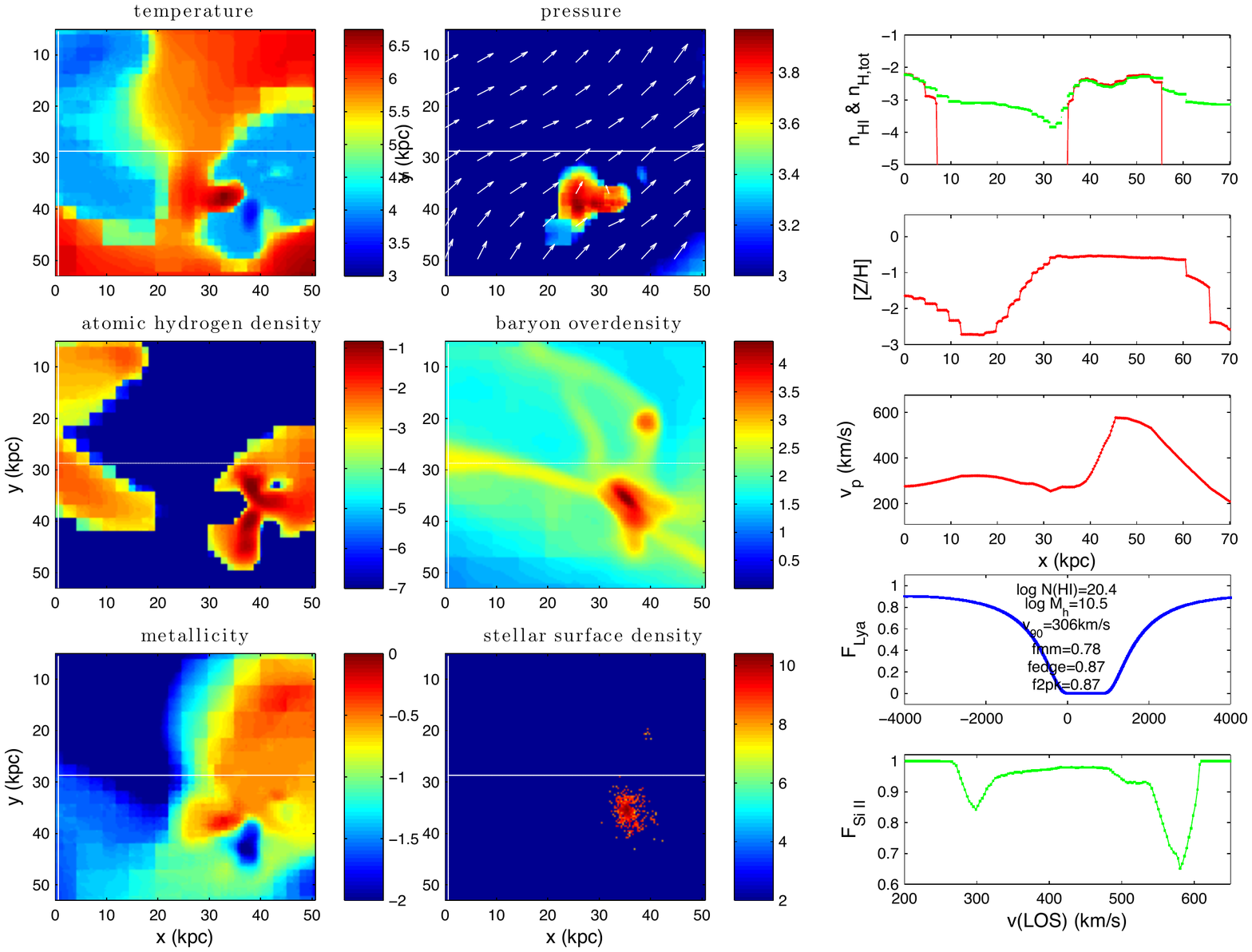}}
\vskip \picturecapskip cm
\caption{\footnotesize 
Top left: temperature (K); 
middle left: atomic hydrogen density (cm$^{-3}$); 
bottom left: metallicity (solar units); 
top middle: pressure (Kelvin cm$^{-3}$);
the above maps have a thickness of $1.3$kpc.
Middle middle: baryonic overdensity;
bottom middle: SDSS U band luminosity surface density ($\lsun$/kpc$^2$);
these two maps are projected over the virial diameter of the galaxy. 
Included in pressure map  
is peculiar velocity field with $5$kpc corresponding to $500$km/s.
The five panels on the right column, from top to bottom, are:
atomic hydrogen density (cm$^{-3}$; red solid curve)
with total hydrogen density (dotted green curve), 
gas metallicity (solar units), 
LOS proper peculiar velocity,
$\lya$ flux and \ion{Si}{2} $\lambda$1808 flux.
The top three panels are plotted against physical distance,
whereas the bottom two versus LOS velocity.
Indicated in the second from bottom panel
are properties of the DLA: $\log$ N(HI),
$\log {\rm M}_h$, $v_{90}$, $f_{\rm mm}$, $f_{\rm edg}$, $f_{\rm 2pk}$.
}
\label{fig:pica3}
\end{figure}

\begin{figure}[ht]
\centering
\vskip \pictureskip cm
\resizebox{\picturesize in}{!}{\includegraphics[angle=0]{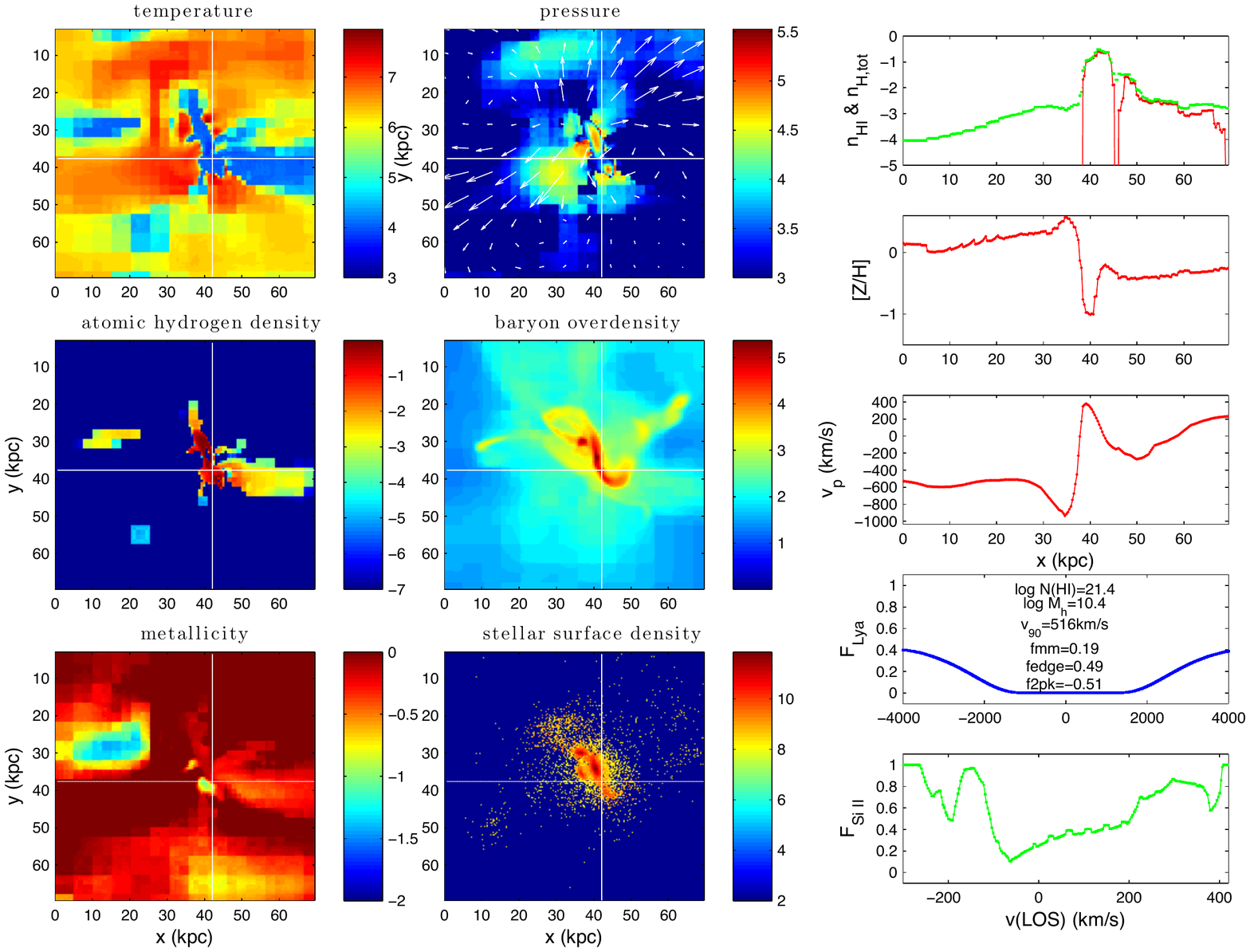}}
\vskip \picturecapskip cm
\caption{\footnotesize 
Top left: temperature (K); 
middle left: atomic hydrogen density (cm$^{-3}$); 
bottom left: metallicity (solar units); 
top middle: pressure (Kelvin cm$^{-3}$);
the above maps have a thickness of $1.3$kpc.
Middle middle: baryonic overdensity;
bottom middle: SDSS U band luminosity surface density ($\lsun$/kpc$^2$);
these two maps are projected over the virial diameter of the galaxy. 
Included in pressure map  
is peculiar velocity field with $5$kpc corresponding to $500$km/s.
The five panels on the right column, from top to bottom, are:
atomic hydrogen density (cm$^{-3}$; red solid curve)
with total hydrogen density (dotted green curve), 
gas metallicity (solar units), 
LOS proper peculiar velocity,
$\lya$ flux and \ion{Si}{2} $\lambda$1808 flux.
The top three panels are plotted against physical distance,
whereas the bottom two versus LOS velocity.
Indicated in the second from bottom panel
are properties of the DLA: $\log$ N(HI),
$\log {\rm M}_h$, $v_{90}$, $f_{\rm mm}$, $f_{\rm edg}$, $f_{\rm 2pk}$.
}
\label{fig:pica4}
\end{figure}

\begin{figure}[ht]
\centering
\vskip \pictureskip cm
\resizebox{\picturesize in}{!}{\includegraphics[angle=0]{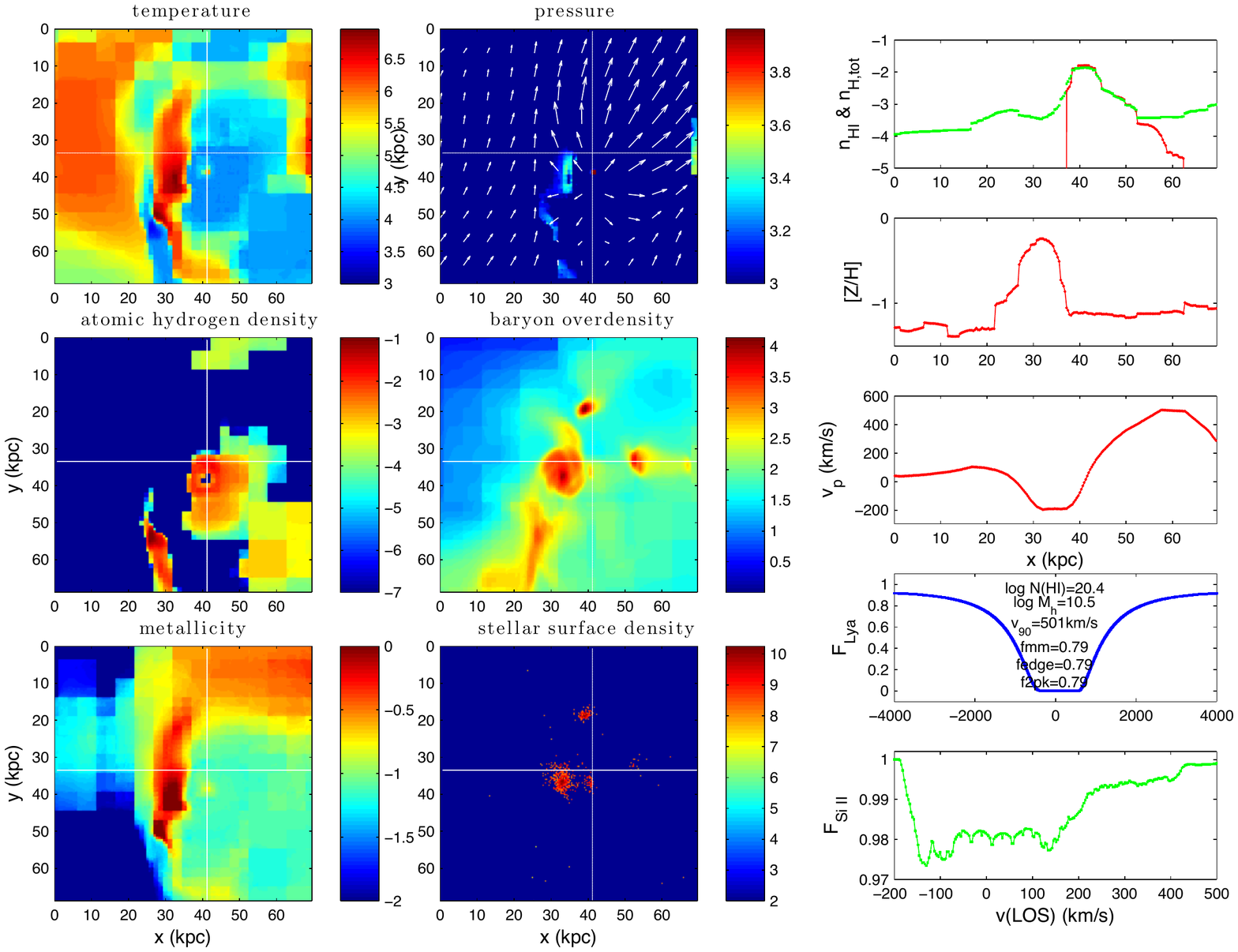}}
\vskip \picturecapskip cm
\caption{\footnotesize 
Top left: temperature (K); 
middle left: atomic hydrogen density (cm$^{-3}$); 
bottom left: metallicity (solar units); 
top middle: pressure (Kelvin cm$^{-3}$);
the above maps have a thickness of $1.3$kpc.
Middle middle: baryonic overdensity;
bottom middle: SDSS U band luminosity surface density ($\lsun$/kpc$^2$);
these two maps are projected over the virial diameter of the galaxy. 
Included in pressure map  
is peculiar velocity field with $5$kpc corresponding to $500$km/s.
The five panels on the right column, from top to bottom, are:
atomic hydrogen density (cm$^{-3}$; red solid curve)
with total hydrogen density (dotted green curve), 
gas metallicity (solar units), 
LOS proper peculiar velocity,
$\lya$ flux and \ion{Si}{2} $\lambda$1808 flux.
The top three panels are plotted against physical distance,
whereas the bottom two versus LOS velocity.
Indicated in the second from bottom panel
are properties of the DLA: $\log$ N(HI),
$\log {\rm M}_h$, $v_{90}$, $f_{\rm mm}$, $f_{\rm edg}$, $f_{\rm 2pk}$.
}
\label{fig:pica5}
\end{figure}

\begin{figure}[ht]
\centering
\vskip \pictureskip cm
\resizebox{\picturesize in}{!}{\includegraphics[angle=0]{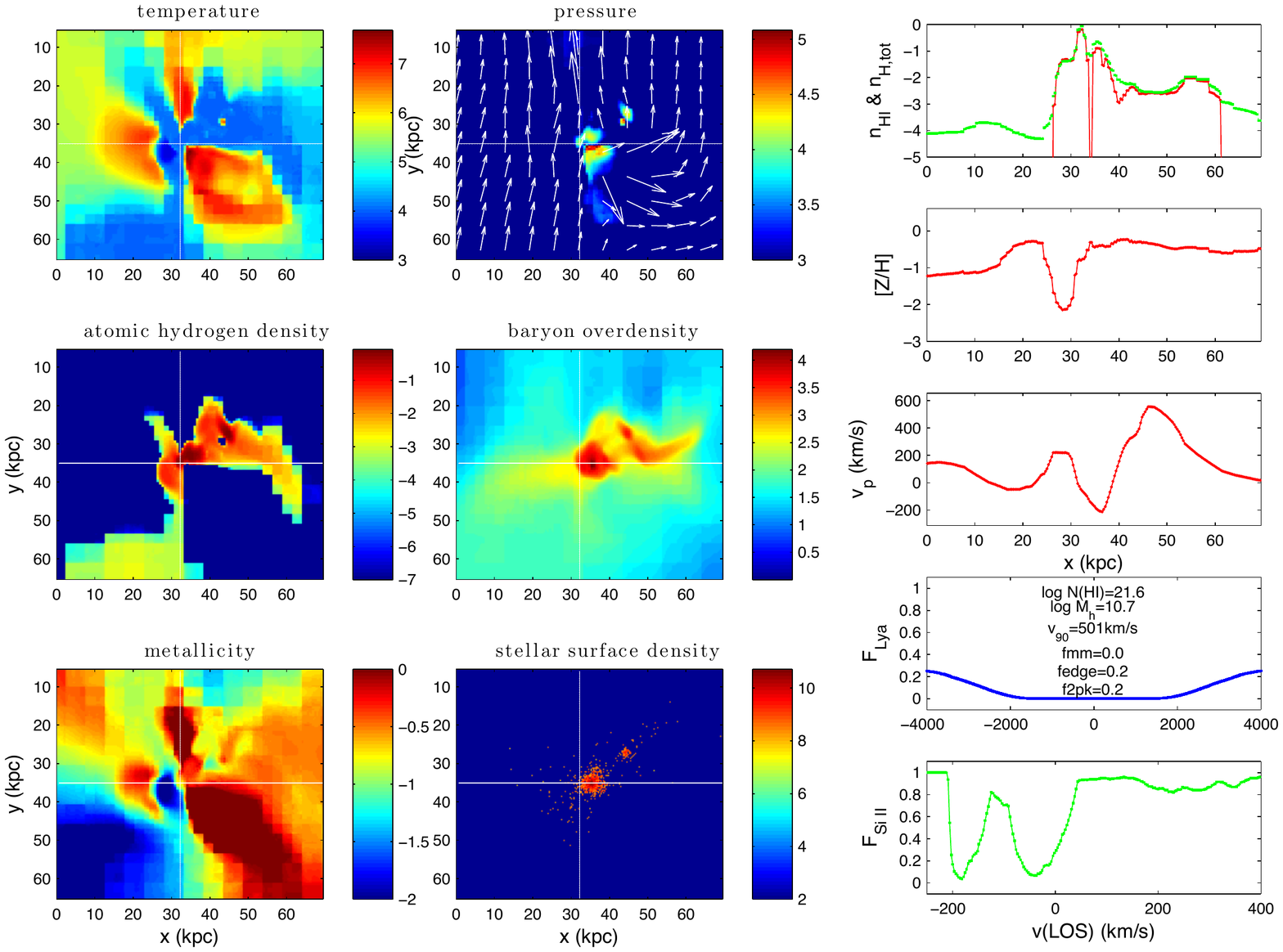}}
\vskip \picturecapskip cm
\caption{\footnotesize 
Top left: temperature (K); 
middle left: atomic hydrogen density (cm$^{-3}$); 
bottom left: metallicity (solar units); 
top middle: pressure (Kelvin cm$^{-3}$);
the above maps have a thickness of $1.3$kpc.
Middle middle: baryonic overdensity;
bottom middle: SDSS U band luminosity surface density ($\lsun$/kpc$^2$);
these two maps are projected over the virial diameter of the galaxy. 
Included in pressure map  
is peculiar velocity field with $5$kpc corresponding to $500$km/s.
The five panels on the right column, from top to bottom, are:
atomic hydrogen density (cm$^{-3}$; red solid curve)
with total hydrogen density (dotted green curve), 
gas metallicity (solar units), 
LOS proper peculiar velocity,
$\lya$ flux and \ion{Si}{2} $\lambda$1808 flux.
The top three panels are plotted against physical distance,
whereas the bottom two versus LOS velocity.
Indicated in the second from bottom panel
are properties of the DLA: $\log$ N(HI),
$\log {\rm M}_h$, $v_{90}$, $f_{\rm mm}$, $f_{\rm edg}$, $f_{\rm 2pk}$.
}
\label{fig:pica6}
\end{figure}

\begin{figure}[ht]
\centering
\vskip \pictureskip cm
\resizebox{\picturesize in}{!}{\includegraphics[angle=0]{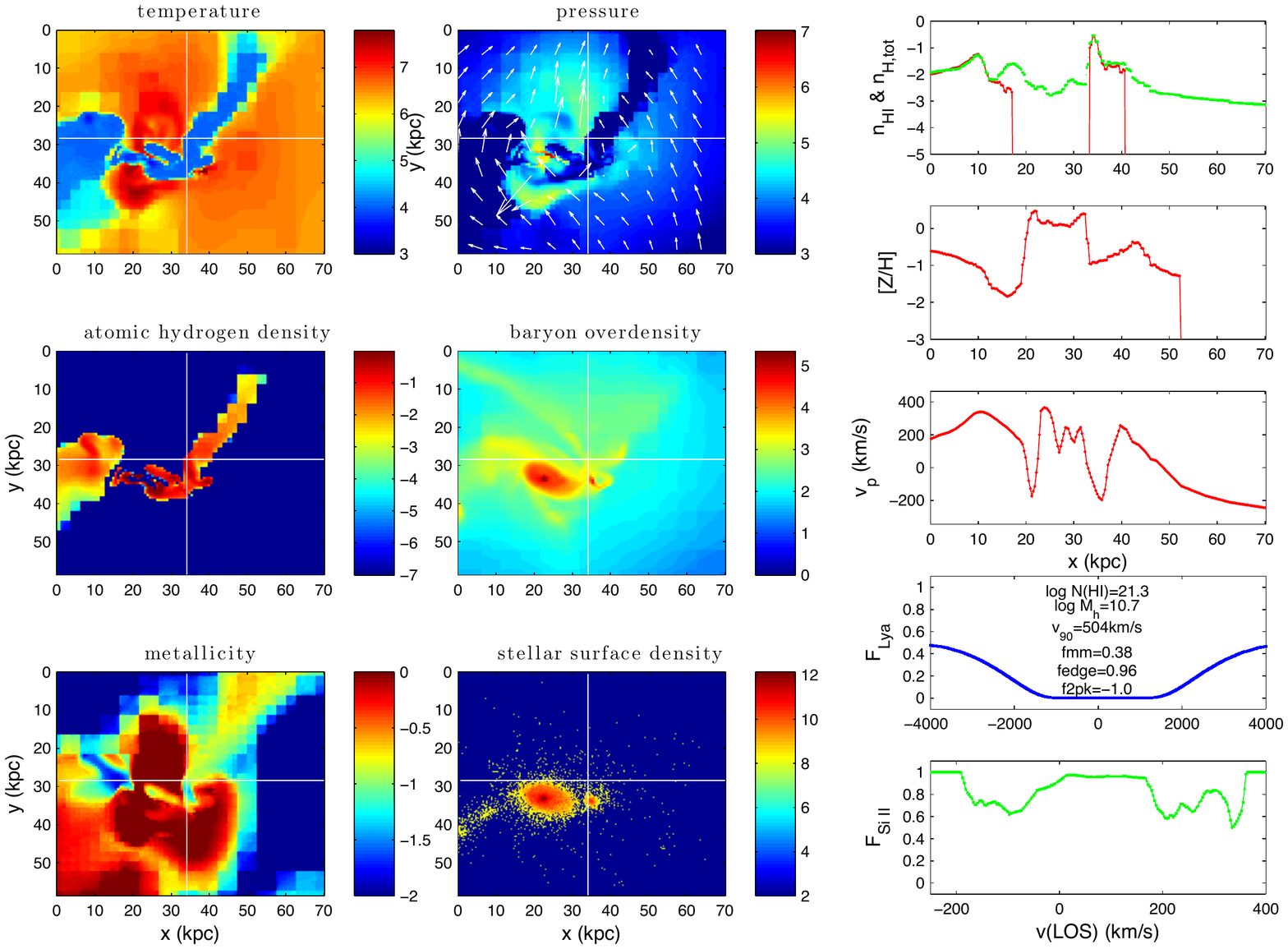}}
\vskip \picturecapskip cm
\caption{\footnotesize 
Top left: temperature (K); 
middle left: atomic hydrogen density (cm$^{-3}$); 
bottom left: metallicity (solar units); 
top middle: pressure (Kelvin cm$^{-3}$);
the above maps have a thickness of $1.3$kpc.
Middle middle: baryonic overdensity;
bottom middle: SDSS U band luminosity surface density ($\lsun$/kpc$^2$);
these two maps are projected over the virial diameter of the galaxy. 
Included in pressure map  
is peculiar velocity field with $5$kpc corresponding to $500$km/s.
The five panels on the right column, from top to bottom, are:
atomic hydrogen density (cm$^{-3}$; red solid curve)
with total hydrogen density (dotted green curve), 
gas metallicity (solar units), 
LOS proper peculiar velocity,
$\lya$ flux and \ion{Si}{2} $\lambda$1808 flux.
The top three panels are plotted against physical distance,
whereas the bottom two versus LOS velocity.
Indicated in the second from bottom panel
are properties of the DLA: $\log$ N(HI),
$\log {\rm M}_h$, $v_{90}$, $f_{\rm mm}$, $f_{\rm edg}$, $f_{\rm 2pk}$.
}
\label{fig:pica7}
\end{figure}

We first present a gallery of twelve DLAs at $z=3.1$ 
(Figure~\ref{fig:pic2}-\ref{fig:pica7}) 
to show the richness 
of their physical properties.
For each DLA six maps and five quantitative panels are displayed.
In each of the six maps the line of sight (LOS) intercepting the DLA is shown
as a white horizontal line and the exact location of 
the primary component of the DLA is at the intersection with another, white vertical line.
In cases with multiple components along the LOS, the primary component coincides with
the highest neutral density.
Four of the maps - top left (temperature in Kelvin), 
middle left (atomic hydrogen density in cm$^{-3}$), 
bottom left (metallicity in solar units) and 
top middle (pressure in units of Kelvin cm$^{-3}$) -
have a physical thickness of $1.3$kpc. 
Also indicated in top middle (pressure)
is the peculiar velocity field with a scaling of $5$kpc corresponding to
$500$km/s.
The remaining two maps - middle middle (baryonic overdensity) and bottom middle (stellar surface density in $\msun$/kpc$^2$) -
are projected over the entire galaxy of depth of order of the virial diameter of the 
primary galaxy. 
While these two projected maps give an overall indication of relative projected location of the DLA 
respect to the galaxy, the exact depth of the DLA inside the paper is, however, not shown.
When we quote distance from the galaxy, we mean the projected distance on the paper plane.

The five panels on the right column show various physical quantities
along the line of sight (i.e, along the white horizontal line shown
in the maps on the left two columns).
From top to bottom they are:
atomic hydrogen density (in cm$^{-3}$; red solid curve with a narrower shape)
along with total hydrogen density (dotted green curve with a more extended shape), 
gas metallicity (in solar units), 
line-of-sight proper peculiar velocity (in km/s),
$\lya$ flux
and flux for \ion{Si}{2} $\lambda$1808 line.
The top three panels are plotted against physical distance,
whereas the bottom two panels are plotted versus the LOS velocity.
Also indicated in the $\lya$ flux panel (second from bottom)
are several quantitative measures of the DLA,
including the neutral hydrogen column density ($\log$ N(HI)),
the halo mass of the primary galaxy in the system ($\log {\rm M}_h$),
the velocity width of the associated \siii line ($v_{90}$) 
and three kinetic measures of the \siii line, 
$f_{\rm mm}$, $f_{\rm edg}$, $f_{\rm 2pk}$.
We now describe in turn each of the twelve DLA examples.

Figure~\ref{fig:pic2} shows a DLA produced by the LOS intersecting 
the tip of a long chimney at a distance of $\sim 30$kpc, 
The velocity structure suggests that it is still moving away (upwards) 
from the galaxy at a velocity of $\sim 500$km/s, 
likely caused by galactic winds.
The metallicity at the interception is $[Z/H]\sim -1.5$ 
but there are large gradients and variations of metallicity 
(we found that some 
other very nearby DLA systems intersecting different parts of the chimney 
have metallicity $[Z/H]<-3$, not shown),
suggesting very inhomogeneous enrichment process by galactic winds.
While the primary galaxy has a mass of $4\times 10^{12}\msun$, i.e., 
a 1-d velocity dispersion of $500$km/s,
the kinetic width of this line is only $129$km/s with $N_{\rm HI}=21.7$.
Although the $\lya$ flux appears as a single component,
as will be the case in all subsequent examples,
the \siii absorption has several separate features, reflecting
the two-peak structure of the absorbing column and complex velocity structure within.
We note that the nearby satellite galaxies may have triggered the starburst
and the galactic winds.
The responsible gas for this DLA is probably cooling and 
confined by external pressure likely due to thermal instability, as seen 
in the pressure panel.

In Figure~\ref{fig:pic3} a DLA of width $219$km/s is created jointly by 
two major components along the sightline, one at $x\sim 70$kpc
of metallicity of $[Z/H]\sim [-1.0, -0.5]$ and size of $\sim 20$kpc
at an impact parameter of $\sim 30$kpc and the other
at $x\sim 110$kpc of metallicity of $[Z/H]\sim 0.0$ and size of $\sim 30$kpc.
What is striking is many long gaseous structures in this galaxy. 
As we will see frequently, 
there are often long gaseous structures connected with galaxies
that seem always coincidental with visible galaxy interactions
of multiple galaxies or galaxies and satellites in close proximity.
We shall call these features ``galactic filaments" hereafter.
It seems likely that some of these galactic filaments are cold streams
\citep[][]{2005Keres, 2006Dekel}.
However, galactic filaments found in our simulations appear to be very rich in 
variety and disparate in metallicity (spanning 3 decades or more in metallicity).
In other words, they are not necessarily primordial cold streams.
In the case of this DLA, the filaments 
are likely made of gas pre-enriched, having cooled (as the pressure panel shows) and 
now rotating about the galaxy (roughly counter-clockwise).
In contrast, note that the DLA in 
Figure~\ref{fig:pic2} was still moving away from the galaxy.
Like in Figure~\ref{fig:pic2}, the rich galactic filaments 
appear to be associated with significant satellite structures in close proximity.
The \siii absorption has several separate features, reflecting
the two separate physical components as well as substructures within each component.

Figure~\ref{fig:pic5} shows a DLA 
that is associated with a low metallicity ($[Z/H]\sim [-2.0, -1.5]$) 
filament that is feeding a small satellite, which in turn appears to be
interacting and possibly feeding the primary galaxy at a projected distance of $\sim 20$kpc.
This is yet another example of interacting galaxies producing rich gas-feeding 
filaments, as already seen in Figures~\ref{fig:pic2} and \ref{fig:pic3}.
The relatively large width of $303$km/s is produced by steep velocity gradient
in the region from $x\sim 45$ to $50$kpc.
One could see that galactic winds are blowing to the upper left corner by the
primary galaxy,
whose starburst is likely triggered by the interaction.

Figure~\ref{fig:pic7} shows a DLA that is made up by 
several filaments at distances of $30-40$kpc from the galaxy.
The metallicity of all the components is near solar, indicating that
these are probably pre-enriched gas cooling due to thermal instability.
The velocity structures show that they are falling back towards the galaxy,
in a fashion perhaps similar to galactic fountains \citep[][]{1976Shapiro}, on somewhat extended scales.
Once again, it appears that galaxy-galaxy interactions 
may be responsible for the rich gas filaments, 
as seen in Figures~ \ref{fig:pic2}, \ref{fig:pic3} and \ref{fig:pic5}. 
There is evidence that winds are blowing upwards from the galaxy.

Figure~\ref{fig:pic8} shows another example of a DLA arising from 
a galaxy with a very rich filament system due to galaxy interactions.
The primary galaxy is the same as the one shown in Figure~\ref{fig:pic3} and
we are now looking at its south side.
These filaments that are responsible for the neutral column of the DLA
appear to have been enriched to a level of $[Z/H]\sim -1.0$ and have 
cooled to low temperature.
The large width of $420$km/s is due to the multiple
components spanning a spatial range of $\sim 40$kpc each of physical depth of several
kpc and individual velocity width $\le 100$km/s.
Interestingly, for this DLA system, 
while most of the gas filaments are now falling back toward the galaxy 
(not necessarily radially),
galactic winds are still blowing towards the upper right corner.
Comparison of Figure~\ref{fig:pic8} and Figure~\ref{fig:pic3} 
indicates that the metallicity in the upper right quadrant 
of the galaxy is somewhat more metal enriched 
($[Z/H]\sim 0$)
than other regions 
($[Z/H]\sim -1$ or lower),
consistent with the directions of ongoing galactic winds.
This is strongly suggestive that metallicity enrichment process not only is
episodic, multi-generational, anisotropic, but also in general possesses no parity.

Figure~\ref{fig:pic9} shows a DLA intercepting two filaments at a small inclined angle,
giving rise to a broad physical extension of $\sim 25$kpc.
All the visible filaments appear to run roughly top-left to bottom-right,
whereas the metal enriched regions seem to spread out 
like a butterfly in the direction roughly perpendicular to the filaments.
This is the classic picture that galactic winds tend to blow in directions
that are perpendicular to the filaments that are feeding the galaxy.
The inner regions of the filaments appear to be less enriched ($[Z/H]\sim -2$) 
than the outer regions of the filaments ($[Z/H]\sim [-0.5, 0]$), 
strongly indicative of galactic winds tending to circumvent the denser filaments.
The large width of $510$km/s appears to be caused by oppositely 
moving (i.e., converging) flows at $x\sim 15-40$kpc,
probably caused by the bipolar winds interacting with the complex filament structures.
This galaxy has a mass of $2\times 10^{11}\msun$ and we notice that most of its
surrounding regions is relatively cold,
whereas in 
Figures~(\ref{fig:pic2}-\ref{fig:pic8}) we consistently see
a hot atmosphere permeating the circumgalactic regions.
The galaxies in Figures~(\ref{fig:pic2}-\ref{fig:pic8}) all have
mass $\ge 10^{12}\msun$, consistent with the mass demarcation of cold and hot accretion modes
\citep[][]{2005Keres, 2006Dekel}.
Nevertheless, the existence of cold galactic filaments seen in 
Figures~(\ref{fig:pic2}-\ref{fig:pic8}) is consistent with the suggested
cold mode of accretion of massive galaxies at high redshift \citep[][]{2009Dekel}.

Figure~\ref{fig:pic10} shows 
a ``normal" DLA where a relatively quiet galactic disk 
is pierced edge-on. Its large width of $522$km/s
is simply due to the large halo that the galaxy is residing in
of mass $3\times 10^{12}\msun$ (at $z=3.1$).
The surrounding environment seems relatively ``pristine" with no widespread
metal enrichment at a level of $[Z/H]\ge -1$.
However, the temperature panel indicates that there is a hot halo
permeating the entire region and embedding and 
pressure-confining (see the pressure panel) the cold neutral clouds.
It seems likely that this hot gaseous halo is produced by gravitational shocks
rather than galactic wind shocks.
There are several filaments attached to the galaxy.
This is a good example of cold streams feeding a massive galaxy 
by penetrating a hot atmosphere.

Figure~\ref{fig:pica3} shows a DLA with 
a large velocity width arising from a relatively small galaxy 
of total mass $3\times 10^{10}\msun$.
The galactic winds are blowing in the north-east direction that
entrain cold neutral clouds with it.
The LOS of the DLA  intercepts a high velocity component at $x=35-55$kpc.
The combination of this high velocity component with the low velocity
component at $x\sim 0$ produces the relatively large width of $306$km/s.
Note that an isotropic Maxwellian velocity distribution of dispersion equal to
that of the halo velocity dispersion would only yield a width of $v_{90}=2.33v_{vir}=176$km/s.
Clearly, galactic winds are directly responsible for the large width of this DLA,
by entraining cold gas clouds to a high velocity.

Figure~\ref{fig:pica4} shows another DLA produced by a small galaxy 
of mass $2\times 10^{10}\msun$ with a large velocity width.
The galaxy system have multiple, interaction galaxies at close distances.
The galactic winds are blowing primarily, in a bipolar fashion,
in north-east and south-west direction, roughly perpendicular to the galactic disk,
 that entrain the cold neutral cloud at $x\sim 40$kpc to a broad velocity
of $v_x=0-400$km/s relative to the galaxy itself.
In combination with another complex structure 
at $x=45-70$kpc the galactic winds produce a very large width of $516$km/s.
Note that for this galaxy $2.33v_{vir}=160$km/s.

Figure~\ref{fig:pica5} shows yet another DLA arising from a small galaxy 
but having a large velocity width of $501$km/s.
This is an interesting case where the galactic winds are blowing, 
by the primary galaxy,
towards and passing through the satellite galaxy
at $(x,y)\sim (40,20)$kpc in the north-east direction.
The SDSS u band luminosity map suggests that the satellite galaxy itself
is experiencing a starburst and likely blowing and enhancing the north-east/north winds. 
A very large positive velocity gradient in the positive x-direction (downstream)
of $\sim 700$km/s over an LOS physical interval of $\Delta x\sim 20$kpc
is produced, resulting in a very large width.
Note that $2.33v_{vir}=162$km/s for this galaxy.
The entrained neutral gas cloud and its downstream appear to have escaped 
the metal enrichment by ongoing winds and remain at $[Z/H]\sim -1$
- ``shadowing" effect due to dense clouds and filaments.

Figure~\ref{fig:pica6} shows another wide DLA from a small galaxy.
For this DLA the majority of the column is due to 
the intersection with the disk of the galaxy, which 
would have produced a velocity width of $\sim 200$km/s on its own.
The galactic winds blowing at the south-east direction 
entrain some cold clouds at $x=35-45$kpc to a velocity up to $500$km/s.
Together, a large width of $501$km/s is produced.
Given the small mass of the galaxy the surrounding regions
are not embedded in a hot gravitationally shock heated atmosphere.
There are some solid angles with low gas column that have been heated by galactic winds,
probably triggered by the binary galaxy interaction,
as can be seen by comparing temperature map and the density map.
Note that $2.33v_{vir}=204$km/s for this galaxy.

Finally, Figure~\ref{fig:pica7} shows the last example of a wide DLA from a small galaxy.
Two galactic filaments make up this DLA,
one at $x\sim 10$kpc and the other at $x\sim 35$kpc.
The galactic system is a primary-satellite binary that is interacting,
which has likely caused both to experience starbursts.
The primary galaxy at $(x,y)\sim (23,36)$kpc is blowing bipolar galactic winds
mainly in the north-south direction,
whereas the satellite at $(x,y)\sim (35,35)$kpc is blowing bipolar galactic winds
in the east-west direction.
Together they produce a very complex, multi-stream velocity structure.
The total velocity width of this DLA is $500$km/s,
although each of the two components individually has a velocity width of $\le 200$km/s.
Note that $2.33v_{vir}=200$km/s for the primary galaxy.

In summary, we see that DLAs arise in a wide variety of cold gas clouds,
from galactic disks to cold streams to cooling gas from galactic winds
to cold clouds entrained by hot galactic winds
at a wide range of distances from galaxies, with a wide range of metallicity
and in galaxies of all masses from $10^{10}-10^{-12.5}\msun$ at $z\sim 3$.  
Inspection of the gallery has already hinted
that many large velocity width DLAs
may be produced directly or indirectly by galactic winds.
That is, directly by entraining cold gas clouds and compressing cold gas clouds with
high pressure and indirectly by enhancing cooling and thermal instability with added
metals and shock compression.
In addition, the composite nature of many large width DLA systems 
should also help remove the perceived failure of the standard LCDM model
with respect to producing large width systems
\citep[e.g.,][]{1997Prochaska}.
Quantitative results later prove this is indeed the case.

\subsection{Kinematic Velocity Width Distribution Functions}

\begin{figure}[ht]
\centering
\resizebox{4.5in}{!}{\includegraphics[angle=0]{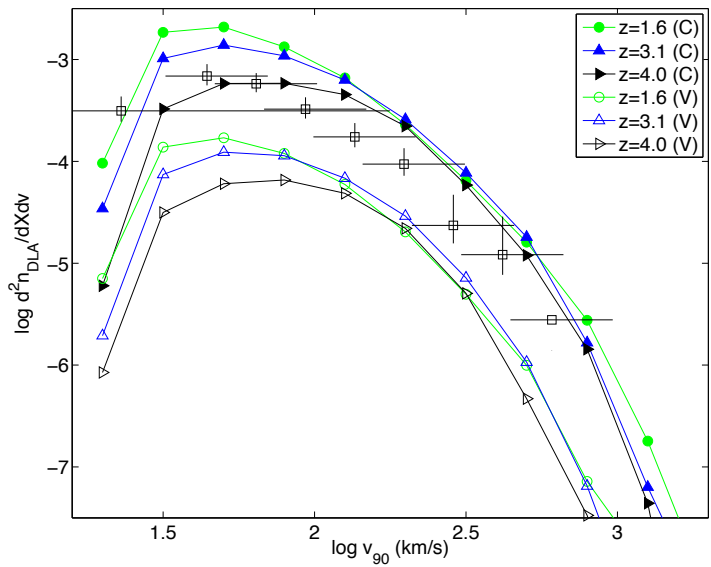}}
\caption{
Velocity distribution functions, defined to be the number 
of DLAs per unit width velocity per unit absorption length,
at $z=1.6,3.1,4.0$. 
Two sets of simulation results are shown,
one for the ``C" run (solid symbols) and ``V" run (open symbols).
The corresponding observational data for each of the individual redshifts
\citep[][]{2005Prochaska} are shown as open squares, which span
the redshift range of $z=1.7-4.5$.
}
\label{fig:v90his}
\end{figure}

We now present more quantitative statistical results on the velocity width distribution
functions of DLAs at several redshifts.
Since the velocity structure in $\lya$ flux of a DLA is ``damped" and does not provide 
the kinematic information of the underlying physical cloud.
Following \citet[][]{1997Prochaska},
the velocity width, $v_{90}$, 
is defined to be the velocity interval of 90\% of the optical depth
of the \ion{Si}{2} $\lambda$1808 absorption line associated with the DLA.
Figure~\ref{fig:v90his} shows the velocity width distribution at three redshifts
($z=1.6,3.1,4.0$), covering most of the observed redshift range.
We see a factor of $\sim 10$ variation from ``C" to ``V" run,
indicating the need to have 
a larger statistical set of simulations covering, more densely, 
different environments,
before a more precise comparison can be made with observations.
Insofaras the observed velocity width distribution function lies inbetween
the two bracketing runs, ``C" and ``V",
and the shape of the functions are in excellent agreement with observations,
including the high velocity tail ($v_{90}\ge 300$km/s),
this should be considered a success for the LCDM model - 
there is no lack of large width DLAs with
$v_{90}\ge 300$km/s in the LCDM simulation. 
This conclusion is consistent with that of 
\citet[][]{2010Hong}, who studied this issue with a different code and 
a different feedback implementation.
There is a significant difference between our results and theirs in the 
that we find galactic winds are directly responsible for many of 
the large width DLAs, by entraining neutral dense clouds to large velocities.
In addition, they conclude that a large halo mass ($\ge 10^{11}\msun$) 
is a necessary condition for producing large velocity widths,
while we find that a non-negligible fraction of large velocity width DLAs arise in halos less massive than
$10^{11}\msun$. 

\begin{figure}[ht]
\centering
\resizebox{5.5in}{!}{\includegraphics[angle=0]{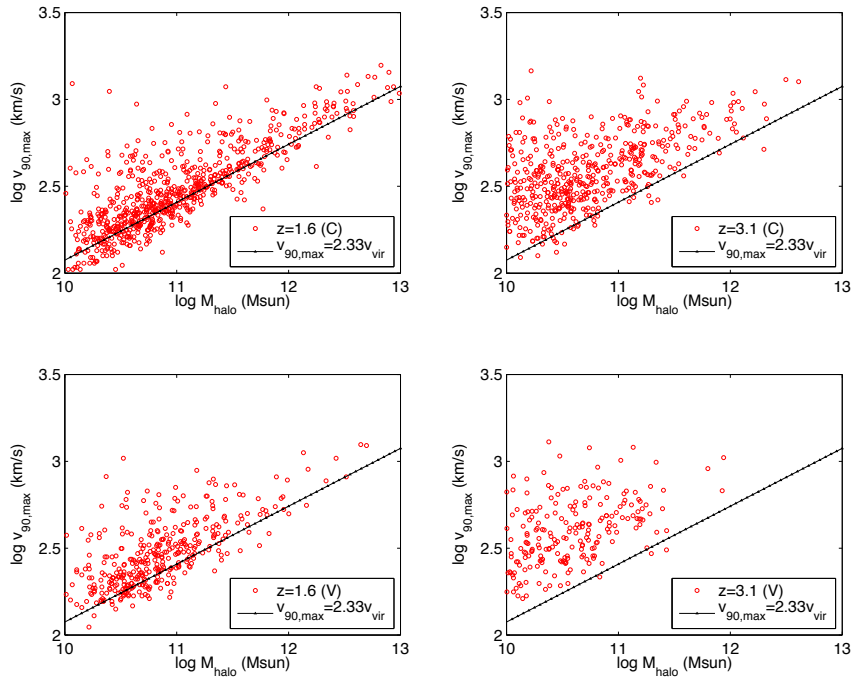}}
\caption{
The maximum $v_{90}$ of all DLAs
associated with each galaxy against the halo mass of the galaxy $M_{\rm halo}$
for $z=1.6$ and $z=3.1$ for ``C" and ``V" run. 
The black line $v_{90}=2.33v_{vir}$ is what $v_{90}$ would be 
if the velocity distribution is an isotropic and Maxwellian distribution
with its dispersion equal to $v_{vir}$,
and the \siii gas density is constant across the DLA.
}
\label{fig:v90max}
\end{figure}

\begin{figure}[ht]
\centering
\resizebox{5.5in}{!}{\includegraphics[angle=0]{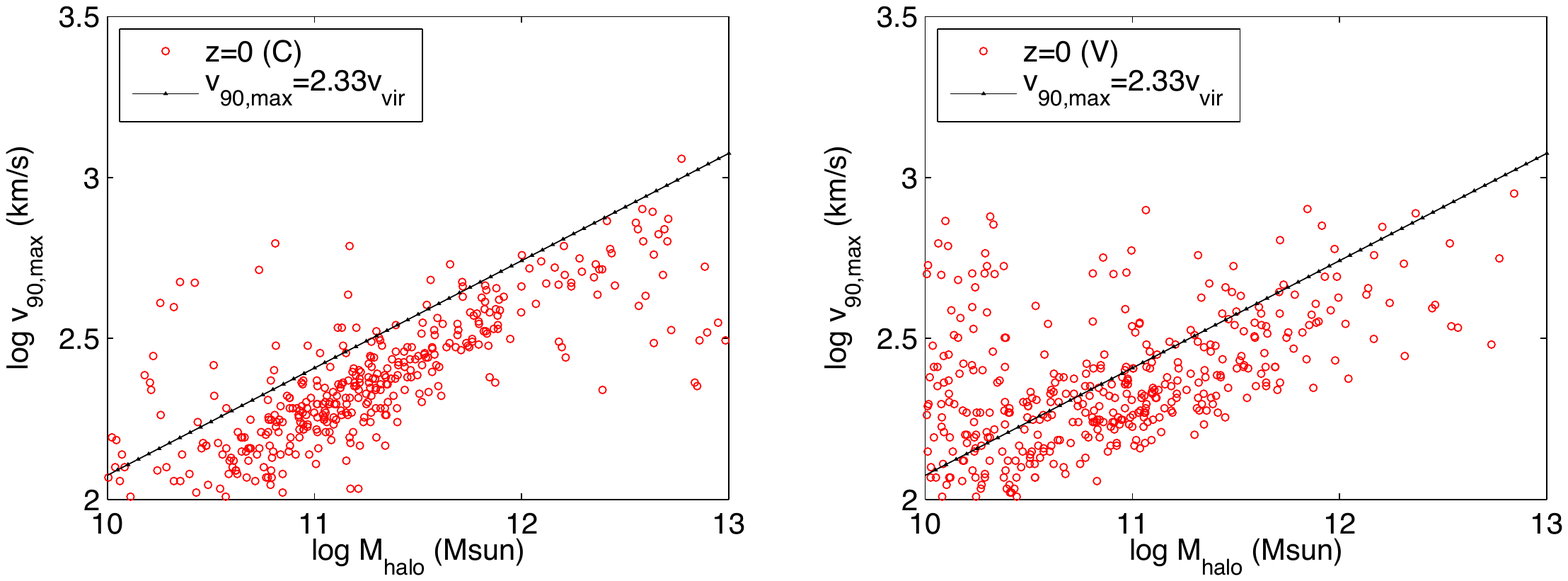}}
\vskip -2in
\caption{
The maximum $v_{90}$ of all DLAs
associated with each galaxy against the halo mass of the galaxy $M_{\rm halo}$
at $z=0$ for ``C" and ``V" run. 
The black line $v_{90}=2.33v_{vir}$ is what $v_{90}$ would be 
if the velocity distribution is an isotropic and Maxwellian distribution
with its dispersion equal to $v_{vir}$,
and the \siii gas density is constant across the DLA.
}
\label{fig:v90maxz0}
\end{figure}

To help understand the large velocity width DLAs,
we plot in Figure~\ref{fig:v90max} the maximum $v_{90}$ of all DLAs, $v_{\rm 90,max}$,
associated with each galaxy against the halo mass of the galaxy, $M_{\rm halo}$.
The black line $v_{90}=2.33v_{vir}$ is what $v_{90}$ would be 
if the velocity distribution is an isotropic and Maxwellian distribution
with its dispersion equal to $v_{vir}$,
and the \siii gas density is constant across the DLA.
We see that the Maxwellian velocity distribution (the black line)
approximately provides a lower bound to $v_{90}$,
although there is, unsurprisingly, some fraction of systems that lie below
(see Figure~\ref{fig:pic2} for an example).
What is very interesting is that at $z=3.1$ there is a large number 
of galaxies whose $v_{\rm 90,max}$ are substantially larger than 
what $v_{vir}$ could produce, i.e.,
``super-gravitational motion" in the terminology of \citet[][]{2010Hong}.
This super-gravitational motion is produced by galactic winds,
as we have seen clearly in 
Figures~\ref{fig:pica3}, \ref{fig:pica4}, \ref{fig:pica5}, \ref{fig:pica6} and \ref{fig:pica7} in \S 3.1.
We also note that at $z=1.6$ for both ``C" and ``V" run (and especially the ``C" run), 
the correlation between $v_{90}$ and $v_{vir}$ becomes
substantially better with much reduced scatter,
and the excess of DLAs with large $v_{90}/v_{vir}$ is much removed.
This is circumstantial but strong 
evidence that galactic winds are responsible for
most of the large $v_{90}/v_{vir}$ DLAs,
because of higher star formation activities hence galactic winds
at $z=3.1$ than $z=1.6$.
Figure~\ref{fig:v90maxz0} below will further strengthen this point.

It appears that the redshift evolution at a fixed environment
is relatively mild in the redshift range $z=4.0$ to $z=1.6$.
We speculate that the weak evolution of the velocity width distribution
from $z=4.0$ to $z=1.6$ 
may be coincidental and attributable to two countering processes:
growth of halo mass hence virial velocity 
with time and diminution of supergravitational motion produced
by galactic winds with time (due to reduced star formation activities with time at $z\le 2$).
This prediction of a weak evolution of velocity width distribution with redshift
is verifiable with future larger DLA sample and 
is a powerful test for the non-gravitational origin of a large fraction
of the large width systems.

\begin{figure}[ht]
\centering
\resizebox{4.0in}{!}{\includegraphics[angle=0]{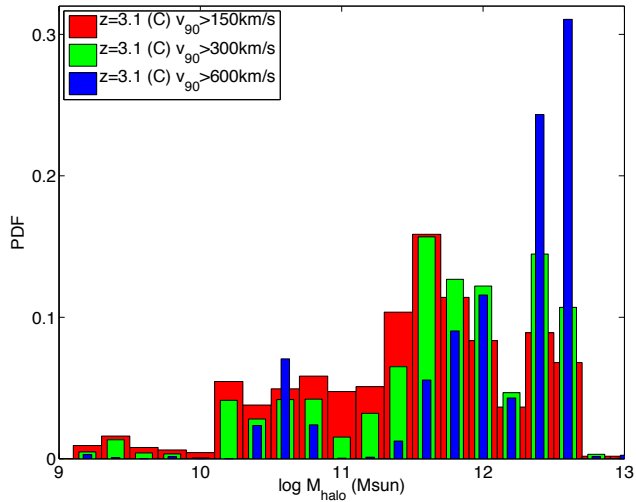}}
\caption{
shows the DLA incidence weighted halo mass probability distribution function
for DLAs above three velocity width cuts, $v_{90}\ge 150, 300, 600$km/s
at $z=3.1$ for the ``C" run.
Note that a DLA associated with a satellite galaxy or any gas cloud
within the virial radius is given the halo mass of the primary galaxy.
}
\label{fig:v150his}
\end{figure}

Figure~\ref{fig:v90max} does not, however, fairly characterize 
the relative contribution of halos of different masses to
velocity width distribution function,
because it does not specify the number of DLAs at a given halo mass.
In Figure~\ref{fig:v150his} 
we show the halo mass probability distribution function
for DLAs above three velocity width cuts, $v_{90}\ge 150, 300, 600$km/s,
respectively.
We see a clear trend that larger halos make larger contribution
to larger width DLAs,
as one would have expected.
For example, about one half of all DLAs with $v_{90}\ge 600$km/s
arise in halos of mass greater than $10^{12}\msun$ at $z=3.1$,
whereas that division line drops to $2\times 10^{11}\msun$ for 
$v_{90}\ge 150$km/s.
It should be noted that the ratio of the virial velocity of a halo of mass
$2\times 10^{11}\msun$ to that of $10^{12}\msun$ is $0.58$, significantly
greater than $0.25=150/600$,
indicating an overweight of DLA cross-section by large galaxies.
For moderate to large velocity width of $v_{90}\ge 150$km/s,
halos of mass $1\times 10^{11}\msun$ dominate the contribution to DLA incidence,
largely in agreement with \citet[][]{2010Hong}.
Slightly at odds with \citet[][]{2010Hong}, however,
we find a significant fraction of these relative wide systems arising
in galaxies of mass less than $1\times 10^{11}\msun$:
$(24\%, 18\%, 12\%)$ of DLAs with velocity width larger than
$(150,300,600)$km/s are due to galaxies with mass less than $1\times 10^{11}\msun$.
We note that our definition of associating DLAs with galaxies
biases associating them to larger galaxies; Figure~\ref{fig:pic5} gives an example,
where the DLA is defined to arise from the larger galaxy of mass $8\times 10^{11}\msun$,
even though it is more closely related to a much smaller satellite galaxy that
is orbiting around the larger galaxy.
Our results are perhaps unsurprising in the sense that 
one would have expected that galactic winds, when they are blowing,
should be stronger, or at least not weaker,
in dwarf starburst galaxies than larger galaxies 
thanks to shallow gravitational potential wells in the former, when cold gas 
is still abundant at high redshift.
Both Figure~\ref{fig:v150his} and gallery pictures in \S 3.1 confirm this point.
Galactic winds, however, could be weaker in dwarf galaxies if star formation is 
inproportionally less vigorous.
This may be the case at lower redshift, as shown in Figure~\ref{fig:v90maxz0}.
What is interesting, and further evidence,
is that at $z=0$ the dwarf galaxies in ``V" run appear to have more
super-gravitational motion than in the ``C" run,
simply because the former are gas richer and have higher star formation rate than the latter.
Thus, it seems that galactic winds are a bivariate function of galaxy mass
and star formation rate, in a fashion that is consistent with observations
\citep[e.g.,][]{2005Martin}.

\subsection{\siii Line Profile Shape Measures}

\begin{figure}[h]
\centering
\resizebox{3.4in}{!}{\includegraphics[angle=0]{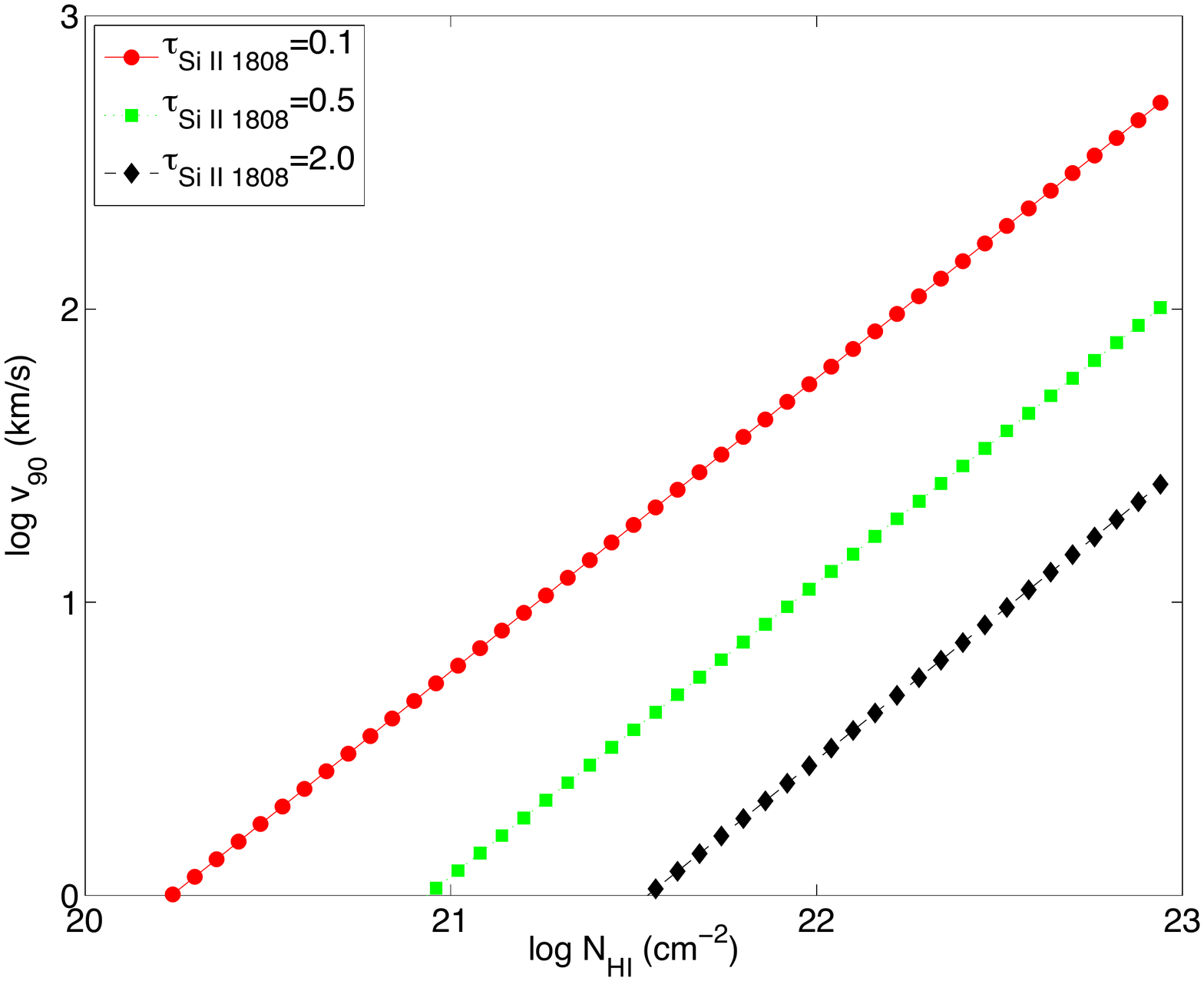}}
\hskip -1cm
\resizebox{3.4in}{!}{\includegraphics[angle=0]{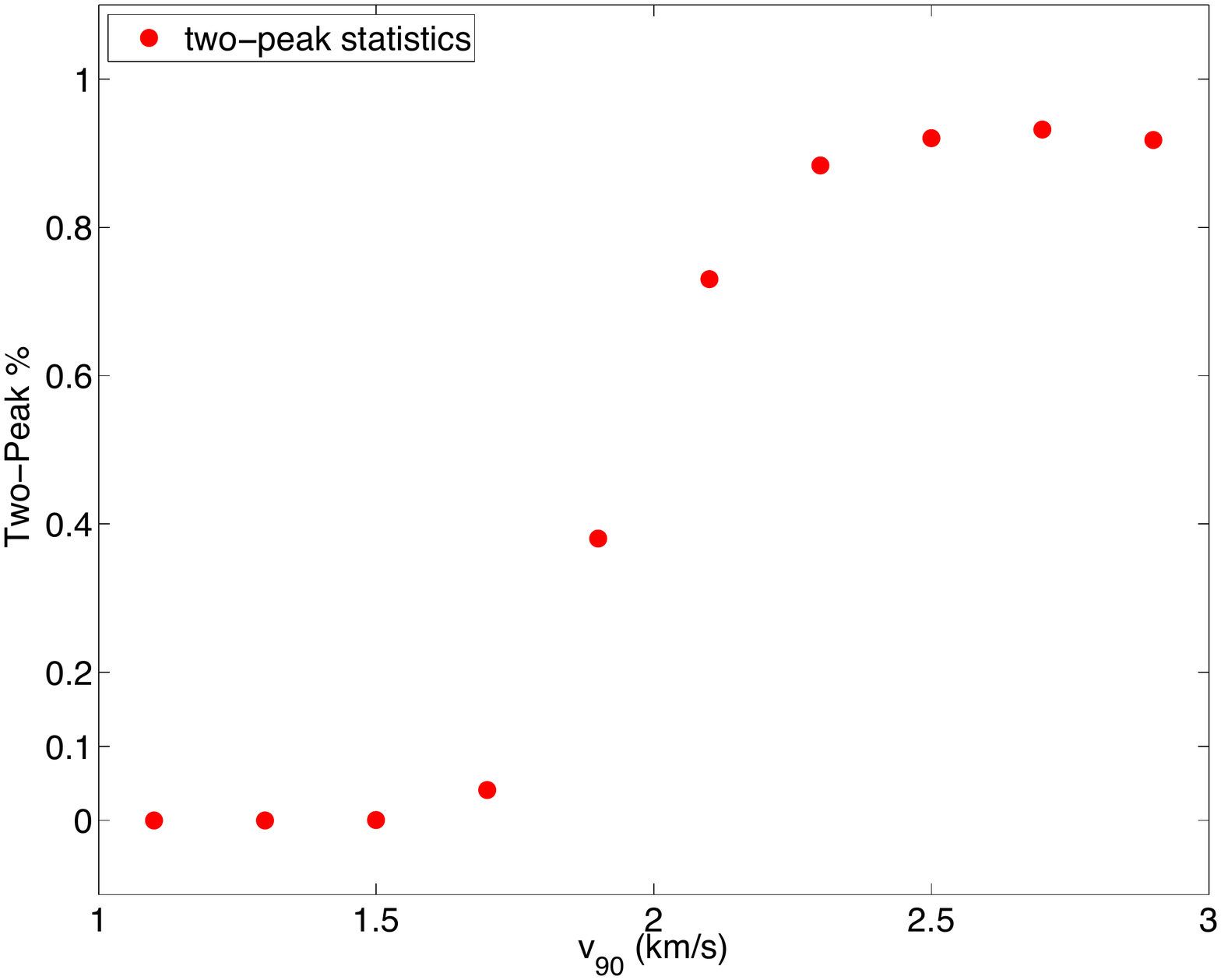}}
\caption{
Left panel: $v_{90}$ as a function of $\log N_{\rm HI}$ assuming $Z=0.1\zsun$.
Right panel: the percentage of DLAs that have multiple components, as a function of $v_{90}$.
}
\label{fig:itwopeaks}
\end{figure}

\begin{figure}[h]
\hskip -0.7cm
\centering
\resizebox{3.71in}{!}{\includegraphics[angle=0]{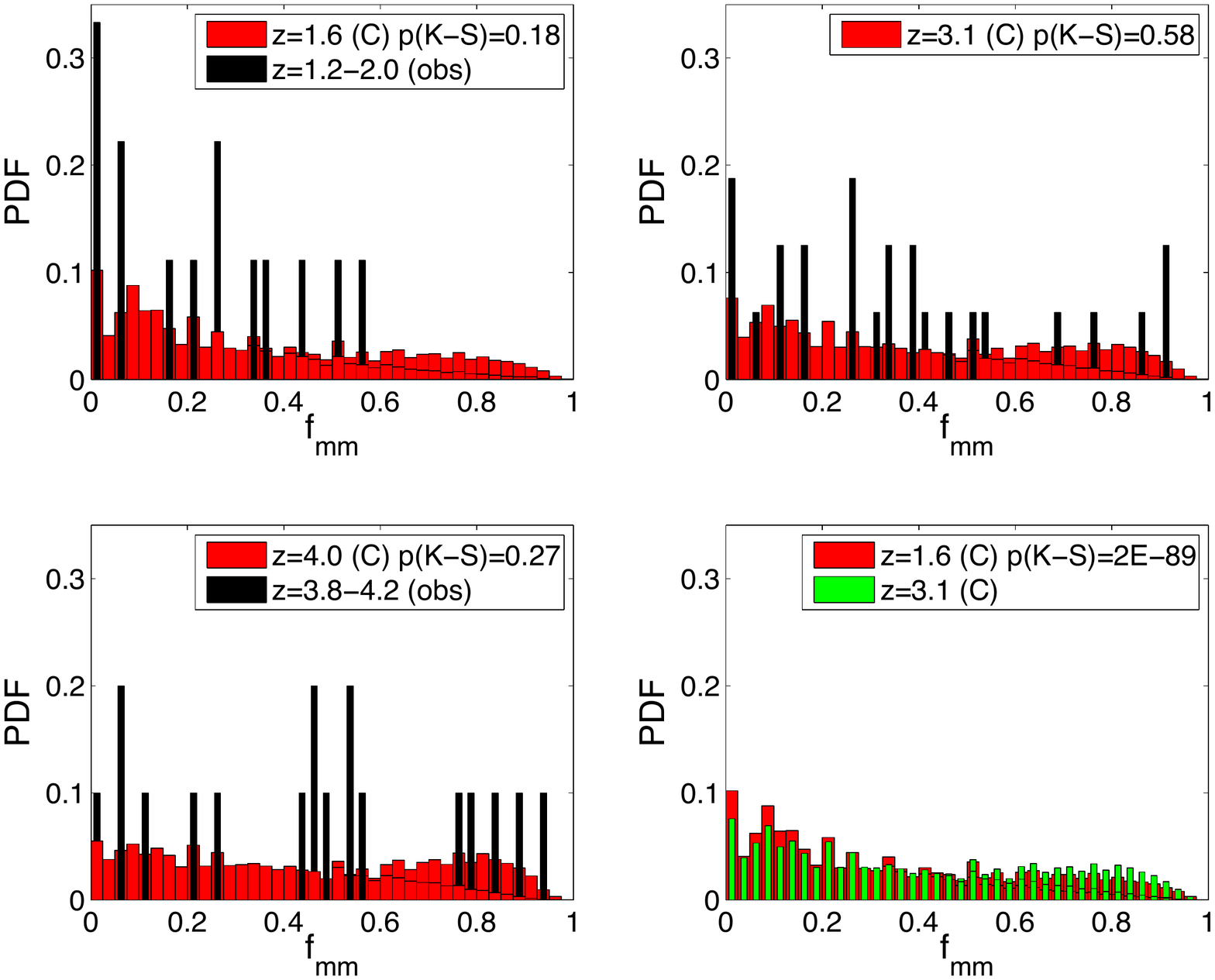}}
\hskip -1.8cm
\resizebox{3.71in}{!}{\includegraphics[angle=0]{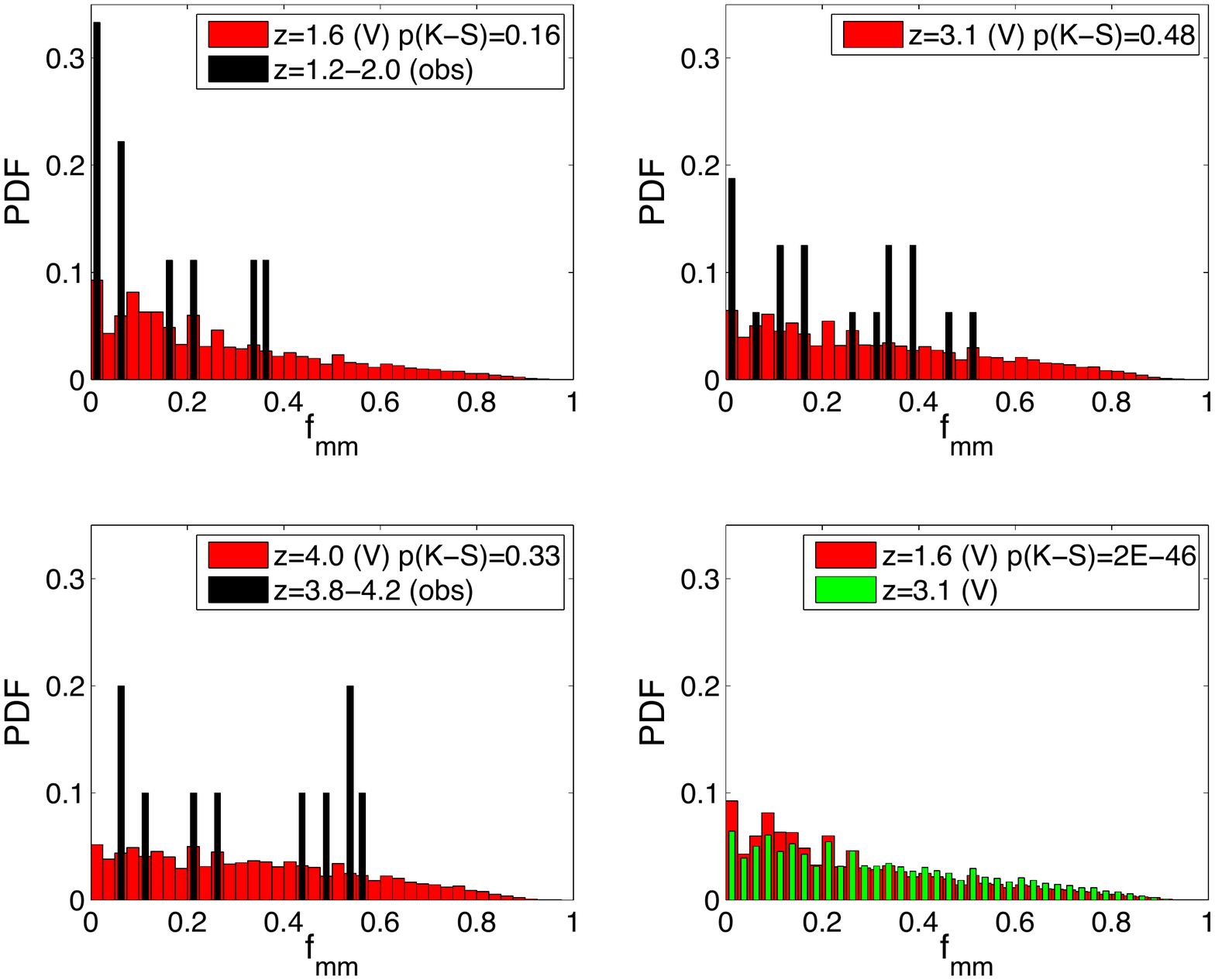}}
\caption{
Left set of four panels:  $f_{\rm mm}$ distributions for ``C" run 
at redshift $z=1.6$ (top left), $z=3.1$ (top right) and $z=4.0$ (bottom left).
For $z=1.6$ we compared to observed DLAs at the redshift range $z=1.2-2.2$;
for $z=3.1$ we compared to observed DLAs at $z=2.9-3.3$;
for $z=4.0$ we compared to observed DLAs at $z=3.8-4.2$.
Observed sample is an updated version of \citet[][]{1997Prochaska}, 
shown as the black histogram.
Also shown in each K-S test probability that the two distributions
(computed and observed) are drawn form the same underlying distribution.
In the bottom right panel, we compare computed $z=1.6$ and $z=3.1$ distributions
along with the K-S test probability
to show a significant evolution of this shape distribution function with redshift.
Right set of four panels:  $f_{\rm mm}$ distributions for ``V" run.
}
\label{fig:fmmhis}
\end{figure}

\begin{figure}[t]
\hskip -0.7cm
\centering
\resizebox{3.71in}{!}{\includegraphics[angle=0]{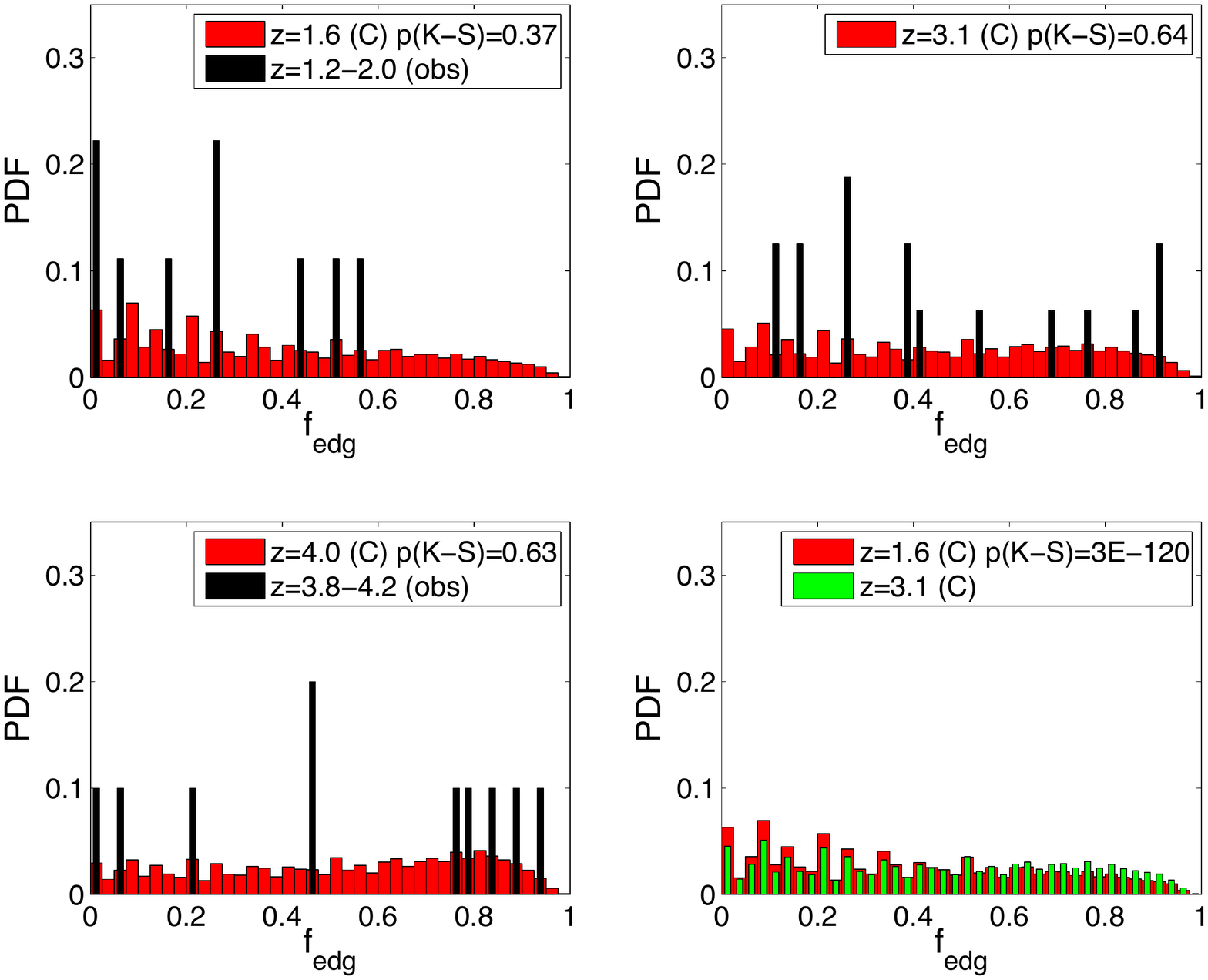}}
\hskip -1.8cm
\resizebox{3.71in}{!}{\includegraphics[angle=0]{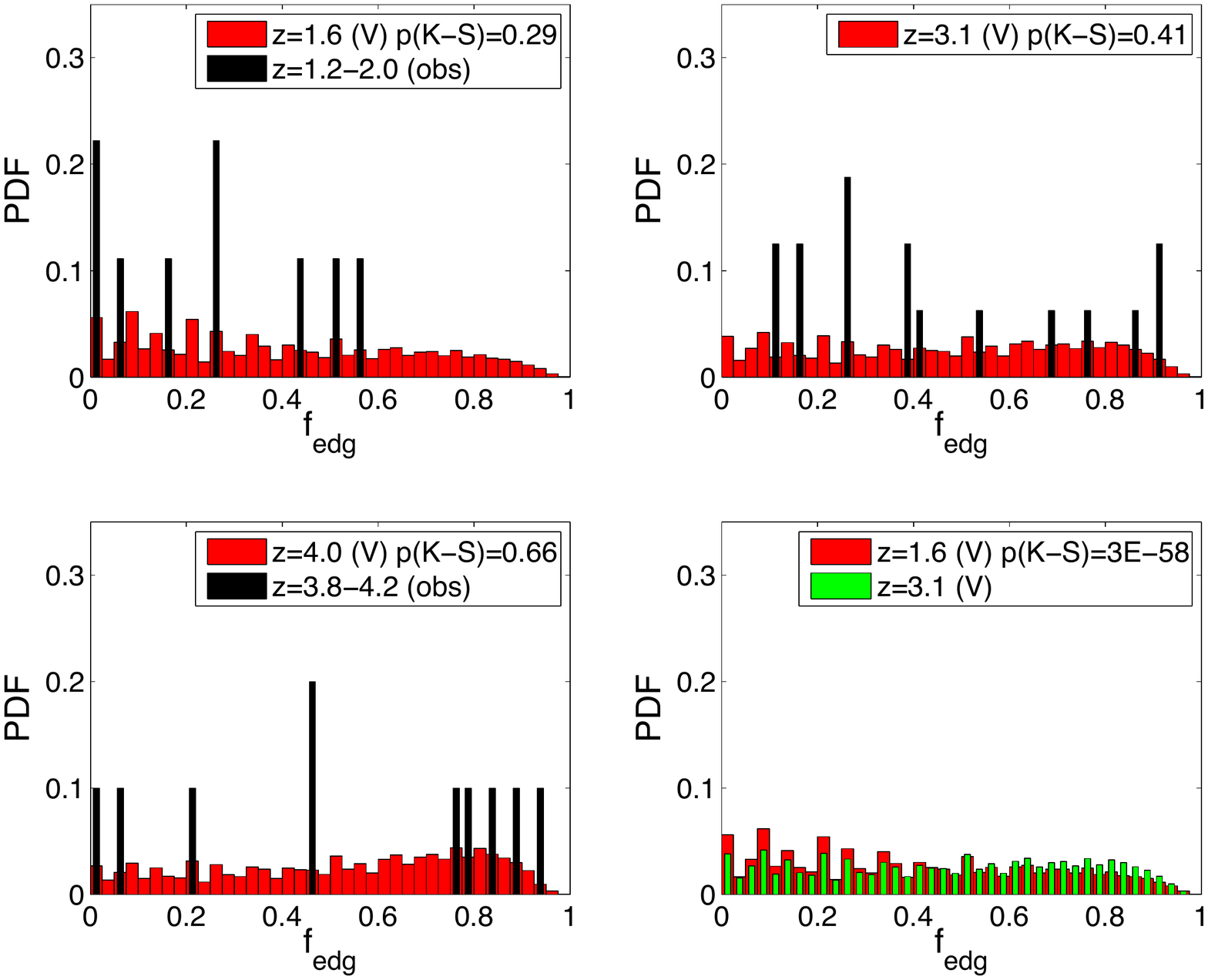}}
\caption{
Left set of four panels:  $f_{\rm edg}$ distributions for ``C" run 
at redshift $z=1.6$ (top left), $z=3.1$ (top right) and $z=4.0$ (bottom left).
For $z=1.6$ we compared to observed DLAs at the redshift range $z=1.2-2.2$;
for $z=3.1$ we compared to observed DLAs at $z=2.9-3.3$;
for $z=4.0$ we compared to observed DLAs at $z=3.8-4.2$.
Observed sample is an updated version of \citet[][]{1997Prochaska}, 
shown as the black histogram.
Also shown in each K-S test probability that the two distributions
(computed and observed) are drawn form the same underlying distribution.
In the bottom right panel, we compare computed $z=1.6$ and $z=3.1$ distributions
along with the K-S test probability
to show a significant evolution of this shape distribution function with redshift.
Right set of four panels:  $f_{\rm edg}$ distributions for ``V" run.
}
\label{fig:fedgehis}
\end{figure}

\begin{figure}[t]
\hskip -0.7cm
\centering
\resizebox{3.71in}{!}{\includegraphics[angle=0]{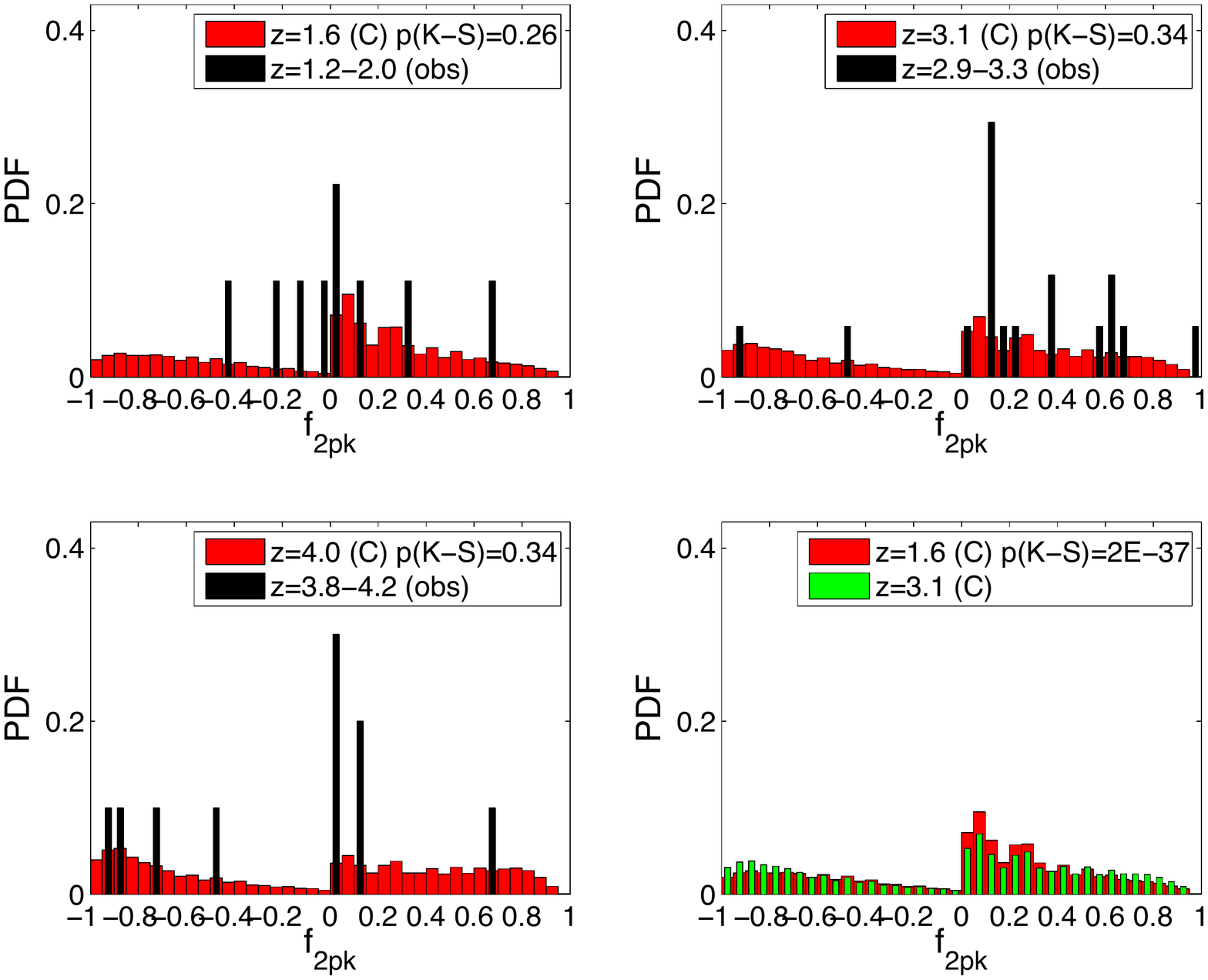}}
\hskip -1.8cm
\resizebox{3.71in}{!}{\includegraphics[angle=0]{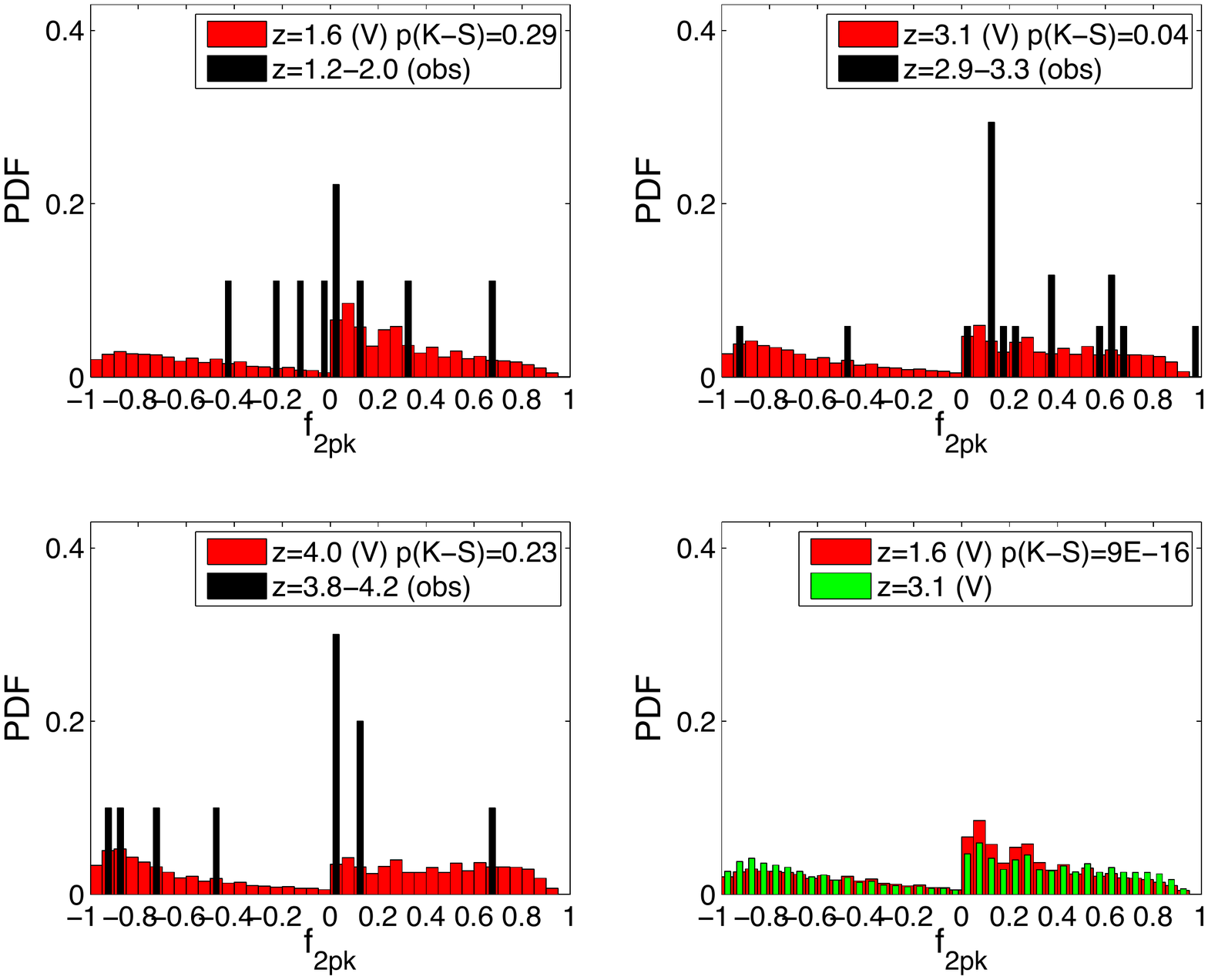}}
\caption{
Left set of four panels:  $f_{\rm 2pk}$ distributions for ``C" run 
at redshift $z=1.6$ (top left), $z=3.1$ (top right) and $z=4.0$ (bottom left).
For $z=1.6$ we compared to observed DLAs at the redshift range $z=1.2-2.2$;
for $z=3.1$ we compared to observed DLAs at $z=2.9-3.3$;
for $z=4.0$ we compared to observed DLAs at $z=3.8-4.2$.
Observed sample is an updated version of \citet[][]{1997Prochaska}, 
shown as the black histogram.
Also shown in each K-S test probability that the two distributions
(computed and observed) are drawn form the same underlying distribution.
In the bottom right panel, we compare computed $z=1.6$ and $z=3.1$ distributions
along with the K-S test probability
to show a significant evolution of this shape distribution function with redshift.
Right set of four panels:  $f_{\rm 2pk}$ distributions for ``V" run.
}
\label{fig:f2pkhis}
\end{figure}

Having found an overall good agreement with observations with
respect to the velocity width distribution,
we now turn to shape measures of \ion{Si}{2} $\lambda$1808 
absorption line profile. 
Before comparing to observational data from
\citet[][]{1997Prochaska}, we shall first try to understand
the relationship among the optical depth of a \siii line, HI column density
metallicity and velocity width.
Assuming that the optical depth profile of the \siii line is a simple top-hat
(assuming a different profile such as a gaussian makes no material difference for our purpose),
it can be shown that
\begin{equation}
\label{eq:tauSiII}
\tau_{\siii} = 0.01 ({N_{\rm HI}\over 2\times 10^{20}~{\rm cm}^{-2}}) ({Z\over \zsun}) ({v_{90}\over 100\kms})^{-1},
\end{equation}
\noindent
where $Z$ is the metallicity of DLA in solar units.
Left panel of Figure~\ref{fig:itwopeaks} shows $v_{90}$ as a function of $\log N_{\rm HI}$ for $Z=0.1\zsun$.
As expected, an increase in velocity width requires
a corresponding increase in column density to produce
a same optical depth. 
More important is that, quantitatively,
in order to achieve an optical depth of $0.1$, 
with a width $v_{90}\sim 100$km/s it requires a DLA column of $\sim 2\times 10^{22}$cm$^{-2}$,
if the DLA is composed of one single component with $[Z/H]=-1$.
Since the abundance of DLAs with $N_{\rm HI}\ge 10^{22.5}$cm$^{-2}$ declines rapidly (see Figure~\ref{fig:colhis})
but the abundance of \siii line peaks near $v_{90}\sim 100$km/s (Figure~\ref{fig:v90his}),
this suggests that a significant fraction of \siii lines must have multiple components.
To quantitatively illustrate this, we define a new simple two-component measure as follows.
If there are at least two peaks in the optical depth profile that are separated 
by more than $0.5v_{90}$ and the ratio of the peak heights 
is greater than $1/15$, we define the DLA to be a two-component DLA.
The ratio, $1/15$, comes about such that the lower peak is guaranteed
to be included in the accounting of $v_{90}$ interval,
although changing it to say $1/10$ makes no dramatic difference in the results.
Note that DLAs with more than two components are included as two-component systems. 
Right panel of Figure~\ref{fig:itwopeaks} shows the percentage of two-component DLAs
as a function of $v_{90}$.
In good agreement with the simple expectation,
we see that at $v_{90}=100$km/s, about 50\% of DLAs have more than one component
and that number increases to $\sim 90\%$ at $v_{90}=300$km/s.
This result is also consistent with the anecdotal evidence shown in the gallery examples in \S 3.1,
where most of large width DLAs contain more than one physical component.

We now turn to the three kinematic shape measures defined in 
\citet[][]{1997Prochaska}, $f_{\rm mm}$, $f_{\rm edg}$, $f_{\rm 2pk}$,
representing, respectively, measures of the symmetry, leading-edgeness
and two-peakness of the profile of \ion{Si}{2} $\lambda$1808 absorption lines associated with DLAs
(see the bottom right panels of the gallery pictures in \S 3.1).
Figures (\ref{fig:fmmhis},\ref{fig:fedgehis},\ref{fig:f2pkhis})
show comparisons of simulation results with observations at
three redshifts $z=1.6$, $z=3.1$ and $z=4.0$.
We see the overall agreement between simulations and observations is excellent,
with K-S tests (indicated in the figures) for both runs (``C" and ``V") at three compared redshifts ($z=1.6,3.1,4.0$)
all being at acceptable levels.
Our results are in good agreement with one of the models with feedback
in \citet[][]{2010Hong}, except for the case of $f_{\rm 2pk}$:
our simulations find acceptable K-S test values of 26-29\%, 4-34\% and 23-34\% at $z=1.6$, $3.1$ and $z=4.0$,
 respectively,
whereas they find none of their models have probability higher than 5\% at $z=3.1$.
We speculate that difference in the detailed treatments of metal transport 
process as well as feedback prescription
between our simulations and theirs may have partly contributed to this difference;
with detailed metal transport we find very inhomogeneous
metallicity distributions across space and among DLAs 
in our simulations (see Figure \ref{fig:mtlhis} below),
whereas they assume a constant metallicity of $[Z/H]=-1$ for all DLAs.
It is also noted that the metallicity distributions of our simulations
at redshift range $z=1.6-4.0$ are in excellent agreement with observations
(Figure~\ref{fig:mtlhis}).
At the bottom-right panels of each four-panel set in
Figures (\ref{fig:fmmhis},\ref{fig:fedgehis},\ref{fig:f2pkhis})
we show a comparison between $z=1.6$ and $z=3.1$ distributions
for each of the shape statistics and find that there is significant evolution in all three shape measures.
Current small observational sample does not allow for such a test.
Our results demonstrate that the standard LCDM model,
with a proper modeling of astrophysical processes, including galaxy formation
and feedback in the forms of mechanical feedback and metal enrichment,
can successfully 
produce \siii line shapes that are in good agreement with observations.

\subsection{Column Density Distribution, Line Density and $\Omega_g(\rm DLA)$ Evolution}

\begin{figure}[h]
\centering
\resizebox{6.0in}{!}{\includegraphics[angle=0]{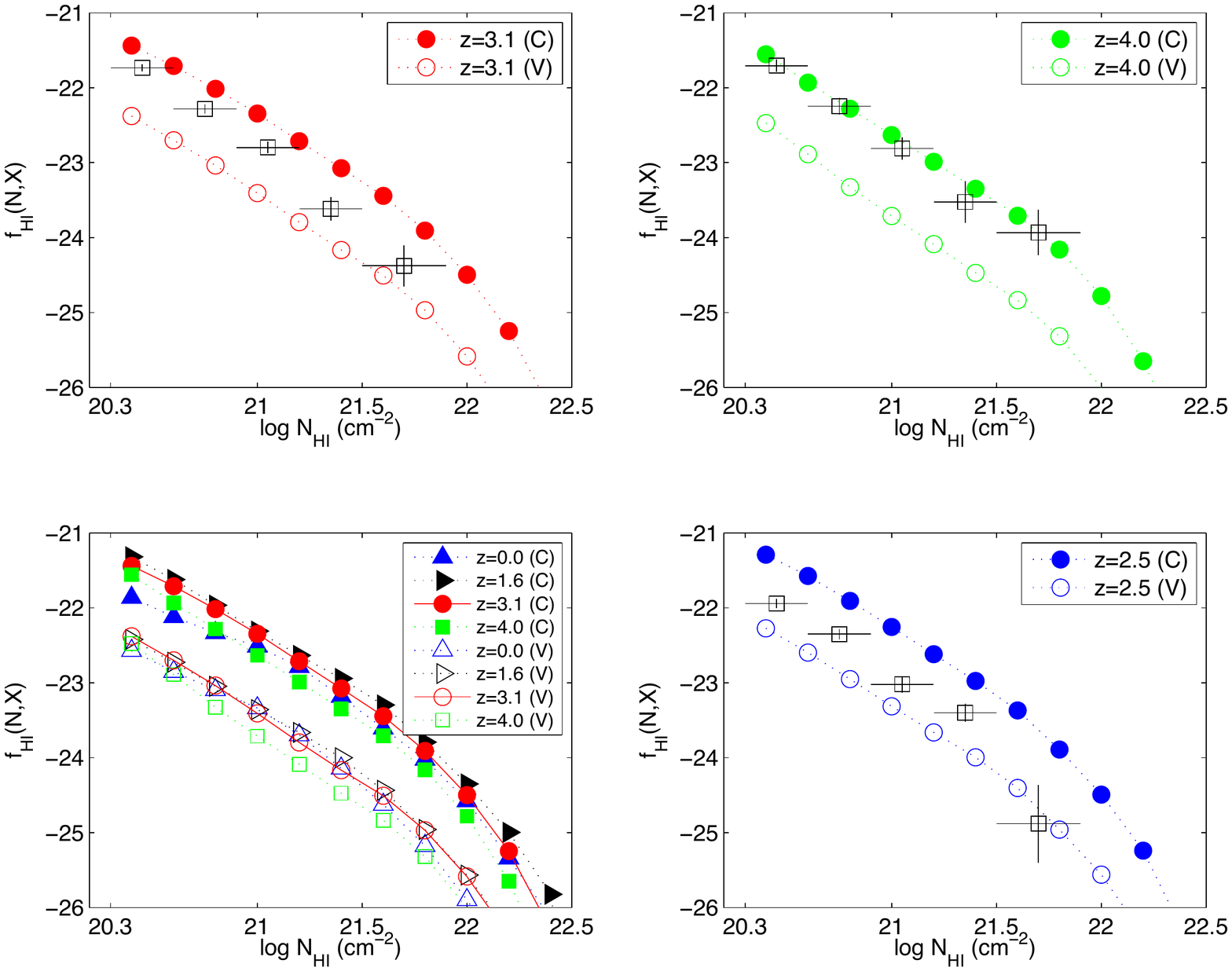}}
\vskip -0.5cm
\caption{
Column density distributions, defined to be the number of
of DLAs per unit column density per unit absorption length,
at $z=2.5$ (lower right),
at $z=3.1$ (upper left),
at $z=4.0$ (upper right),
separately,
and together for $z=0,1.6,3.1,4.0$ (lower left).
In each panel, two sets of simulation results are shown,
one for the ``C" run (solid dots) and ``V" run (open circles).
The corresponding observational data for each of the individual redshifts
are an updated version with SDSS DR7 
from \citet[][]{2005Prochaska}, shown as open squares.
}
\label{fig:colhis}
\end{figure}

Let us now address the fundamentally important observable: 
the column density distribution of DLAs and its evolution.
Figure~\ref{fig:colhis} shows the column density distribution at several redshifts
from $z=0$ to $z=4$.
Where comparisons can be reliably made with observations, at $z=2.5$, $z=3.1$ and $z=4$,
we see that the overdense run ``C" and underdense run ``V" appropriately
bracket the observational data in amplitude.
Similar to the situation for the velocity distribution function (Figure~\ref{fig:v90his}),
the strong environmental dependence of the column density distribution 
renders it impractical to make vigorous comparisons between the simulations
and observations. 
Given that the amplitude of observed column density distribution lies between
that of ``C" and that of ``V" run,
and the shapes of both simulated functions are in reasonable agreement with observations
we tentatively conclude that the standard LCDM model
can reasonably reproduce the observed the column density distribution.
Note that the shape at the highest column end depends on the treatment
of high density regions, for which we have used an empirical relation.
Ultimately, when pc resolution is reached, we can make more definitive tests. 
What is also interesting is that the variations between different environments are larger than
the redshift evolution of the column density distribution in each run.
It is further noted that, seen in the lower-left panel of 
Figure~\ref{fig:colhis},
the evolution of the column density distribution in ``C" and ``V" run 
is different.
In ``C" run, we see weak evolution from $z=4$ to $z=1.6$ and
then a relatively large drop in amplitude at $z=0$.
In ``V" run, on the other hand,
we see practically little evolution from $z=3.1$ to $z=0$.
This likely reflects the dynamical stage of a simulated sample,
where the ``C" run is more dynamically advanced than the ``V" at a same redshift,
consistent with the behavior seen in Figure~\ref{fig:dishis} below.

\begin{figure}[h]
\centering
\resizebox{3.4in}{!}{\includegraphics[angle=0]{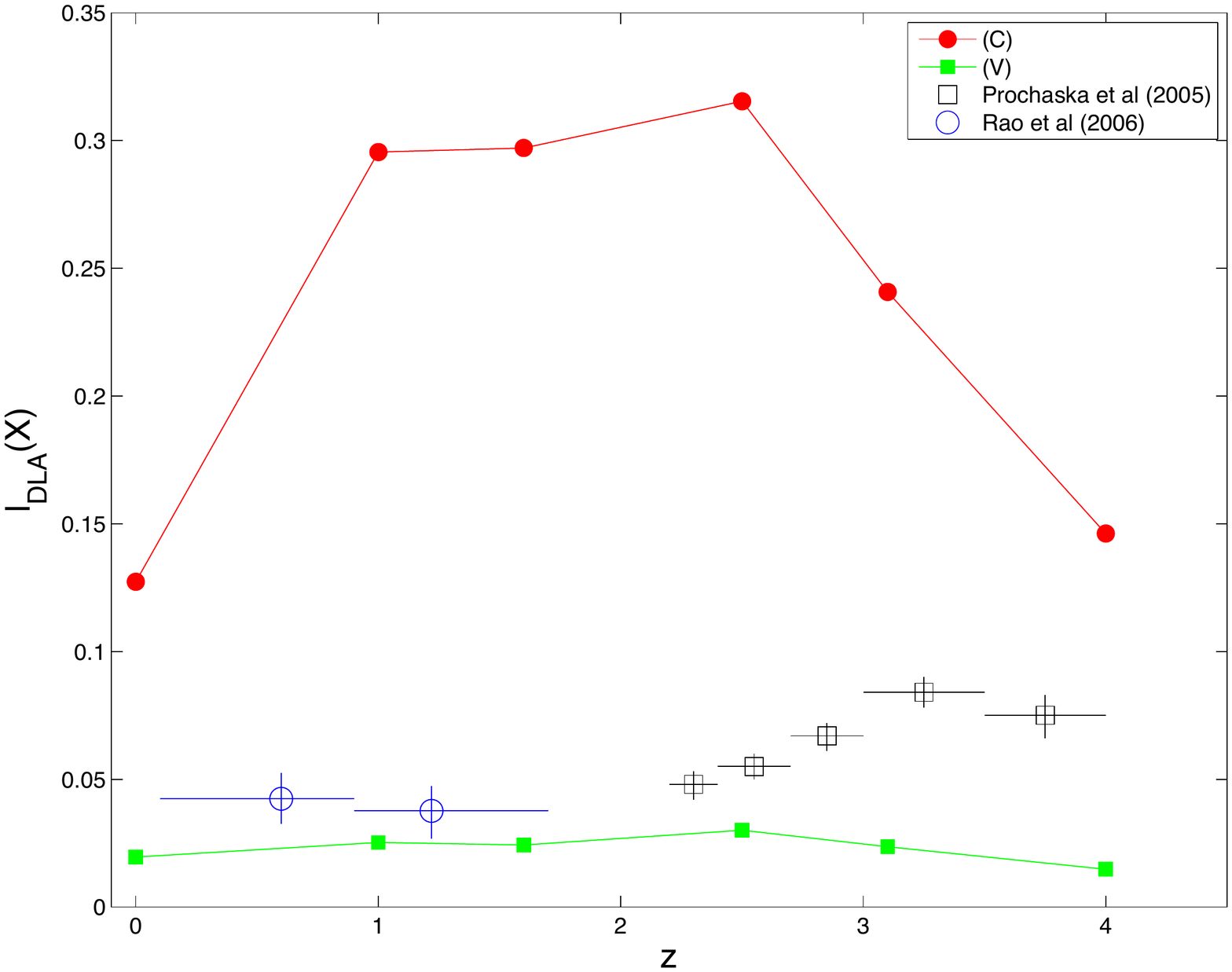}}
\hskip -1cm
\resizebox{3.4in}{!}{\includegraphics[angle=0]{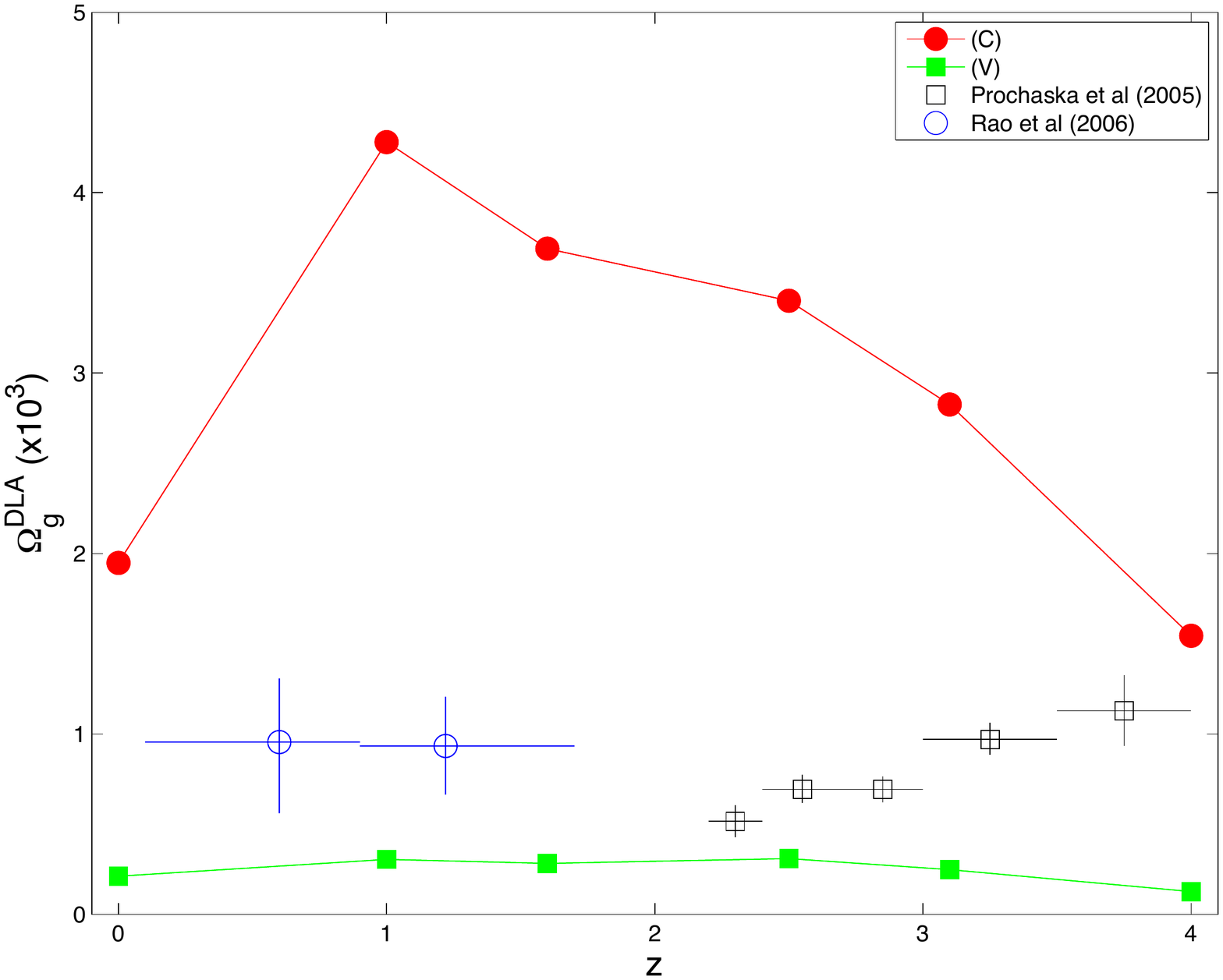}}
\caption{
Left panel: the redshift evolution of the DLA line density
for the ``C" run (solid dots) and ``V" run (solid squares).
The observational data at $z>2$
are an updated version with SDSS DR7 
from \citet[][]{2005Prochaska}, shown as open squares,
the observational data at $z<2$
are from \citet[][]{2006Rao}, shown as open circles.
Right panel: the redshift evolution of the neutral gas density in DLAs
for the ``C" run (solid dots) and ``V" run (solid squares).
The observational data at $z>2$
are an updated version with SDSS DR7 
from \citet[][]{2005Prochaska}, shown as open squares,
the observational data at $z<2$
are from \citet[][]{2006Rao}, shown as open circles.
}
\label{fig:Omega}
\end{figure}

Left panel of Figure~\ref{fig:Omega}
shows the redshift evolution of DLA line density,
defined to be the number of DLAs per unit absorption length.
Right panel of Figure~\ref{fig:Omega}
shows the redshift evolution of neutral gas density in DLAs.
Inherited from the situation shown in Figure~\ref{fig:colhis},
there is a large variation of both plotted quantities 
between the two (``C" and ``V") runs.
What is reassuring is that the observed data lie sensibly
between results from these two bracketing environments.
If one assumes that the cosmic mean of each of the two plotted quantities
should lie between ``C" and ``V" run,
reading the range spanned by the two runs suggests that 
the LCDM model is likely to agree with observations
to within a factor of $\sim 2$ with respect to both quantities,
although what the overall temporal shape will look like is difficult to guess.

To firmly quantify these important observables and to more precisely 
assess the agreement/disagreement between the predictions of the LCDM model and observations,
a larger set of simulations sampling, more densely, different environments
in a statistically correct fashion will be necessary,
so is a more accurate treatment of the transition from atomic to molecular
hydrogen in very high density regions (that likely affects the shape at the high column density end).
We reserve this for future work.

\subsection{Metallicity Distribution and Evolution}

\begin{figure}[h]
\centering
\resizebox{5.5in}{!}{\includegraphics[angle=0]{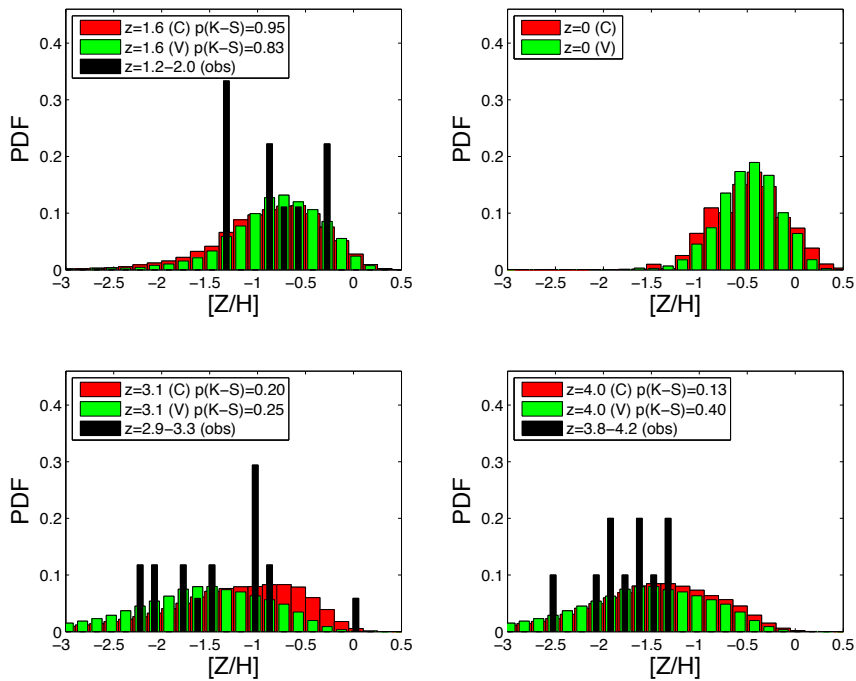}}
\vskip -1cm
\caption{
shows the DLA metallicity distributions
at four redshift, $z=0,1.6,3.1,4.0$ for both ``C" (red histograms) and ``V" (green histograms) run.
The observational data are from \citet[][]{2005Prochaska}, shown as black histograms.
Because there is non-negligible evolution,
the comparisons between simulations at a given redshift 
are only made with observed DLAs within a narrow redshift window, as shown.
Probabilities that simulated and observed samples are drawn from the same underlying distribution
are indicated in each panel, separately for ``C" and ``V" run.
}
\label{fig:mtlhis}
\end{figure}

The current set of simulations is vastly superior to those used in our
earlier work addressing the observed relatively weak but
non-negligible evolution of DLA metallicity \citep[][]{2003dCen}
and here we return to this critical issue.
Figure~\ref{fig:mtlhis} shows the DLA metallicity distributions
at four redshift, $z=0,1.6,3.1,4.0$.
For the three redshifts, $z=1.6,3.1,4.0$,
where comparisons can be made,
we find that the agreement between simulations and observations 
at $z=1.6$ to $z=0$ is excellent,
as K-S tests show.
This is a non-trivial success, given that our feedback prescription has
essentially one free parameter, that is, the supernova energy that
is driving galactic winds transports energy, metals and mass throughout
interstellar (ISM), circumgalactic (CGM) and intergalactic space (IGM). 
Furthermore, the absolute amount of metals is totally fixed 
by requiring that 25\% of stellar mass with metallicity equal to 
$10\zsun$ returning to the ISM, CGM and IGM. 
The agreement indicates that our choices of both the supernova ejecta mass and its metallicity 
and the explosion energy, which are inspired by theories of stellar interior and direct observations,
may provide a reasonable approximation of truth.

We see that the peak of the DLA metallicity distribution evolves
from $[Z/H]=-1.5$ at $z=3-4$, 
to $[Z/H]=-0.75$ at $z=1.6$,
and to $[Z/H]=-0.5$ at $z=0$.
Thus, both simulations and nature indicate that there is a weak but real evolution in DLA metallicity.
What is also important to note is that, in agreement with observations,
simulations indicate that the distribution of metallicity 
is very wide, spanning three or more decades at $z\ge 1.6-4$.
This wide range reflects the rich variety of neutral gas that composes
the DLA population, from relatively pristine gas clouds falling
onto or feeding galaxies, to metal-enriched cold clouds
that are falling back to (galactic fountain) or still moving away from (due to entrainment of galactic winds)
galaxies, to cold neutral gas clouds in galactic disks.
There is a metallicity floor at $[Z/H]\sim -3$ at $z=1.6-4$ and
that floor moves up to $[Z/H]\sim -1.5$ by $z=0$, 
consistent with observations \citep[][]{2003Prochaska}.
The distribution at $z=0$ is significantly narrower, partly reflecting the overall enrichment of
the IGM and partly due to much reduced variety of DLAs with galactic disks becoming
a more dominant contributor to DLAs (see discussion below).

\citet[][]{2010Ellison} find that proximate DLAs (PDLAs), those within a velocity distance from the QSO $\Delta v<3000$km/s,
seem to have metallicity higher than the more widely studied, intervening DLAs.
It seems conceivable that the total sample of PDLAs plus conventional (intervening) DLAs may 
somewhat shift the metallicity distribution to the right,
perhaps bringing it to a still better agreement with our simulations.

\begin{figure}[h]
\hskip 1.7cm
\centering
\hskip -2.5cm
\resizebox{3.31in}{!}{\includegraphics[angle=0]{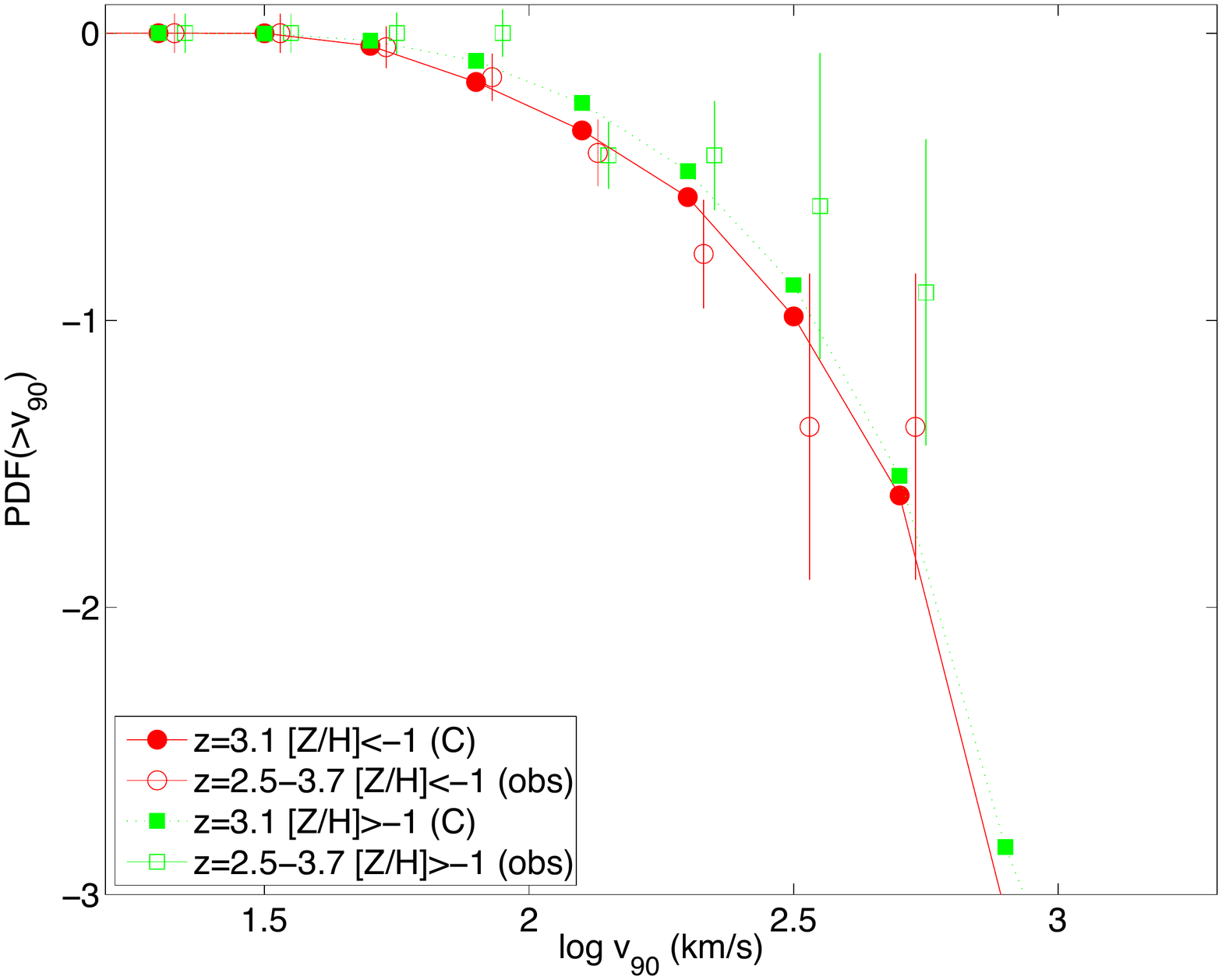}}
\resizebox{3.31in}{!}{\includegraphics[angle=0]{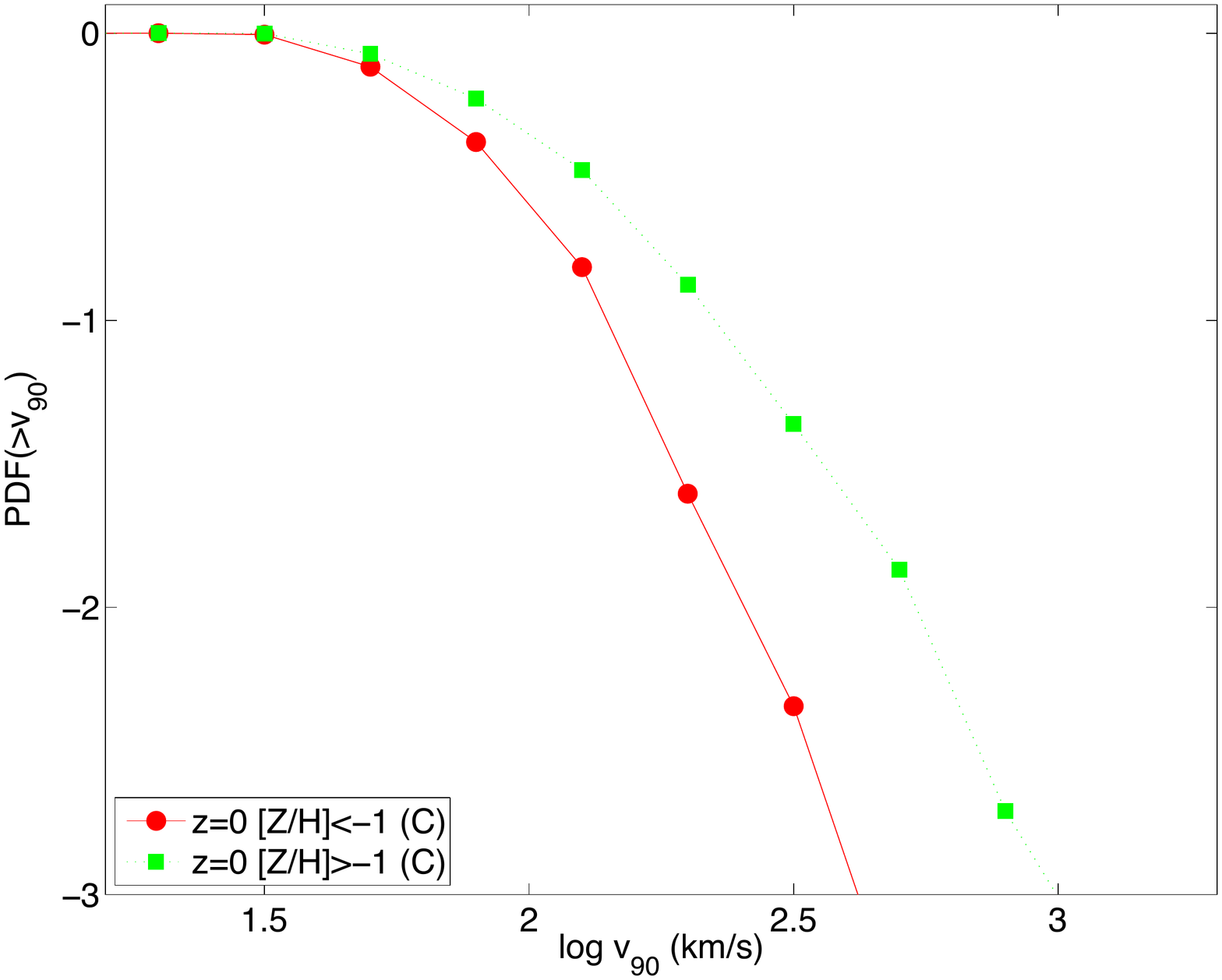}}
\caption{
Left panel: shows the cumulative velocity width probability function for two subsets
of the DLA sample, divided by DLA metallicity at $[Z/H]=-1$, at $z=3.1$ from the ``C" run;
the results with ``V" run, not shown, are nearly identical.
The observational data is an updated version of \citet[][]{1997Prochaska}, 
divided into two subsets such that the ratio of the number of DLAs 
in the two subsets is equal to that of the simulated sample to enable a fair comparison.
The observed data points are slightly shifted to the right by a small amount for
more clear reading.
Right panel: $z=0$ from the ``C" run.
}
\label{fig:v90hisZdiv}
\end{figure}

Observations have found a strong positive correlation between galaxy mass
and metallicity \citep[e.g.,][]{2006Erb}. 
We divide the simulated DLA sample at $z=3.1$ into two subsets,
one with metallicity less than $[Z/H]=-1$ and the other more than $[Z/H]=-1$.
We then compute the velocity width functions separately for each subset,
which are shown as solid dots (lower metallicity) and solid squares (higher metallicity)
in the left panel of Figure~\ref{fig:v90hisZdiv}.
What we see is that there is a small excess of large velocity width DLAs 
for the higher metallicity subset compared to the 
lower metallicity one.
This is of course in the sense that is consistent with the observed metallity-mass
relation.
However, current observational data sample is consistent with simulations,
and the difference between the two simulated subsets
and between the two observed subsets is statistically insignificant.
A larger sample (by a factor of 4) may allow for a statistically significant test.
Do we expect a larger difference in the disk model \citep[][]{1986Wolfe,1997Prochaska}?
We do not have a straight answer to this question, without a very involved modeling.
However, we we suggest that the picture we have presented,
where DLAs arise from a variety of galactic systems, in a variety
of locations of widely varying metallicity (see the gallery in \S 3.1),
would be consistent with the small difference found,
because the velocity widths of large width DLAs do not strongly correlate with galaxy mass 
(see Figure~\ref{fig:v90max}).
In other words, the observed correlation between metallicity
and galaxy mass is largely washed out by DLAs that do not arise in disks
and whose metallicity do not strongly correlate with galaxy mass.
If one combines the information provided by Figure~\ref{fig:v150his} 
and Figure~\ref{fig:v90hisZdiv}, one may reach a similar conclusion.

\begin{figure}[h]
\centering
\resizebox{6.0in}{!}{\includegraphics[angle=0]{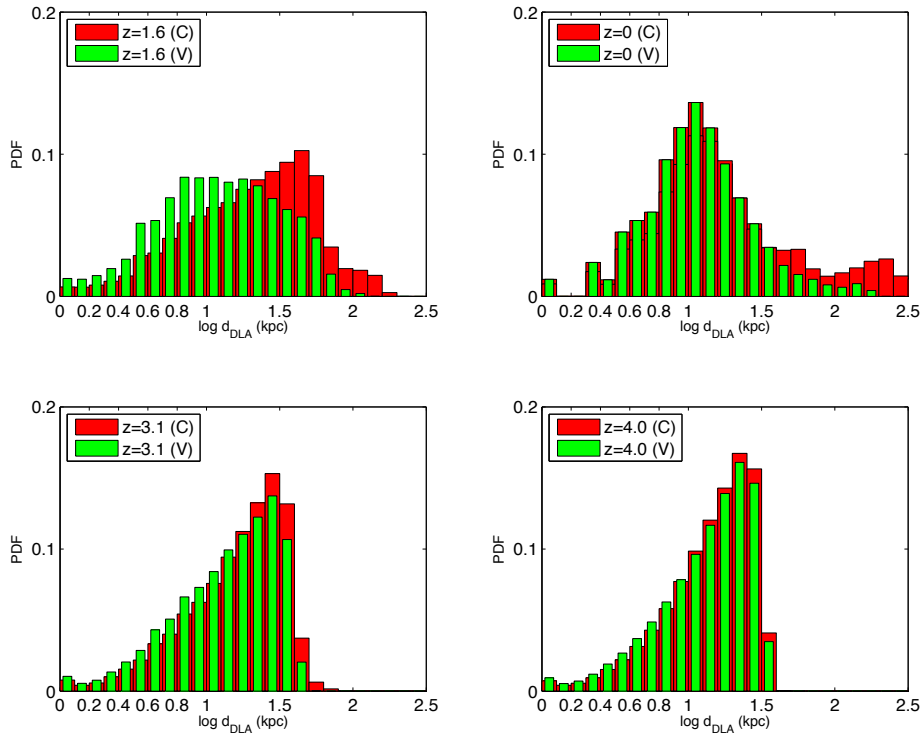}}
\vskip -1cm
\caption{
shows the distributions of distance of DLA from the center of galaxy (i.e., impact parameter)
for ``C" (red histograms)
and ``V" (green histograms) run 
at redshift $z=0$ (top right), $z=1.6$ (top left), 
$z=3.1$ (bottom left) and $z=4.0$ (bottom right).
}
\label{fig:dishis}
\end{figure}

The implication may be that DLAs do not arise predominantly 
in gaseous disks of spiral galaxies at high redshift, 
in agreement with \citet[][]{2001Maller} and \citet[][]{2010Hong}.
We shall elaborate further on this significant point.
In Figure~\ref{fig:dishis} we show the distribution of physical 
distance of DLAs from the galactic center (i.e., impact parameter)
at four redshifts, $z=0,1.6,3.1,4.0$.
Since we have shown in Figure~\ref{fig:v150his} that 
DLA incidence contribution peaks at $\sim 10^{11.5}\msun$,
let us make a simple estimate of their size at $z\sim 3$.
As a reference, let us take the radius of Milky Way (MW) stellar disk to be $15$kpc.
Taking MW to $z=3$ self-similarly would give a radius of $3.8$kpc and
for a $10^{11.5}\msun$ galaxy the stellar disk radius would be $2.5$kpc at $z=3$,
corresponding to $0.40$ in the shown x-axis.
Observed large galaxies (of mass likely in the range $\sim 10^{11}-10^{12}\msun$)
at $z\sim 3$ appear to have sizes of $\sim 1-10$kpc 
\citep[][]{1997Lowenthal,2004Ferguson,2006Trujillo, 2007Toft, 2007Zirm, 2008Buitrago},
roughly consistent with the simple scaling.
The distance distribution peaks at $d_{\rm DLA}\sim 20-30$kpc at $z=3-4$,
which is much larger than a few kpc of the observed (or expected based on $z=0$ galaxies) 
stellar disk size at $z\sim 3$.
It is noted that the virial radius of a Milky Way size galaxy is about $\sim 50$kpc at $z=3$.
Thus, these gaseous structures occur at about half the virial radius at $z=3$,
Thus, we conclude that at $z=3-4$ 
most of the DLAs {\it do not} arise from large galactic stellar disks.
They appear to come from regions that are $\sim 5-8$ larger than the stellar disks.

\begin{figure}[t]
\hskip -0.7cm
\centering
\resizebox{3.71in}{!}{\includegraphics[angle=0]{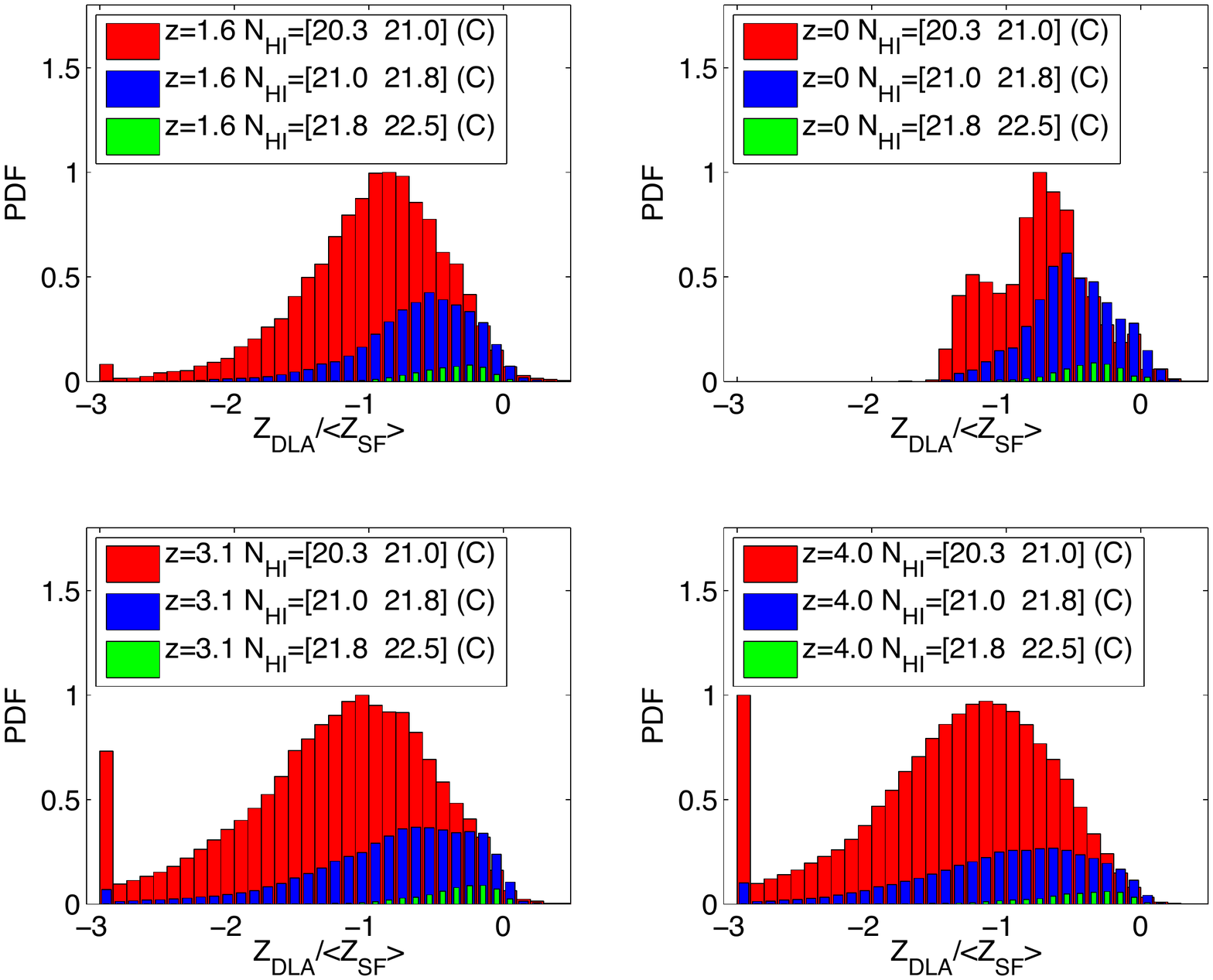}}
\hskip -1.8cm
\resizebox{3.71in}{!}{\includegraphics[angle=0]{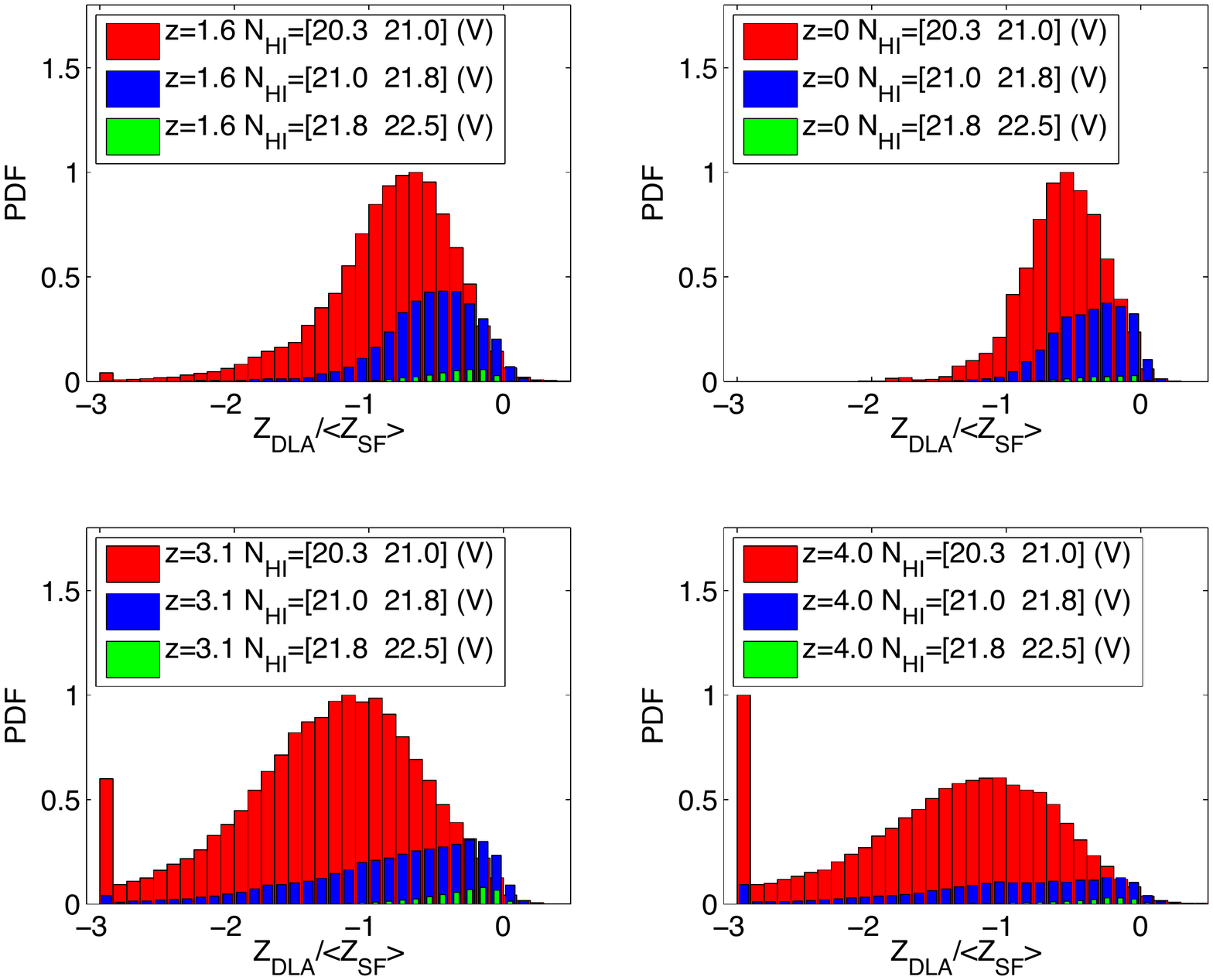}}
\caption{
shows the distribution of the ratio of gas metallicity for DLAs at different column density ranges
to the mean metallicity of ongoing star-forming gas in the ``C" run (left set of four panels)
at $z=0,1.6,3.1,4.0$ and in the ``V" run  (right set of four panels).
We expect that gas that has the x-axis value close or greater
than $0$ may be forming stars.
}
\label{fig:ZratSFRhis}
\end{figure}

The ubiquitous extended structures - galactic filaments -
appear to be at the right distances of $d_{\rm DLA}\sim 20-30$kpc, seen in the gallery examples in \S 3.1.
While the extremely close association of galactic filaments with 
galaxy interactions suggest that the host galaxies are 
likely experiencing starbursts, as seen in the gallery examples,
the clouds that give rise to DLAs 
do not appear to have ongoing {\it in situ} star formation. 
Clearly, most DLAs do not arise in disks and most DLAs have low metallicities,
as we have shown, are self-consistent.
In other words, aside from those DLAs that arise from galactic disks and are metal rich,
the metallicity of the vast majority of more metal poor DLAs do not appear to be forming stars.
It may be that, if and when the gas in the galactic filaments forms stars,
either they are destroyed by star formation feedback and remove themselves from the DLA category
or they have already incorporated into disks of galaxies. 
We suggest that our model gives a natural explanation 
to the apparent puzzle of the lack of obvious star formation of gas-rich DLAs \citep[][]{2006Wolfe}.
On the other hand, the inferred cooling rates of DLAs may be provided, in part, by 
radiative heating from the host galaxy (see Figure~\ref{fig:SFRhis} below) and possibly in part
by compression heating as we frequently see higher external pressure in \S 3.1.
Figure~\ref{fig:ZratSFRhis} shows
the ratio of gas metallicity for DLAs at different subset of DLAs with different column density ranges
to the mean metallicity of ongoing star-forming gas.
It is clear that only the high end of the high column density range ($N_{\rm HI}\ge 10^{22}$cm$^{-2}$) 
DLAs are forming stars; most of the DLAs have little star formation.

Returning to Figure~\ref{fig:dishis}, at $z=1.6$ there is a very interesting divergence between the two 
distributions for ``C" and ``V" run, where the distribution for ``C" run 
peaks at $d_{\rm DLA}\sim 40$kpc and for ``V" run at $d_{\rm DLA}\sim 10$kpc.
This is consistent with the expectation that the overdense region in the ``C" run
and the underdense region in the ``V" run start to ``feel" the difference 
in their respective local large-scale density environment
and evolve differently dynamically.
That is, in ``C" run gravitational shock heating due to large-scale structure formation
begins to significantly affect the cold gas in galaxies, whereas in ``V" run the galaxies have not changed
significantly since $z=3-4$ except that they are now somewhat smaller due to 
lower gas density at lower redshift.
By $z=0$ the two distributions once again become nearly identical;
this is rather intriguing and may reflect the following physical picture:
while galaxies in the ``V" run has by now dynamically ``caught up"
with the field galaxies in the ``C" run, giving rise to the similar
gaussian-like distribution centered at $d_{\rm DLA}=10$~kpc,
the original gas-rich galaxies in the ``C" run have fallen into the cluster, lost gas and 
``disappeared" from the DLA population. 
While there is almost no DLA that is further away than $50$kpc at $z=3-4$,
there is a second bump  at $d_{\rm DLA}=100-300$kpc in the distribution
for the ``C" run at $z=0$.
This bump is likely due to gas rich satellite galaxies orbiting larger galaxies or small groups
of mass $10^{12}-10^{13}\msun$.
Beyond $d_{\rm DLA}=300$kpc, there is no DLA in the ``C" run, which is due to gas-starvation
of galaxies in still larger groups or clusters at $z=0$.
With direct inspection of simulation data 
we find that there is virtually no gas rich galaxies within the virial radius 
of the primary cluster in the ``C" run.

What is also interesting is that the peak distance of $d_{\rm DLA}\sim 10$kpc at
$z=0$ is totally consistent with the notion that gaseous disks of field galaxies, 
like the one in our own Galaxy, significantly contribute to DLAs.
Right panel of Figure~\ref{fig:v90hisZdiv} shows 
the velocity distributions of two subsets of DLAs,
divided at metallicity at $[Z/H]=-1$ at $z=0$.
Here we see a very clear difference between the two distributions:
the higher metallicity subset have large velocity widths,
i.e., there is a strong positive correlation between metallicity and velocity width at $z=0$.
This supports the picture that a large fraction of DLAs arise in gaseous disks of large field galaxies. 
Most of the DLAs at $z=0$ have a higher metallicity of 
$[Z/H]\ge -1.0$ with the overall distribution peaking at $[Z/H]=-0.5$,
also providing support for this picture.
Therefore, by $z=0$ the situation appears to have reversed:
galactic disks of large galaxies make a major contribution to DLAs at $z=0$.
The fact that the peak distance has dropped from $30-40$kpc at $z=3-4$ to $10$kpc at $z=0$
is physically in partly 
due to a large decrease (a factor of $\sim 100$) in the mean gas density of the universe from $z=3-4$ to $z=0$.

\subsection{Size Distribution}

\begin{figure}[h]
\hskip -0.7cm
\centering
\resizebox{3.71in}{!}{\includegraphics[angle=0]{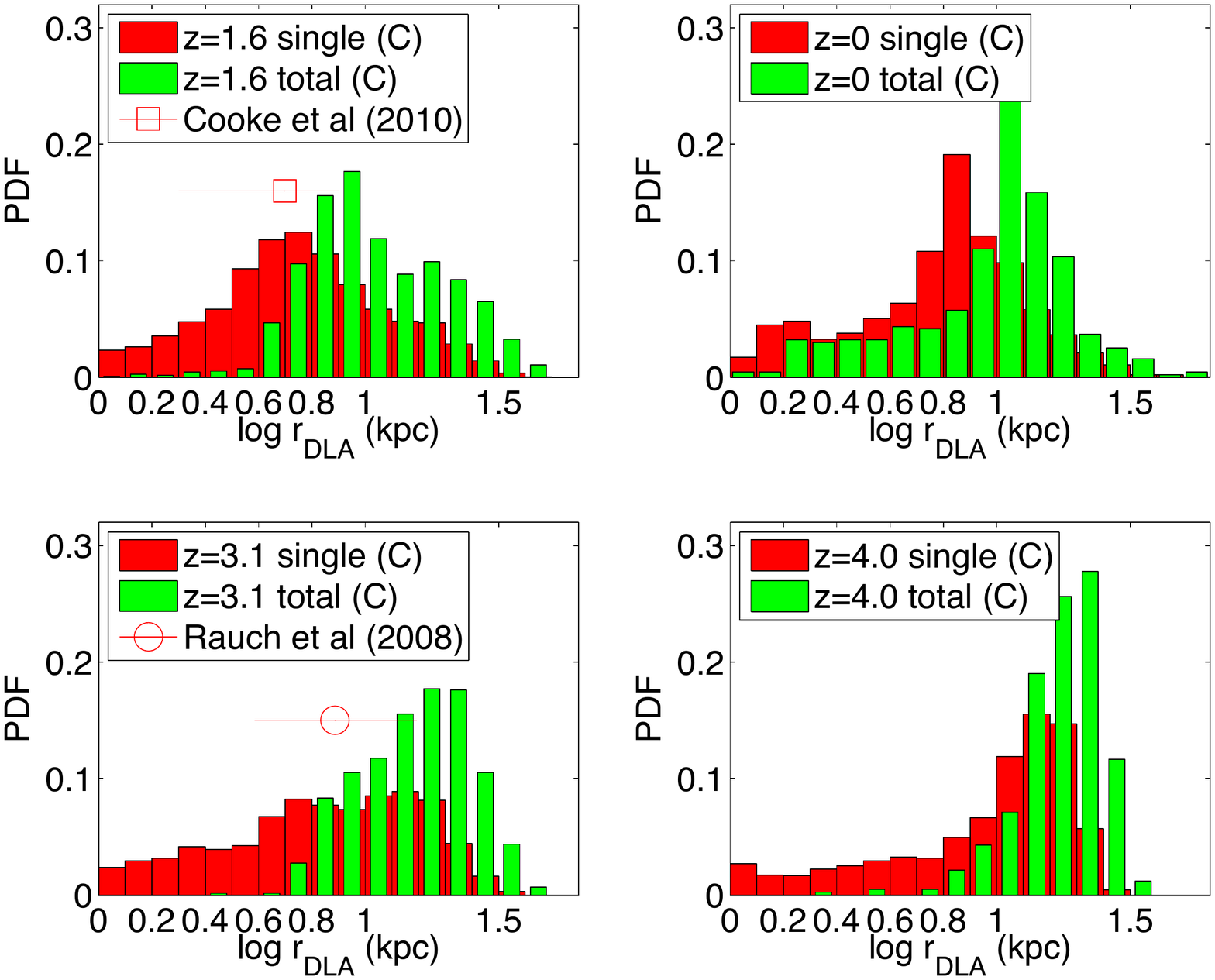}}
\hskip -1.8cm
\resizebox{3.71in}{!}{\includegraphics[angle=0]{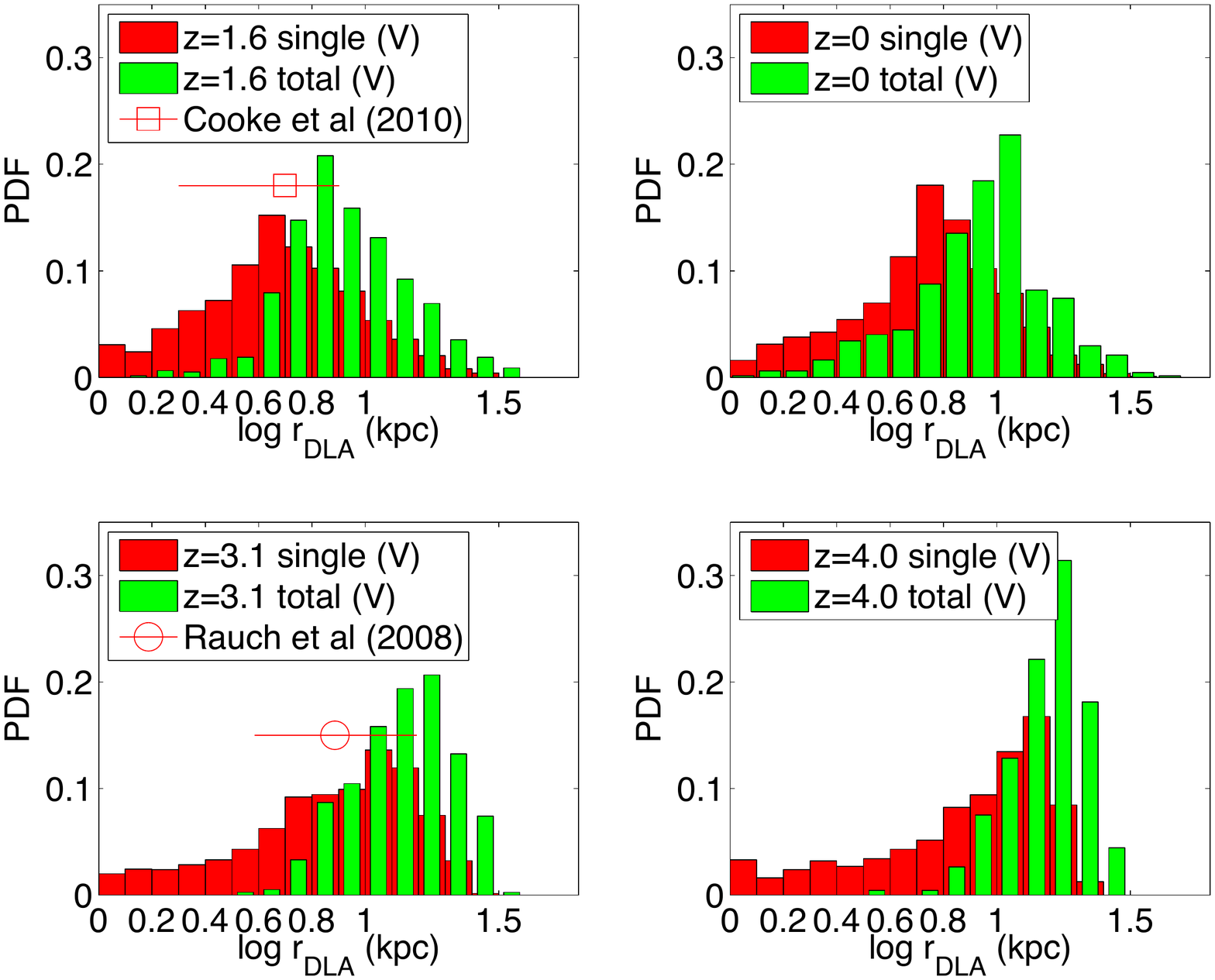}}
\caption{Left set of four panels: 
the DLA size distribution at redshift $z=0,1.6,3.1,4.0$ for ``C" run.
Each individual DLA size $r_{\rm DLA}$ (see text for definition) 
is shown as red histograms, 
whereas the total DLA size of a galaxy $r_{\rm tot}$ (see text for definition)
is shown as green histograms. 
Right set of four panels: 
the DLA size distribution at redshift $z=0,1.6,3.1,4.0$ for ``V" run.
The observationally inferred DLA size, shown as an open square in both $z=1.6$ panels, 
is from \citet[][]{2010Cooke},
and that shown as an open circle in both $z=3.1$ panels is from \citet[][]{2008Rauch}
with the shown dispersion estimated by this author.  
}
\label{fig:sizehis}
\end{figure}

\begin{figure}[h]
\hskip -0.7cm
\centering
\resizebox{3.71in}{!}{\includegraphics[angle=0]{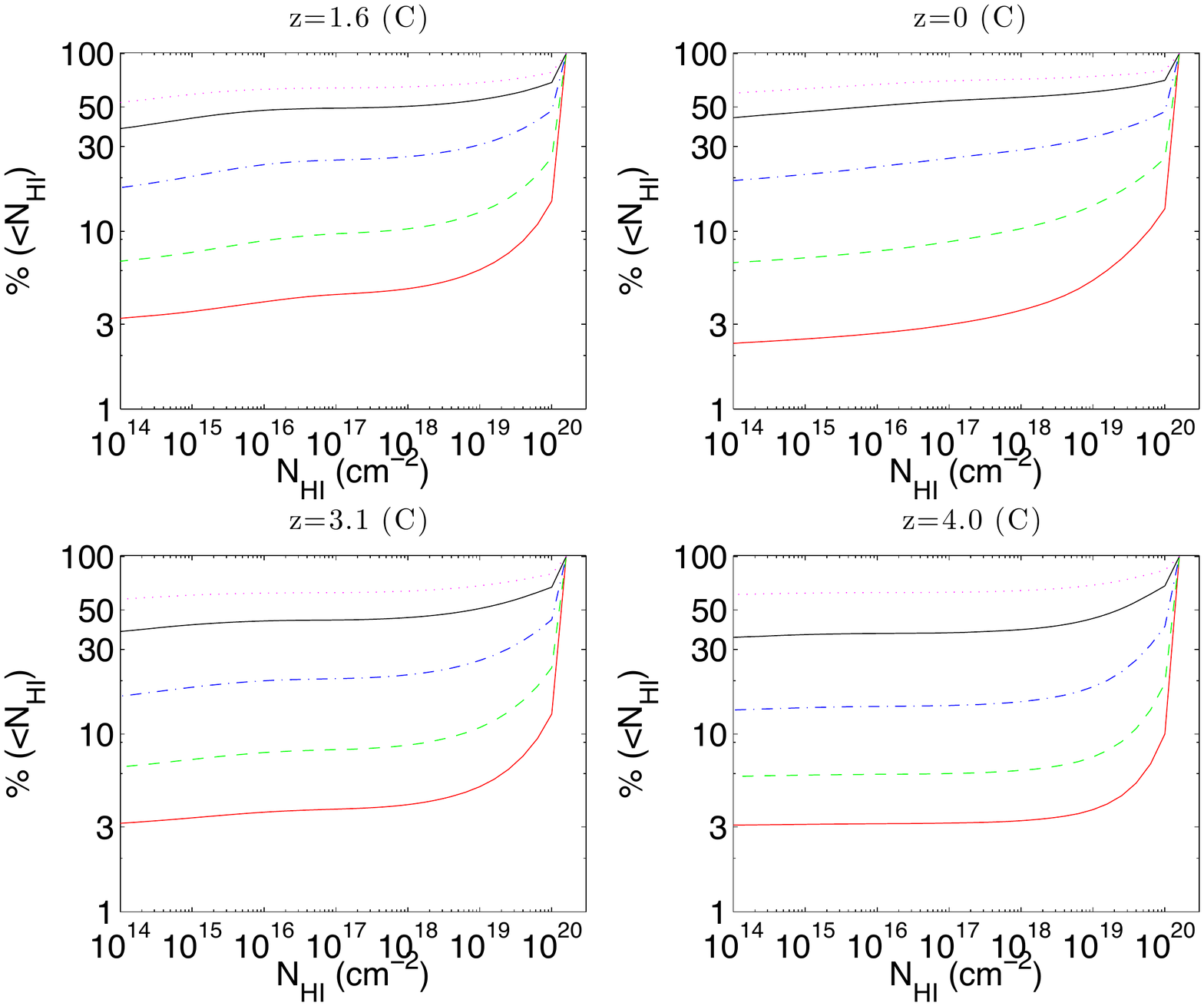}}
\hskip -1.8cm
\resizebox{3.71in}{!}{\includegraphics[angle=0]{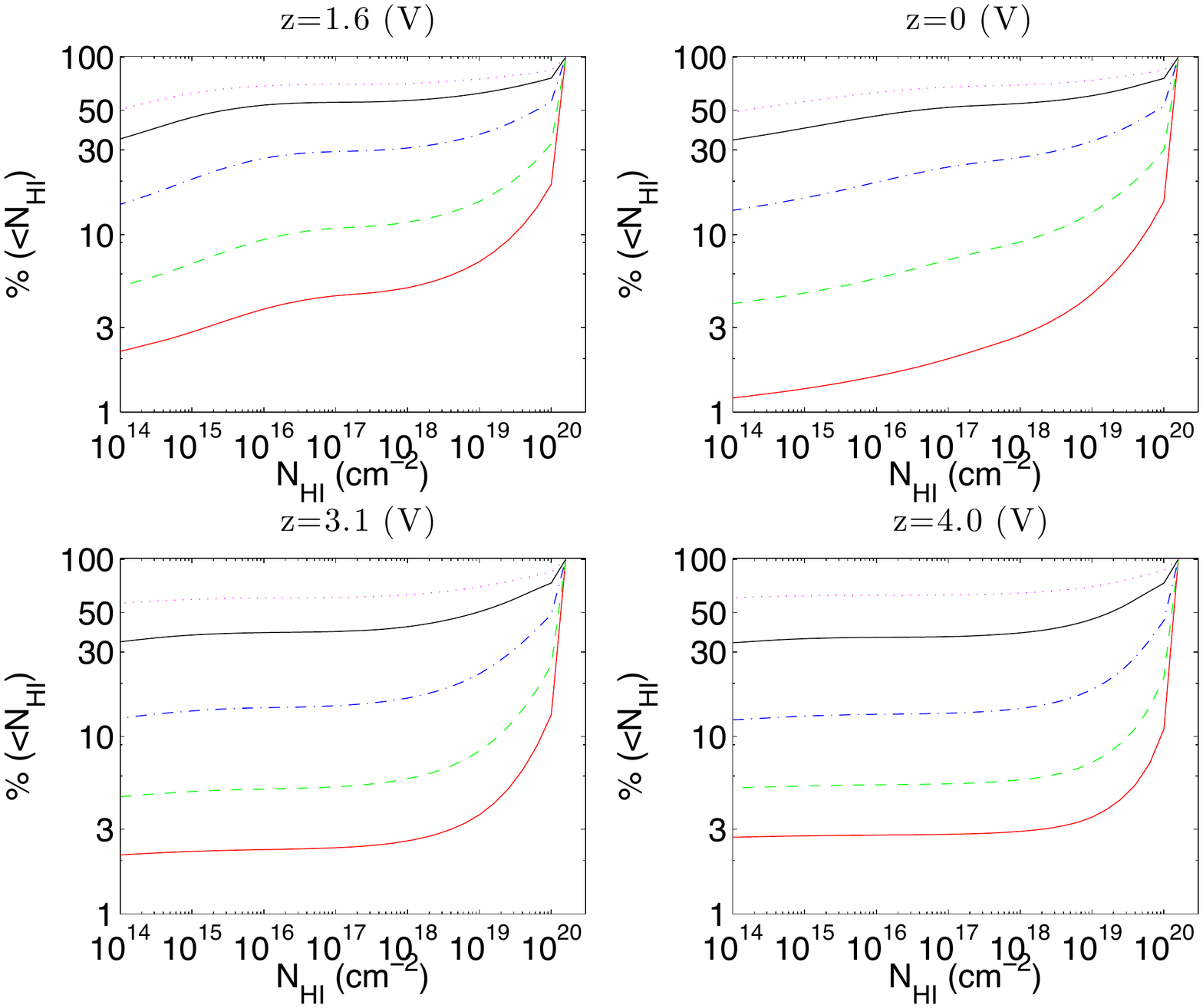}}
\caption{
The probability of the second LOS having a HI column lower than the value shown in the x-axis,
while the first LOS is known to have intercepted a DLA at projected separation  
of $(30,20,10,5,3)$kpc (five curves  from top to bottom shown in each panel) at the redshift in question.
The left set of four panels are at redshift $z=0,1.6,3.1,4.0$ for ``C" run;
the right set of four panels are for ``V" run.
}
\label{fig:PDFcolrad}
\end{figure}

Binary quasars, physical or lensed, provide an unique tool to probe the size of DLAs.
Here we present our predictions of size distributions of DLAs in the LCDM model.
As we have described in \S 2.3,
any cells (of size $0.915h^{-1}$kpc comoving) that are connected by one side in projection
are merged into a ``single isolated" DLA.
The area of each ``isolated" DLA, $A$, is then used to define
the size (radius) of the DLA by $r_{\rm DLA}=(A/\pi)^{1/2}$.
The total area of all isolated DLA associated with a galaxy along three orthogonal directions
($x,y,z$), $A_x$, $A_y$ and $A_z$, are summed to obtain $A_{\rm tot}=\sqrt{A_x^2+A_y^2+A_z^2}$
and the total DLA size (radius) of the galaxy is defined to be 
$r_{\rm tot}=(A_{\rm tot}/\pi)^{1/2}$.
Note that, if DLAs arises from a thin disk, the $A_{\rm tot}$ computed this way
will be the exact size of the disk face on, regardless of its orientation.
On the other hand, if each DLA cloud is a sphere, this method overestimate the size (area)
by a factor of $\sqrt{3}$.

\begin{figure}[h]
\hskip -0.7cm
\centering
\resizebox{3.71in}{!}{\includegraphics[angle=0]{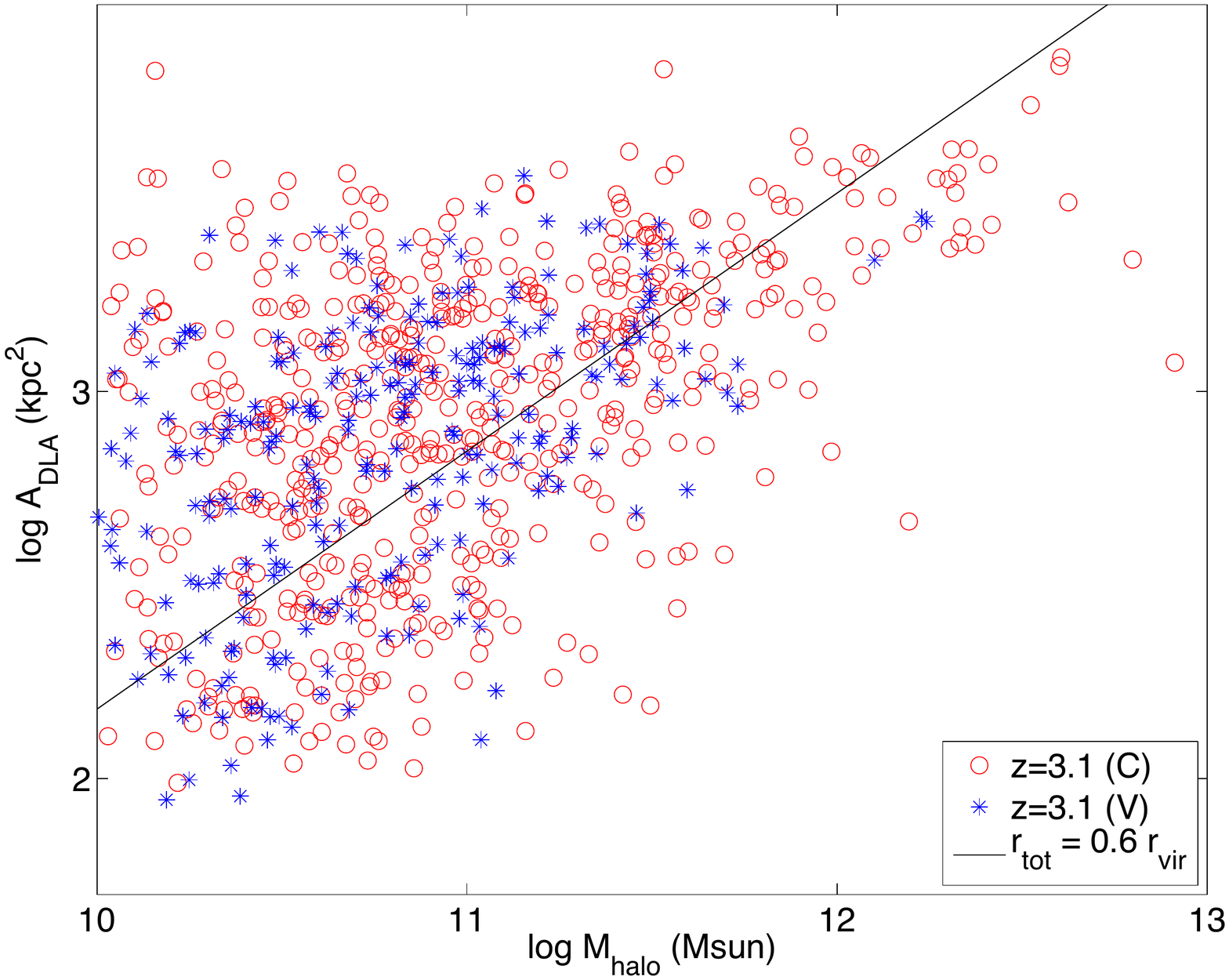}}
\hskip -1.8cm
\resizebox{3.71in}{!}{\includegraphics[angle=0]{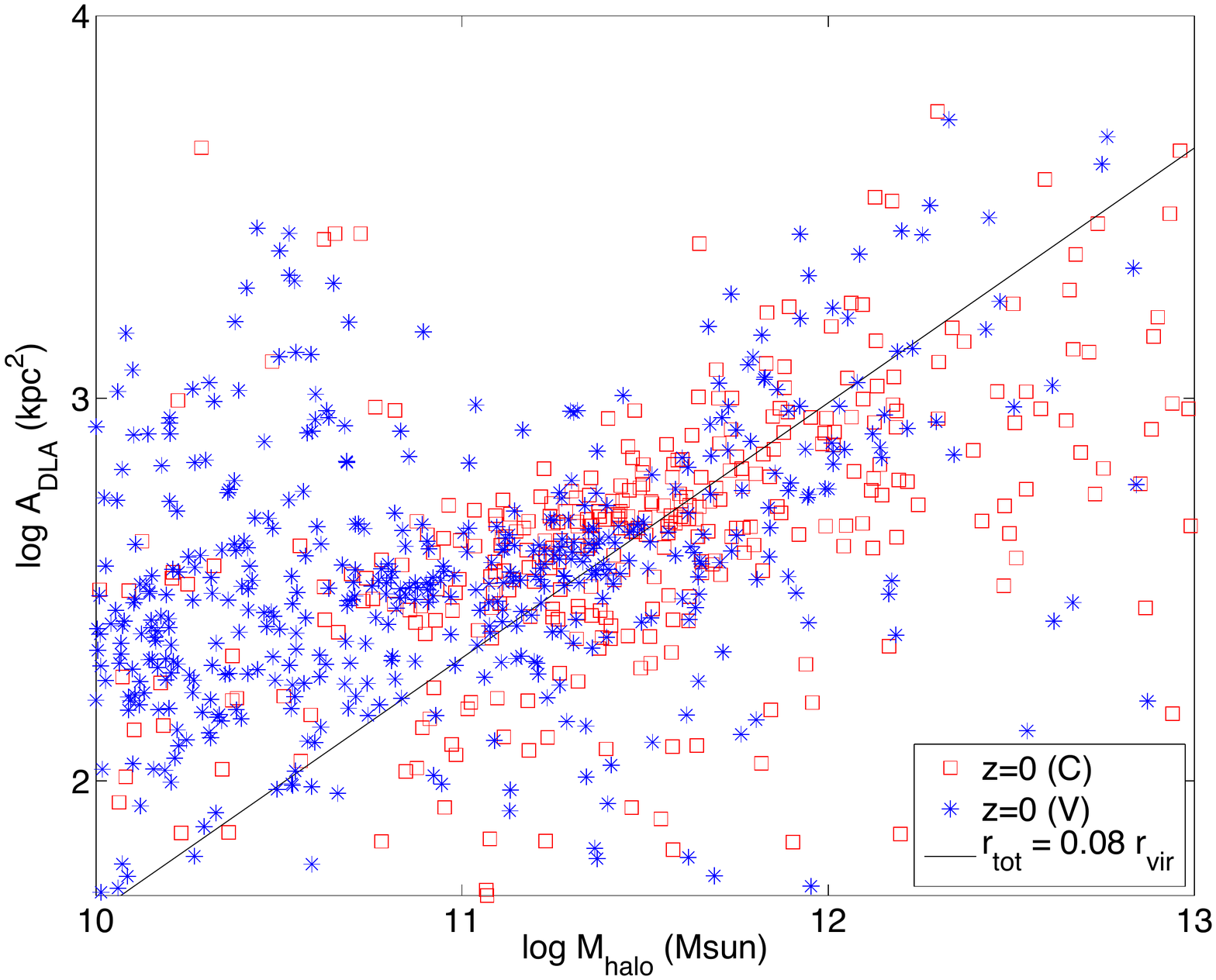}}
\caption{
Left panel: 
a scatter plot of DLA size of each galaxy against its halo mass 
at $z=3.1$.
Also shown as a black straight line is a proposed relation
between the radius of DLA and virial radius of the halo that roughly goes 
through the median at $z=3.1$:
$r_{\rm DLA} = 0.30 r_{\rm vir}$.
Right panel: 
a scatter plot of DLA size of each galaxy against its halo mass 
at $z=0$.
Also shown as a black straight line is a linear relation
between the radius of DLA and virial radius of the halo at $z=0$:
$r_{\rm DLA} = 0.08 r_{\rm vir}$.
}
\label{fig:sizeMtot}
\end{figure}

Figure~\ref{fig:sizehis} shows the size (radius) distribution at redshift $z=0,1.6,3.1,4.0$
for individual DLA size and total DLA size of each galaxy.
Figure~\ref{fig:PDFcolrad} presents the size information in a different way,
where we show the probability that, for a random pair of sightlines separated
by $(30,20,10,5,3)$kpc at the redshift in question,
one sightline intercepts a DLA and the other intercepts
a column density lower than shown in the x-axis.
We see that, in the sense that is consistent with distance distributions shown 
in Figure~\ref{fig:dishis},
on average, individual DLA size as well as
the total DLA size of galaxies 
are larger at high redshift than at lower redshift.
The individual DLA size distribution appears to sharply peak
at $r_{\rm DLA}\sim 15$kpc at $z=4$,
then move left to a broader peak at $r_{\rm DLA}\sim 10$kpc at $z=3.1$,
then sharpen somewhat to peak at $r_{\rm DLA}\sim 5$kpc at $z=1.6$,
and finally move rightward slightly to peak at  $r_{\rm DLA}\sim 7$kpc at $z=0$ for the ``C" run.
The numbers for the ``V" run are largely comparable to the ``C" run:
$(15,12,4,6)$kpc at $z=(4,3.1,1.6,0)$, respectively.
What is quite remarkable is that, at $z=1.6$ where comparison can be made,
the predicted size distribution and the available observation
agree better than anyone would have expected.
This is a testament to the success of the LCDM model and 
physical treatment of our simulations,
especially considering other agreements that we have already found,
for example, with respect to metallicity distribution, column density distribution, kinematics. 

Intriguingly, \citet[][]{2008Rauch} find a new population of faint Lyman alpha emitters which they 
advocate may be the host galaxies of DLAs at $z\sim 3$.
Their estimated size, based on observed $\lya$ line emission profile, shown as the open circle 
in $z=3.1$ panels of Figure~\ref{fig:sizehis},
would be consistent with our model,
although the true nature of this population is relatively uncertain.
Given the reasonable agreement in size, we will make some further speculation 
to gauge if our model can accommodate or explain this population as DLAs.
One possible physical mechanism for the $\lya$ emission of this population, if they are indeed DLA hosts,
would be fluorescence due to ionizing photons from the host galaxies.
Assuming that the distance of each DLA cloud from the host galaxy is $25$kpc (Figure~\ref{fig:dishis}),
the size of each DLA cloud is $8$kpc (Figure~\ref{fig:sizehis})
and 
the star formation rate of host galaxy is $10\msun$/yr (see Figure~\ref{fig:SFRhis} below),
then a typical expected star formation rate inferred from their fluorescence 
would be $(8^2\pi)/(4\pi 25^2)\times 10 = 0.26\msun$/yr.
This very rough estimate curiously falls within the range 
of $0.07-1.5\msun$/yr estimated by \citet[][]{2008Rauch} for the observed emitters
if there are $\lya$ emitters at $z\sim 3$.

\begin{deluxetable}{lcccccccc}
\tablecolumns{9}
\tablewidth{0pc}
\tablecaption{Probability of observed QSO pairs - 
column 1: QSO name;
column 2: DLA redshift;
column 3: pair name;
column 4: separation of two images in the sky in arcseconds;
column 5: physical separation of two sightlines at $z_{abs}$;
column 6: HI column density of the first sightline in units of $10^{20}$cm$^{-2}$;
column 7: HI column density of the second sightline in units of $10^{20}$cm$^{-2}$;
column 8: probability from ``C" run;
column 9: probability from ``V" run.
The observed data are taken from Table 1 of \citet[][]{2010Cooke}, which are all lensed binaries (or multiples).
Note that the physical separations at the redshift of DLAs
are computed correctly with the actual geometry of the lenses.
We have also added in the last row the binary quasar LBQS1429-0053 taken from \citet[][]{2009Monier}.
\label{tab:table1}}
\tablehead{
\colhead{QSO} & \colhead{$z_{abs}$} & \colhead{Pair}  & \colhead{$\theta_{abs}$} & \colhead{$d$ (kpc)} & \colhead{$N_1$} & \colhead{$N_2$}  & \colhead{P(C)}   & \colhead{P(V)}}
\startdata
H1413+1143 & 1.440 & B-A & 0.753 & 3.13 & 60 & 9.0 & 0.84 & 0.80  \\ 
 &  & B-D & 0.967 & 4.02 & 60 & 0.25 & 0.11 & 0.13  \\
 &  & B-C & 1.359 & 5.65 & 60 & 0.20 & 0.17 & 0.21   \\
 &  & A-D & 1.118 & 4.64 & 9.0 & 0.25 & 0.13 & 0.16   \\
 &  & A-C & 0.872 & 3.62 & 9.0 & 0.20 & 0.093 & 0.11   \\
 &  & D-C & 0.893 & 3.71 & 0.25 & 0.20 & 0.097 & 0.12   \\
H1413+1143 & 1.486 & D-A & 1.118 & 4.32 & 2.0 & $<0.05$ & 0.097 & 0.11  \\ 
 &  & D-B & 0.967 & 3.73 & 2.0 & $<0.1$ & 0.084 & 0.097  \\
 &  & D-C & 0.893 & 3.45 & 2.0 & $<0.05$ & 0.070 & 0.079  \\
H1413+1143 & 1.662 & B-A & 0.753 & 2.17 & 6.0 & $1.5$ & 0.92 & 0.89  \\ 
 &  & B-C & 1.359 & 3.91 & 6.0 & $0.6$ & 0.13 & 0.17  \\
 &  & B-D & 0.967 & 2.78 & 6.0 & $0.3$ & 0.064 & 0.078  \\
 &  & A-C & 0.872 & 2.51 & 1.5 & $0.6$ & 0.058 & 0.074  \\
 &  & A-D & 1.118 & 3.22 & 1.5 & $0.3$ & 0.086 & 0.11  \\
 &  & C-D & 0.893 & 2.57 & 0.6 & $0.3$ & 0.052 & 0.063  \\
HE1104-1805 & 1.662 & A-B & 3.0 & 4.47 & 6.3 & $<0.013$ & 0.10 & 0.11  \\ 
UM673 & 1.6265 & A-B & 2.22 & 2.71 & 6.3 & $<0.037$ & 0.036 & 0.036  \\
LBQS1429-0053 & 1.6616 & A-B & 5.1 & 43.9 & 3.0 & $0.1$ & 0.051 & 0.050  
\enddata
\end{deluxetable}

Now we make more direct comparisons between our model
and available observed 18 pairs of QSOs (all lensed pairs/multiples except one real physical binary) 
at $z\sim 1.6$, given in Table 1.
Columns 1-7 list parameters of each observed pair and columns 8-9 
give the probability of each pair occurring the ``C" [P(C))] and ``V" [P(V)] 
run respectively at $z=1.6$.
We have made the following simplification for computing the probability:
we treat all DLA with $N_{\rm HI}\ge 2\times 10^{20}$cm$^{-2}$ the same regardless of 
column density values and
for non-DLA absorbers the probability that we 
present is the probability of having a column equal to or less than the indicated value.
We see from Table 1 that
the the observed QSO pairs are all statistically consistent with our model at $\sim 1.5\sigma$ level
or better, entirely consistent with the agreement shown in 
Figure~\ref{fig:sizehis} between the computed size distribution and observationally
inferred size, which is somewhat model dependent.

Finally, in Figure~\ref{fig:sizeMtot}, we plot the total DLA cross section of each 
galaxy ($A_{\rm tot}$) against its halo mass at $z=3.1$ (left panel) and $z=0$ (right panel).
Also shown as the black line is a linear fit assuming that the effective radius 
$r_{\rm tot}=(A_{\rm tot}/\pi)^{1/2}$ is proportional to the virial radius for simplicity.
Note that a least-square linear fit is slightly flatter than the black curve.
It is noted that there are a small fraction of galaxies that do not have significant neutral gas 
thus do not show up in the plotted range.
What is most striking is that, for gas rich galaxies that give rise to DLAs at $z=3.1$,
on average, the effective area that gives rise to DLAs occupies 
about $1/3$ of the total area within the virial radius, i.e., $r_{\rm tot} = 0.6r_{\rm vir}$.
By $z=0$ the effective radius has much reduced to $r_{\rm tot} = 0.08r_{\rm vir}$,
becoming comparable to the size of the stellar disk.
The total DLA cross section as a function of halo mass
is in agreement with \citet[][]{2008Pontzen} within a factor of $3$
(ours is higher than theirs).
We note that the method we use to compute the total DLA cross section 
gives a correct value for the case of a thin disk but could overestimate
it by a factor of $\sqrt{3}$ in the case of a sphere.

Let us fast forward to Figure~\ref{fig:Mhalohis} and take a note that
at $z=3.1$ the typical halo mass for DLA incidence is $M_{h}\sim 10^{11.5}$ having
a virial radius of $r_{\rm vir}=37$kpc, which then gives $r_{\rm tot} = 22$kpc (see above relation 
between $r_{\rm tot}$ and $r_{\rm vir}$).
From Figure~\ref{fig:dishis} we see that the typical distance of DLA from galaxy center
is $d_{\rm DLA}\sim 25$kpc, thus $r_{\rm tot} \sim  d_{\rm DLA}$.
This near equality between the distance and size suggests that a large fraction of the sky seen
by an object sitting in the galaxy, up to $\pi r_{\rm tot}^2/4\pi d_{\rm DLA}^2\sim 20\%$, 
may be covered by DLAs!
If that object is a QSO, the situation may be different, because the QSO, with its intense radiation
(and possibly other effects, such as winds),
may have significantly modified its surroundings, including the nearby DLAs.
Note that a QSO proximity radius is about $5-10$Mpc.
The evidence we see in the gallery examples 
that the surge in DLA incidence of a galaxy seems always 
associated with starbursts and in gas-rich (probably) spiral like galaxies,
whereas QSOs tend to reside in gas-poor elliptical galaxies.
These two facts together suggest that it is not straight forward to estimate
the fraction of QSOs that have this type of proximity DLAs. 
The situation for gamma-ray bursts (GRBs) may be relatively more straight forward 
to analyze, because GRBs are associated with star formation hence likely coincides
with the surge in DLA cross section of interacting galaxies and
because radiation from GRBs does not significantly affect their surroundings
at a distance of $\sim 25$kpc.
Therefore, we suggest that at high redshift $z\ge 3$,
a significant fraction (up to $20\%$) of GRBs,
which depends on detailed geometry/orientation of DLA clouds, 
have associated DLAs with a typical velocity difference from 
its systemic redshift of $\Delta v \le 100$km/s 
(see Figure~\ref{fig:dvgalDLAhis} below for $\Delta v$).
One complication factor is that some (or possibly most of) GRB-DLAs may be due to dense gas
in the close vicinity of GRBs - nearby DLAs.
Since GRBs are thought to occur in star-forming regions, i.e., embedded in molecular clouds,
one might expect to see a signicant amount of molecules associated with nearby DLAs 
and they should be dusty and relatively more metal enriched.
However, if progenitors of GRBs are biased metal poor, such as preferred in 
the collapsar model \citep[][]{1999MacFadyen,2006Woosley},
then it may become more complicated,
although latest observations seem to indicate otherwise \citep[e.g.,][]{2010Levesque}.
Thus, we suggest that a subset of GRB-DLAs {\it with little molecular hydrogen column density}
and being relatively metal poor and dust poor,
such as those in \citet[][]{2003Hjorth},
\citet[][]{2007bTumlinson} and \citet[][]{2009Ledoux},
may be identified with circumgalactic DLAs proposed here.

This fraction of GRBs with DLAs is, however, likely to decrease rapidly towards lower redshift
because of (1) smaller $r_{\rm tot}/d_{\rm DLA}$ ratio and (2) a smaller fraction of DLAs
outside stellar disk.
The unknown in the above estimate is the geometry of DLAs.
We will reserve a detailed analysis of this issue for a future study.

\subsection{Properties of DLA Host Galaxies}

So far we have presented our simulations and compare to observations of DLAs in many different aspects.
From these detailed comparisons
we conclude that the model is consistent with all extant observations
of DLAs,
including kinematic properties, metallicity, column density distribution,
line density and neutral hydrogen density, size distribution, and their evolution,
wherever comparisons could be made.
Proper modeling of galactic winds along with high numerical resolution
seems to have contributed to the success.
What is clear is that galactic winds play 
an indispensable role in alleviating previous tension between the LCDM model and observations,
especially in producing large velocity width DLAs and disparate and low metallicities of DLA gas.
We now examine the properties of DLA host galaxies.

\begin{figure}[t]
\hskip -0.7cm
\centering
\resizebox{3.71in}{!}{\includegraphics[angle=0]{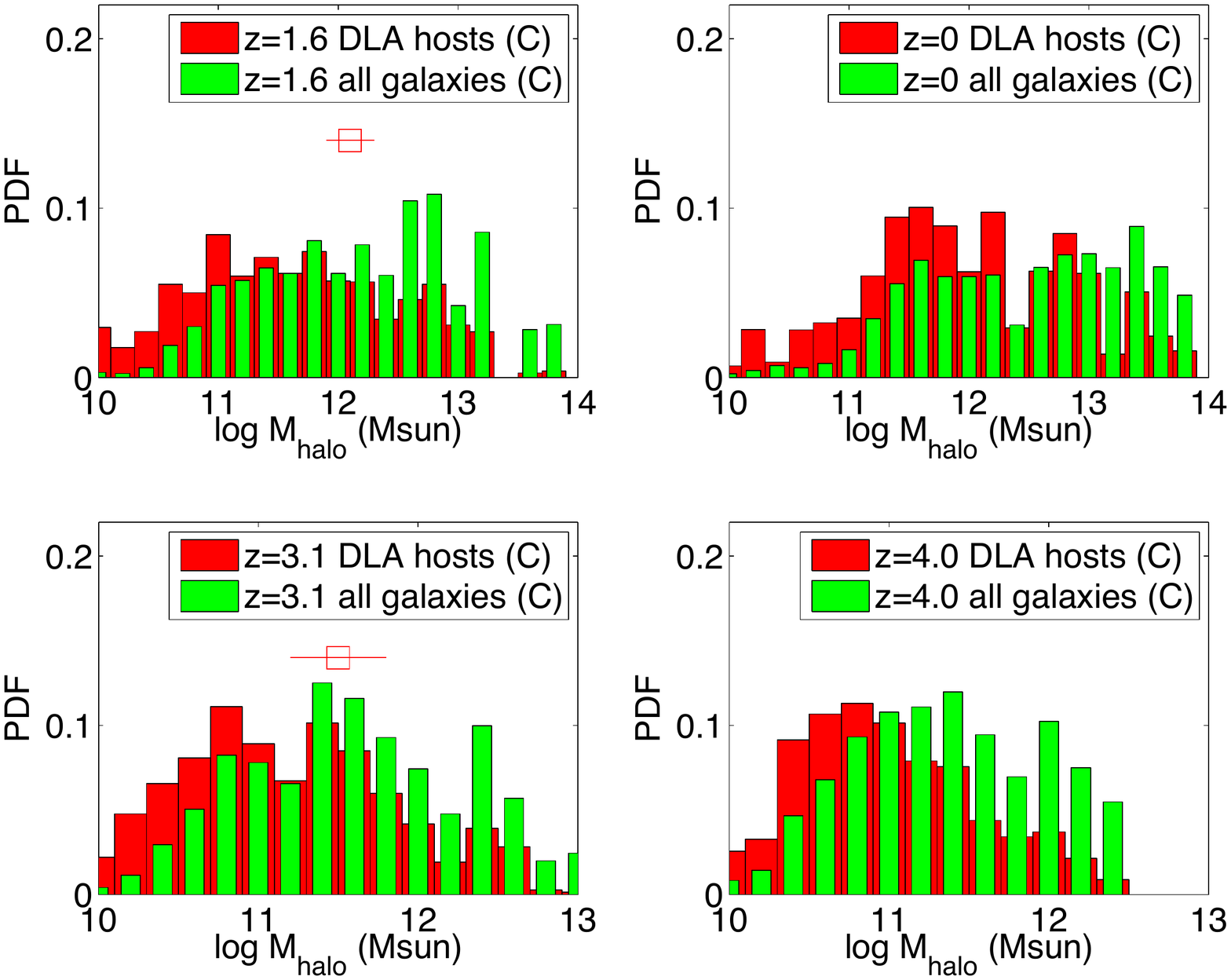}}
\hskip -1.8cm
\resizebox{3.71in}{!}{\includegraphics[angle=0]{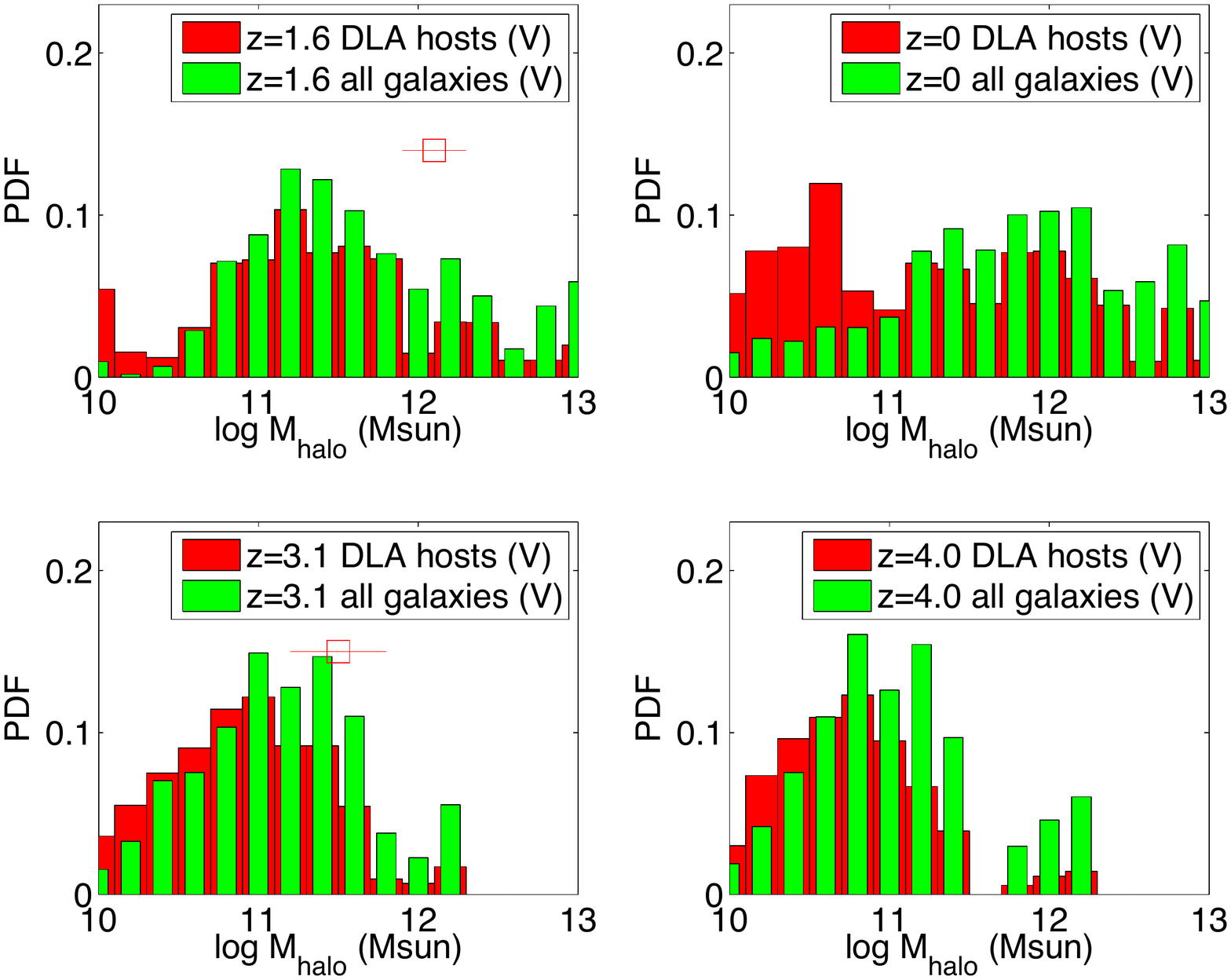}}
\caption{
Left set of four panels: the DLA incidence distribution as a function of host dark matter halo mass
(red histograms) at $z=0,1.6,3.1,4.0$ in the ``C" run.  
Also shown as green histograms are notional distributions,
if the total DLA cross section of a galaxy is proportional to its total mass to the two-third power,
i.e., proportional to its virial radius squared.
Right set of four panels: same for the ``V" run.  
The open squares shown in the $z=1.6$ and $z=3.1$ panels are observationally inferred
halo mass range for LBGs \citep[][]{2005Adelberger}.
}
\label{fig:Mhalohis}
\end{figure}

\begin{figure}[t]
\hskip -0.7cm
\centering
\resizebox{3.71in}{!}{\includegraphics[angle=0]{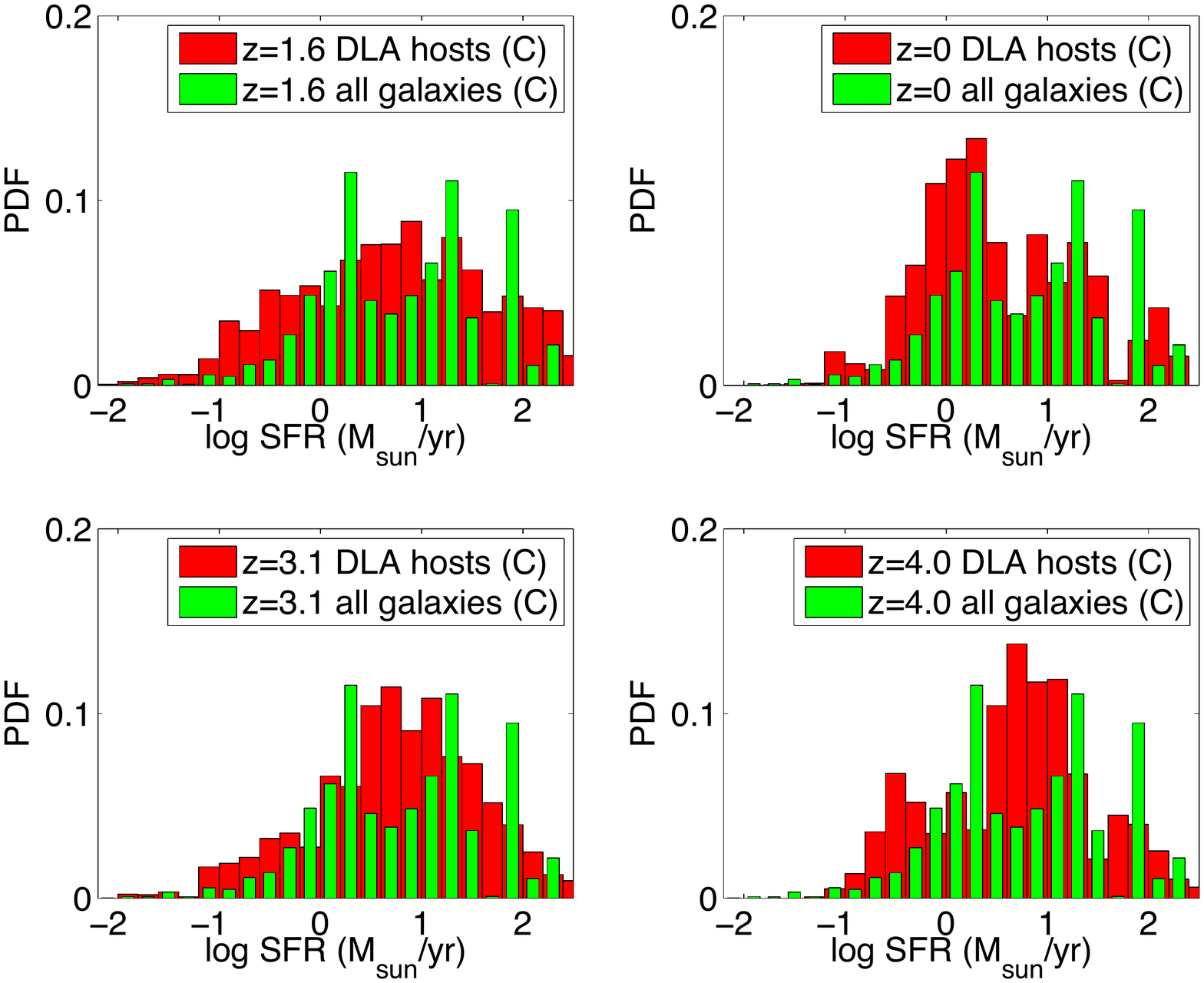}}
\hskip -1.8cm
\resizebox{3.71in}{!}{\includegraphics[angle=0]{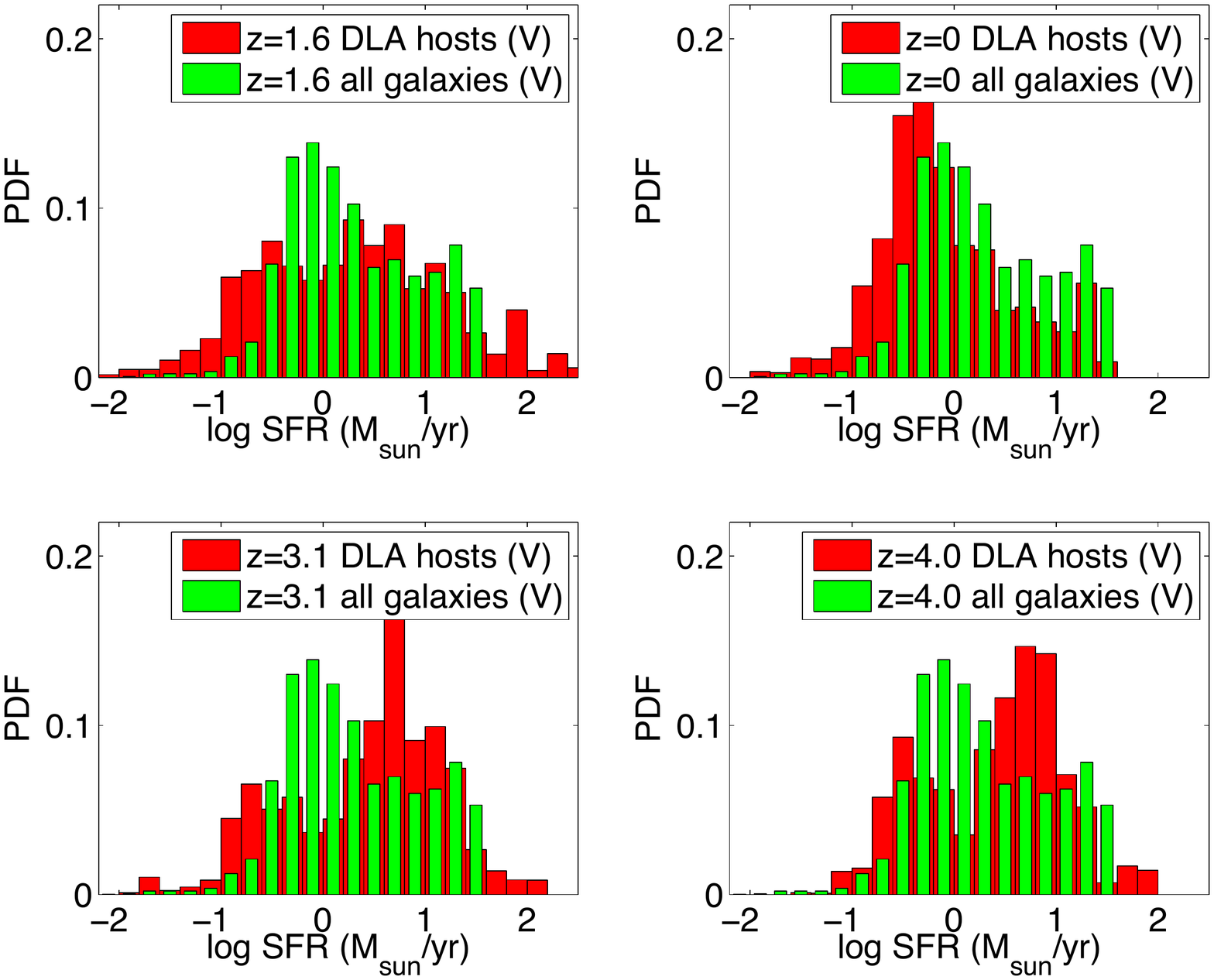}}
\caption{
Left set of four panels: the DLA incidence distribution as a function of host star formation rate 
(red histograms) at $z=0,1.6,3.1,4.0$ in the ``C" run.  
Also shown as green histograms are notional distributions,
if the total DLA cross section of a galaxy is proportional to its stellar mass to the two-third power,
i.e., proportional to its virial radius squared for a constant total mass to stellar mass ratio.
Right set of four panels: same for the ``V" run.  
}
\label{fig:SFRhis}
\end{figure}

Figure~\ref{fig:Mhalohis} shows the DLA incidence as a function of the halo mass.
We see that at redshift $z=0-4$ it appears that the DLA cross section of each galaxy
is roughly proportional to the square of its virial radius,
evidenced by the approximate agreement between the red and green histograms at $z=1.6,3.1,4.0$.
Thus our model makes this rather simple prediction:
DLAs closely trace the overall population of galaxies at $z=0-4$,
with a slight relative bias for lower mass galaxies for DLA hosts compared to the general galaxy population.
The bias is more noticeable at $z=0$.
This is not to say that every galaxy has a DLA in it.
Rather, this simply says that the portion of galaxies that give rise to DLAs at any particular time
has statistically the same mass function as the galaxy population as a whole.
At $z=1.6-4$, the vast majority of DLAs arise in halos of
mass $M_h=10^{10}-10^{12}\msun$, 
as these galaxies dominate the overall population of galaxies. 
We expect the clustering of DLAs to be nearly identical to the overall population of galaxies
at $z\ge 3$ and gradually becomes relatively weaker compared to the galaxy population as a whole with time.

\begin{figure}[h]
\centering
\resizebox{4.5in}{!}{\includegraphics[angle=0]{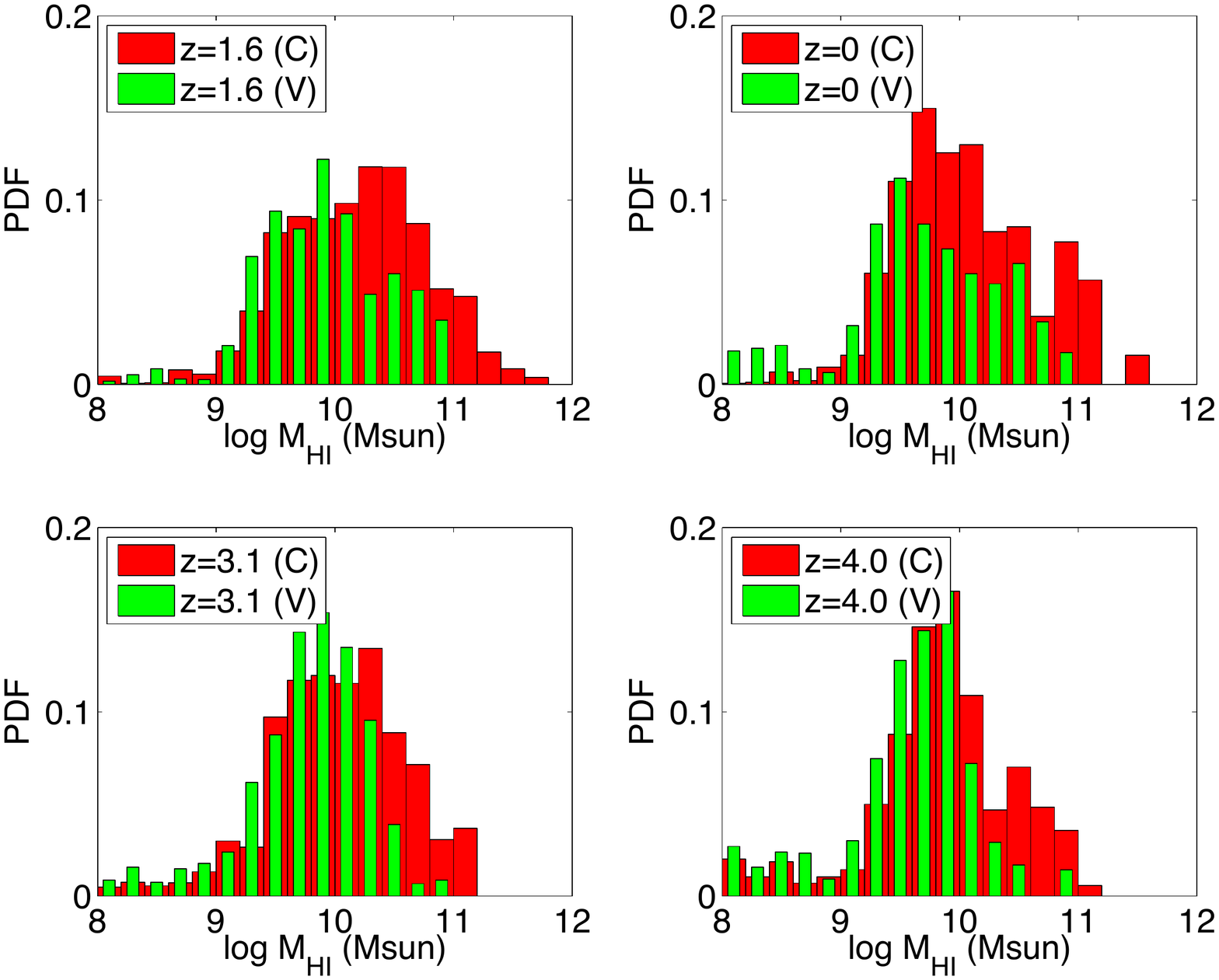}}
\caption{
show the DLA incidence distribution as a function of host HI mass
for the ``C" (red histograms) 
and ``V" (green histograms) run,
at $z=0,1.6,3.1,4.0$.
}
\label{fig:MHIhis}
\end{figure}

Based on clustering analyses, the mass range of the observed LBGs
is inferred to be $10^{11.2}-10^{11.8}\msun$ at z=$2.9$, 
$10^{11.8}-10^{12.2}\msun$ at $z=2.2$, and $10^{11.9}-10^{12.3}\msun$ at z=1.7
\citep[][]{2005Adelberger}.
Comparing these observationally inferred ranges to the histograms in 
Figure~\ref{fig:Mhalohis} we find that 
$20-30\%$ of DLAs at $z\sim 3$ may be LBGs 
and the fraction drops to $10-20\%$  at $z\sim 1.7$.
\citet[][]{2001Schaye} pointed out that, if DLAs have a radius of $27$~kpc,
LBGs could account for all the DLAs at $z\sim 3$.
Working backwards, the estimated $20-30\%$ of DLAs at $z\sim 3$ being LBGs 
would imply a DLA disk radius of $12-15$kpc for LBGs,
which is significantly larger than the size of LBG stellar disks, 
suggesting that the distribution of neutral DLA gas
is in a much more extended region than stellar disk, 
likely a combination of some extended gaseous disk and galactic filaments. 
While a significant overlap between DLAs and LBGs is expected,
the overall clustering of DLAs is expected to be comparable but slightly weaker than LBGs,
since the median halo mass for DLA incidence distribution is slightly smaller than the typical LBG halo mass
(more quantitative comparison will be performed in future work).
\citet[][]{2002Moller} conclude that the properties 
of DLA hosts, including half-light radius, radial profile, optical to near-infrared color, morphology, 
$\lya$ emission equivalent width, and $\lya$ emission velocity structure,
lie within the measured range for the general population of LBGs.
Since those DLA hosts that are sufficiently luminous to be detected 
are large galaxies in the mass range of LBGs
and because DLAs tend to arise in gaseous galactic filaments of galaxies 
that experience starbursts (via galaxy interactions; see \S 3.1), like LBGs,
it is naturally expected in our model that they find similarity between their DLA hosts and LBGs.
Of the remaining 70-80\% that are not covered by LBGs at $z=3-4$,
roughly 10-20\% are due to more massive galaxies and 50-70\% are from smaller galaxies.

When systems become more dynamically advanced with time and large galaxies are no longer
abundant with cold gas,
a slightly more significant shift occurs between DLA hosts 
and the overall population of galaxies at $z=0$.
At $z=0$ we see that the contribution to DLA population is broadly peaked
at $M_h=10^{12}\msun$, whereas most of the sum of the square of virial radius 
is also broadly distributed but, unlike for DLA host,
having a significant contribution from halos of mass $M_h=10^{13-14}\msun$,  in the ``C" run;
For the ``V" run
most of DLAs are in halos of mass $M_h=10^{10-12}\msun$, 
whereas most of the sum of the square of virial radius 
come from halos of mass $M_h=10^{11-13}\msun$.

Figure~\ref{fig:SFRhis} shows the DLA incidence distribution as a function of host star formation rate (SFR)
(red histograms) and the corresponding notional distribution
if the total DLA cross section of a galaxy is proportional to its stellar mass to the two-third power.
The relative distribution between the two seems complex and environment dependent.
To zeroth order, DLA hosts roughly span the same SFR range of the general galaxy population
and most of DLA hosts have SFR in excess of $0.1\msun$/yr at all redshifts.
To first order, there is a bias for DLA hosts to have slighly higher SFR at $z=3-4$
and slighly lower SFR at $z=0$ than the general galaxy population.
The most likely SFR rate for a typical DLA host falls in the range $0.3-30\msun$/yr at $z=3-4$.
LBGs lie at the upper end of this SFR range, consistent with 
the earlier finding that 20-30\% of DLAs overlap with LBGs.
The peak in the distribution move to $\sim 0.5-1\msun$/yr by $z=0$.
It is beyond the scope of this study to investigate the cause for 
the preferred SFR range by DLAs at different redshifts.
We conjecture that the DLA incidence rate likely ``surges"
when mergers or other significant interactions between galaxies
produce the filaments seen in the gallery in \S 3.1.
These same interactions also cause the concerned galaxies to experience
significant star formation, plausibly falling into the found range above.
The trend of moving to lower SFR galaxies at lower redshift 
is perhaps due to the transition from major mergers at high redshift ($z=3-4$)
to minor mergers or smooth accretion and lack of cold gas in very large galaxies at $z=0$.
All evidence -
 size (Figure~\ref{fig:sizehis}),
 distance (Figure~\ref{fig:dishis}),
 metallicity (Figure~\ref{fig:mtlhis})
and now SFR (Figure~\ref{fig:SFRhis})
- seems to point to the picture that by $z=0$ 
most of DLS arise in disks of large galaxies of mass $10^{11}-10^{12.5}\msun$
that are relatively quiet.


\begin{figure}[h]
\hskip -0.7cm
\centering
\resizebox{3.71in}{!}{\includegraphics[angle=0]{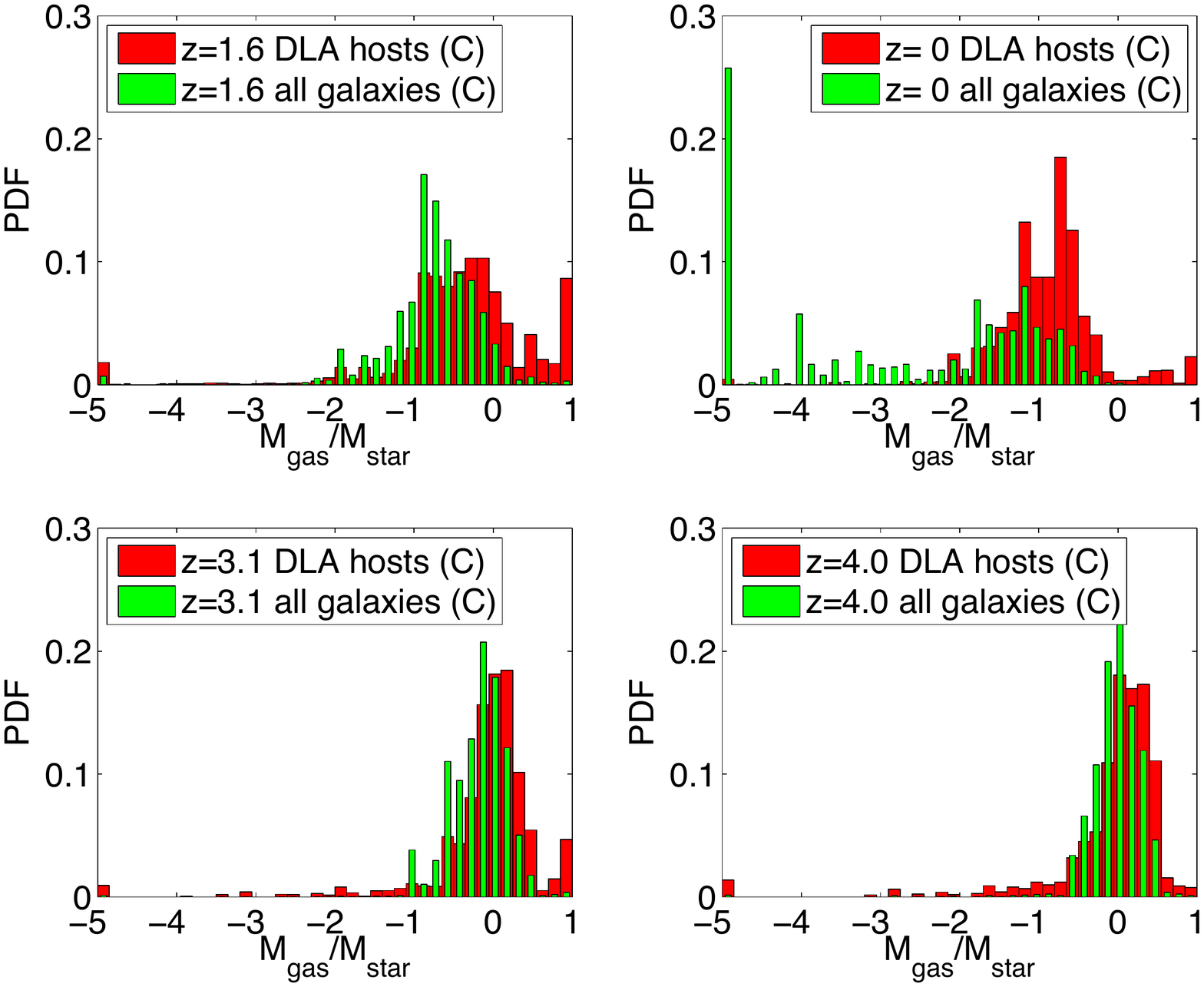}}
\hskip -1.8cm
\resizebox{3.71in}{!}{\includegraphics[angle=0]{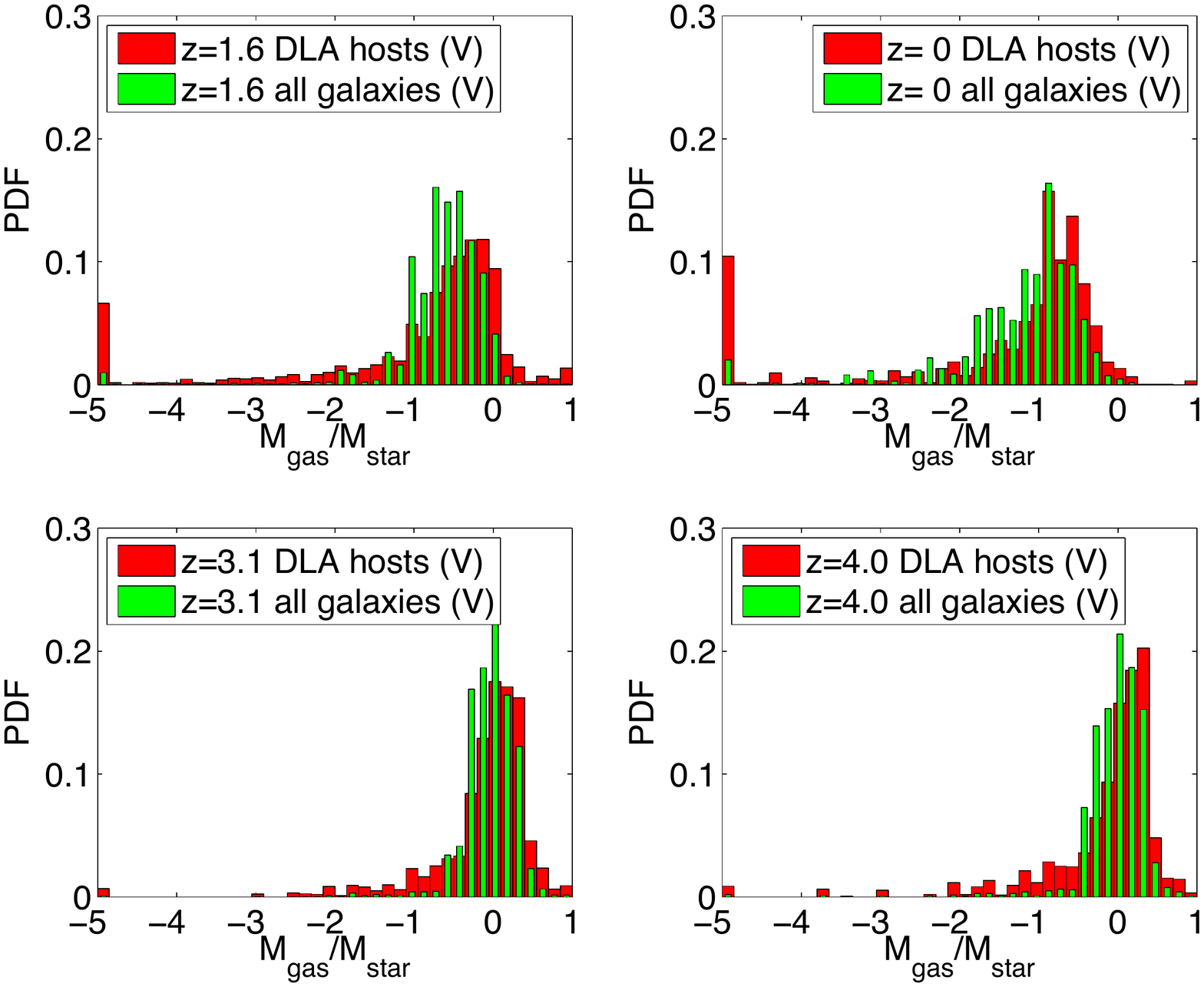}}
\caption{
Left set of four panels: the DLA incidence distribution as a function of host HI mass to stellar mass ratio
at $z=0,1.6,3.1,4.0$ in the ``C" run.  
Also shown as green histograms are notional distributions,
if the total DLA cross section of a galaxy is proportional to its stellar mass to the two-third power,
i.e., proportional to its virial radius squared for a constant total mass to stellar mass ratio.
Right set of four panels: same for the ``V" run.  
}
\label{fig:GShis}
\end{figure}

Figure~\ref{fig:MHIhis} shows the DLA incidence distribution as a function of host HI mass.
It is interesting that the HI mass appears to peak at $10^{10}\msun$ at all redshifts
with $10^9-10^{11}\msun$ covering nearly all the contributions. 
There is a bias for DLA hosts to have slightly more gas rich than the general population of galaxies.
Why the peak position of $M_{\rm HI}$ does not change with time is intriguing but not fully understood
presently.
Figure~\ref{fig:GShis} shows the DLA incidence distribution as a function of host HI mass to stellar mass ratio.
We see that at $z=3-4$ the majority of galaxies are gas-rich ($M_{gas}/M_{star}\ge 0.1$) and
DLA hosts very closely follow the general population of galaxies,
consistent with what we saw in Figure~\ref{fig:Mhalohis}.
This statement remains true for the ``V" run until $z=0$,
whereas for the regions that are more dynamically advanced, such as $z=0$ in the ``C" run,
a significant segregation is evident
in that most of the galaxies are now gas poor 
($M_{gas}/M_{star}\le 0.1$) but majority of DLA hosts are still gas rich ($M_{gas}/M_{star}\ge 0.1$).
Moreover, at $z=0$ in ``C" run, there is a population of old, red and ``dead" galaxies with virtually no cold gas;
these are likely galaxies in clusters of galaxies, which DLAs largely avoid.

\begin{figure}[h]
\hskip -0.7cm
\centering
\resizebox{3.71in}{!}{\includegraphics[angle=0]{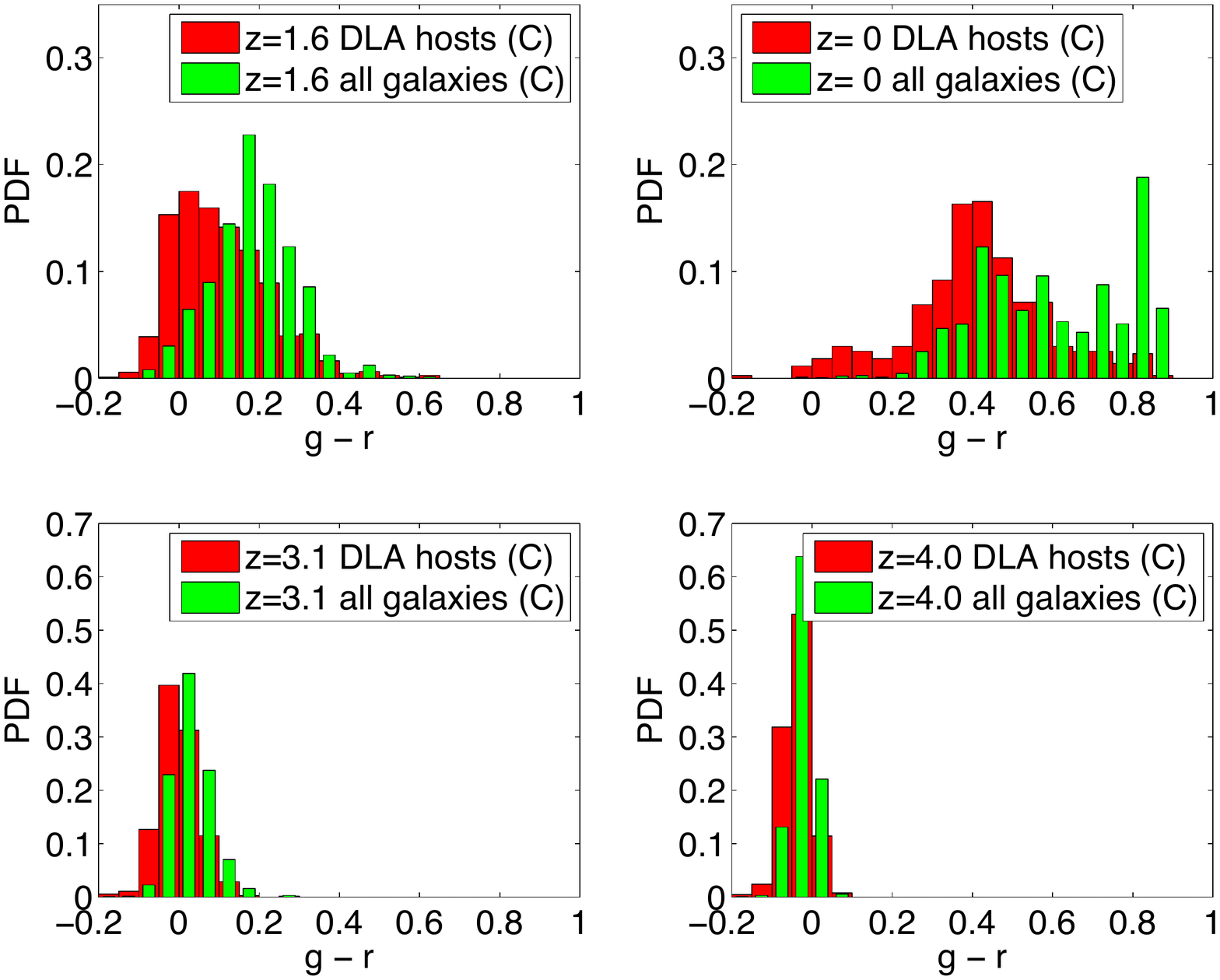}}
\hskip -1.8cm
\resizebox{3.71in}{!}{\includegraphics[angle=0]{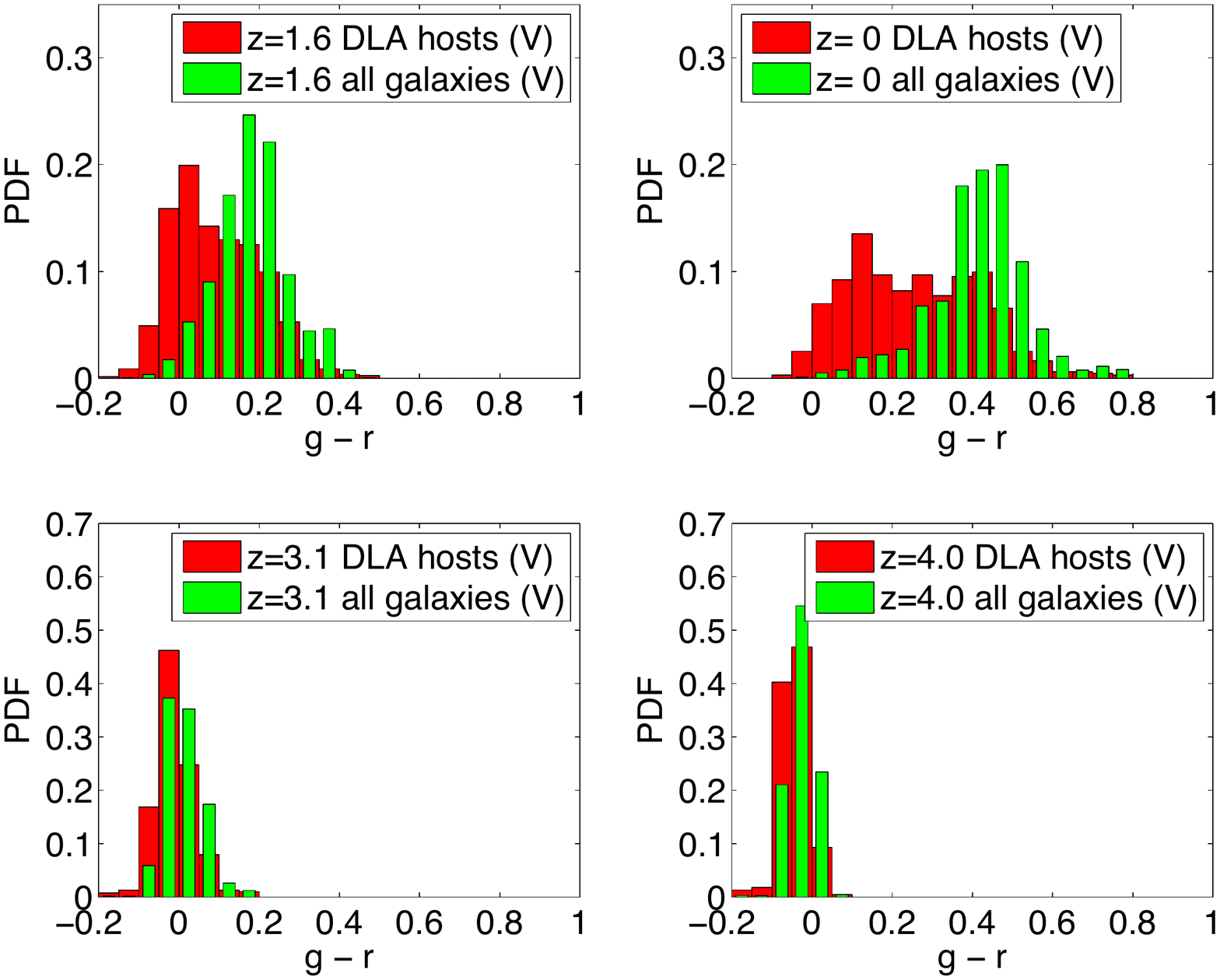}}
\caption{
Left set of four panels: the DLA incidence distribution as a function of host SDSS g-r color
at $z=0,1.6,3.1,4.0$ in the ``C" run.  
Also shown as green histograms are notional distributions,
if the total DLA cross section of a galaxy is proportional to its stellar mass to the two-third power,
i.e., proportional to its virial radius squared for a constant total mass to stellar mass ratio.
Right set of four panels: same for the ``V" run.  
}
\label{fig:colorhis}
\end{figure}

Figure~\ref{fig:colorhis} shows the DLA incidence distribution as a function of host SDSS g-r color.
While the individual color of the simulated galaxies may not be totally consistent with observations
due to its sensitivity to a relatively small variation in the amount of recent star formation
that the simulation may not necessarily properly capture,
a comparative statement is still valid statistically.
We see that DLA hosts trace the general population of galaxies at $z=3-4$ in all environments
with a slight bias to a bluer color peaking sharply at $g-r\sim 0$.
At lower redshift DLA hosts become significantly bluer than average galaxies by roughly $\Delta (g-r)\sim 0.3-0.4$
and they avoid the reddest galaxies that are presumably elliptical galaxies or galaxies in clusters of galaxies.

\begin{figure}[h]
\centering
\resizebox{4.5in}{!}{\includegraphics[angle=0]{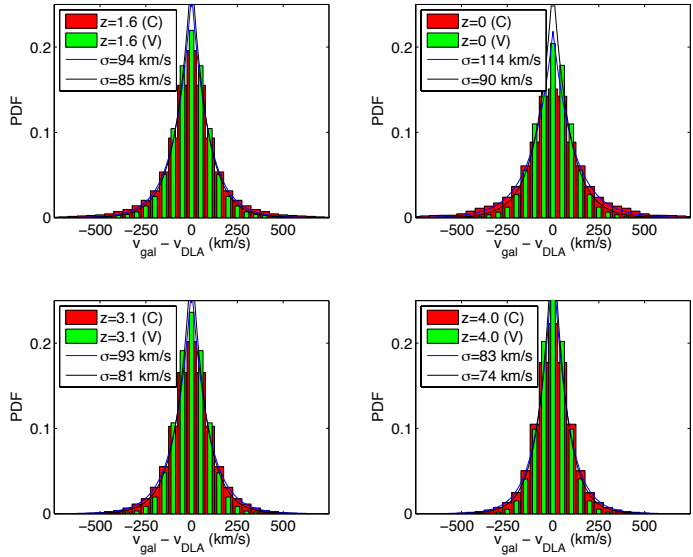}}
\caption{
shows the distribution of the LOS velocity difference between the DLA and its host,
for ``C" (red histograms) and ``V" (green histograms) run.
As shown in blue (for ``C" run) and black (for ``V" run)
are fits to the histograms with the dispersion parameters $\sigma$ indicated.
}
\label{fig:dvgalDLAhis}
\end{figure}

Finally, in Figure~\ref{fig:dvgalDLAhis} we show the distribution 
of the LOS velocity difference between the DLA and its host.
We find the following exponential function provides an excellent fit,
as shown in the figure, while a gaussian fares much worse:
\begin{equation}
\label{eq:tauSiII}
{\rm PDF}(v_{\rm gal}-v_{\rm DLA}) = {1\over 2\sigma} \exp{(-|v_{\rm gal}-v_{\rm DLA}|/\sigma)}.
\end{equation}
\noindent
We find that the dispersion parameter $\sigma$ only increases mildly with decreasing redshift
from $\sigma\sim 75-90$km/s at $z=3-4$ to  $\sim 90-110$km/s at $z=0$.   
The relatively small velocity dispersion and a symmetric distribution 
may make DLAs a useful proxy for the systemic velocity of host galaxies.
This property may be utilized to measure galactic wind velocities,
using a large statistical sample of DLA hosts with measured interstellar absorption lines,
such as Na I line \citep[e.g.,][]{2000Heckman}.

In summary, we show that, at $z=3-4$, DLA hosts approximately follow the general population of galaxies 
with respect to halo mass, star formation rate, HI mass, gas mass to stellar mass ratio,
and color, with a small but noticeable tendency of being more gas rich, slightly smaller and bluer.
It is seen that LBGs make up about 20-30\% of the overall DLAs at $z=3-4$
decreasing to $10-20\%$ by $z=1.7$.
Of the remaining 70-80\% that are not covered by LBGs at $z=3-4$,
roughly 10-20\% are due to galaxies more massive than LBGs
and 50-70\% are from smaller galaxies.
The comparisons at lower redshifts become not as clean cut as at high redshift,
but in a way that seems physically understandable.
Specifically, at lower redshifts, while the general population of galaxies gradually become less cold gas-rich,
DLA hosts seem to ``pick out' those that are gas richer and bluer, but not necessarily of higher star formation rate.

\section{Conclusions}

With high resolution, physically sound treatment of all relevant physical processes,
our state-of-the-art, adaptive mesh-refinement Eulerian cosmological hydrodynamic simulations 
can reproduce a whole array 
of observables of damped Lyman alpha systems,
including the distributions of the following quantities and their evolution with redshift: 
column density, \ion{Si}{2} $\lambda$1808 absorption line velocity width,
kinematic shape measures of the \siii line, 
metallicity, size, line density, HI mass density.
This allows to examine, in addition, with significant confidence, the properties of DLA host galaxies.
We are able to reach the following conclusions.

(1) DLA hosts roughly trace the overall population of galaxies at all redshifts,
with respect to halo mass, star formation rate, HI mass and color, 
with a noticeable tendency of being smaller and bluer, on average.
The standout selection criterion seems being gas rich.
Most of DLA hosts appear to have HI mass peaked at $10^{10}\msun$ at all redshifts
with $10^9-10^{11}\msun$ covering nearly all DLA hosts.
The majority of DLA hosts are gas-rich ($M_{gas}/M_{star}\ge 0.1$) 
closely following the general population of galaxies at high redshift;
by $z=0$ most of the stars are in systems with ($M_{gas}/M_{star}\le 0.1$),
but majority of DLA hosts are still gas rich ($M_{gas}/M_{star}\ge 0.1$).

(2) It seems that the history of DLA evolution is cosmological in nature and 
reflects the underlying cosmic density evolution, galaxy evolution and galaxy interactions. 
With higher density and more interactions at high redshift
DLAs are larger in both absolute terms and in relative terms with respect to virial radii of halos.
At $z=3-4$ galactocentric distance of DLAs is, 
on average, $0.6r_{vir}$, whereas at $z=0$ it decreases to $0.08r_{vir}$.
The typical DLA impact parameter is  $d=20-30$kpc at $z=3-4$ and  $10$kpc at $z=0$.
The typical size (radius) of individual DLAs is $\sim 10$kpc at $z\sim 3-4$
dropping to $\sim 5$kpc at $z\sim 1.6$, in agreement with observations,
then increasing to $\sim 7$kpc at $z=0$.

(3) The variety of DLAs at high redshift is richer with 
a large contribution coming from galactic filaments - filamentary gaseous structures 
sometimes extending to as far as virial radius -
that are created through close galaxy interactions.
The portion of gaseous disks of galaxies where most stars reside makes relatively small contribution
to DLA incidence at $z=3-4$.
By $z=0$ galaxy interactions have become rare, cosmic mean density has decreased dramatically,
so what is left to contribute to DLA incidence is gas rich disks of galaxies in
the mass range of $10^{10}-10^{12}\msun$.

(4) Galactic winds play an indispensable role in shaping the kinematic properties of DLAs. 
Specifically, the high velocity width DLAs are a mixture of those arising 
in high mass, high velocity dispersion halos and those arising
in smaller mass systems where cold gas clouds are entrained to high velocities by galactic winds.
Closer examination reveals that most of the large width DLAs are due to multiple physically
distinct components of varying LOS velocities within a $100$kpc separation (along the LOS).

(5) Quantitatively, the vast majority of DLAs arise in halos of mass $M_h=10^{10}-10^{12}\msun$
at $z=1.6-4$, as these galaxies dominate the overall population of galaxies then.
At $z=3-4$, 20-30\% of DLA hosts are Lyman Break Galaxies (LBGs), 
10-20\% are due to galaxies more massive than LBGs and 50-70\% are from smaller galaxies.
The fraction of LBG DLA hosts drops to $10-20\%$ by $z\sim 1.7$.

(6) In agreement with observations, we see a weak but noticeable evolution in DLA metallicity.
The metallicity distribution centers at $[Z/H]=-1.5$ to $-1$ 
and spans more than three decades at $z=3-4$,
with the peak moving to $[Z/H]=-0.75$ at $z=1.6$ and $[Z/H]=-0.5$ by $z=0$.
The overall metallicity seems floored at $[Z/H]\sim -3$ at $z=1.6-4$ and at $[Z/H]\sim -1.5$ at $z=0$.
When most of the DLA incidence is due to cold gas outside the stellar disks at high redshift,
not only the metallicity is relatively low, but also there is little star formation within.
This explains self-consistently their low metallicity and lack of star formation activity.
If and when significant internal star formation occurs, there are two possible scenarios.
If that takes place outside galactic disk, the DLA will be destroyed and remove itself from the DLA category.
If it takes place on galactic disk, the DLA will either be destroyed permanently,
or temporarily and then cool back down to become a more metal-enriched one, 
after star formation has stopped.
The former occurs predominantly at high redshift, which is why most of DLAs are metal poor.
The latter occurs at low redshift, which explains the significant increase of metallicity.

(7) The star formation rate of DLA hosts is, however, quite strong,
heavily concentrated in the range $0.3-30\msun$/yr at $z=3-4$,
gradually shifting lower to peak at $\sim 0.5-1\msun$/yr by $z=0$.
The finding that the typical size and galactocentric distance of DLAs
are both large and comparable gives a moderate, ``apparent" star formation rate seen
in $\lya$ emission due to fluorescence of ionizing photons from the host galaxies.
Both the size of $\lya$ emission and the magnitude of this fluorescence is, curiously,
consistent with the population of faint $\lya$ emitters 
observed by \citet[][]{2008Rauch}, if they are  at $z\sim 3$.

(8) Finally, we suggest that a significant fraction of sky seen by gamma-ray bursts
should be covered by circumgalactic DLAs at high redshift ($z\ge 3$). 
They may be identified with those GRB-DLAs
that have little H$_2$ column densities and are relatively metal and dust poor.

\vskip 1cm

I would like to thank Dr. M.K.R. Joung for help on
generating initial conditions for the simulations and running a portion
of the simulations and Greg Bryan for help with Enzo code.
I would like to thank Dr. Jason X. Prochaska and Dr. Andrew Pontzen for kindly
providing the observational data,
Dr. Sara L. Ellison for sending a paper draft before publication and for discussion,
Dr. Edward Jenkins for discussion on atomic data.
Computing resources were in part provided by the NASA High-
End Computing (HEC) Program through the NASA Advanced
Supercomputing (NAS) Division at Ames Research Center.
This work is supported in part by grants NNX08AH31G and NAS8-03060. 
The simulation data are available from the author upon request.


\appendix
\section{Resolution Convergence Tests}
\label{sec:convergence}

\begin{figure}[h]
\centering
\resizebox{4.5in}{!}{\includegraphics[angle=0]{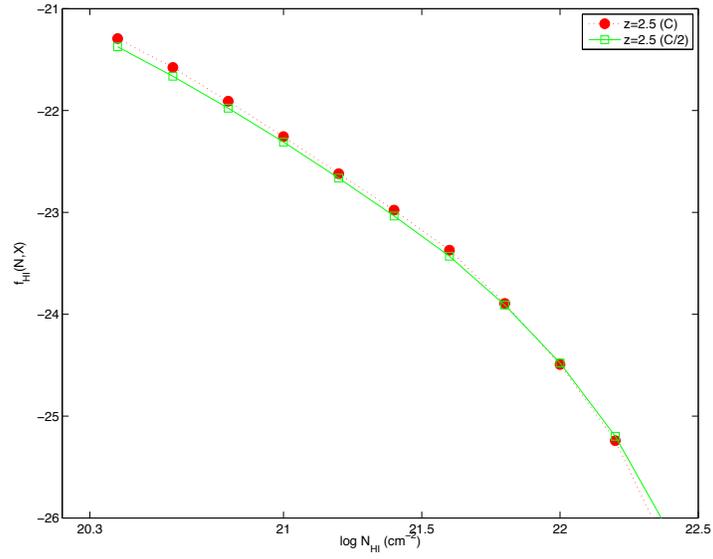}}
\caption{
a comparison of DLA column density distributions at $z=2.5$ for ``C" run (solid dots)
and ``C/2" run (open squares).
}
\label{fig:colhisconverge}
\end{figure}

\begin{figure}[h]
\centering
\resizebox{4.5in}{!}{\includegraphics[angle=0]{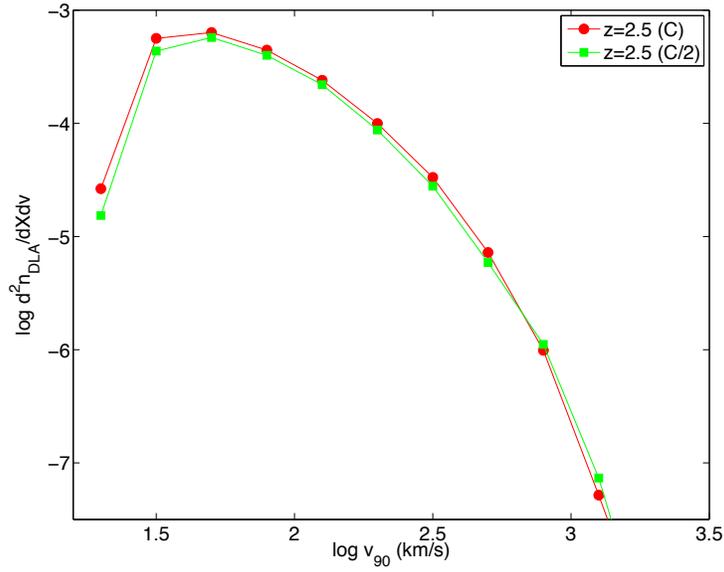}}
\caption{
a comparison of DLA velocity width distributions at $z=2.5$ for ``C" run (solid dots)
and ``C/2" run (open squares).
}
\label{fig:v90hisconverge}
\end{figure}

\begin{figure}[h]
\centering
\resizebox{4.5in}{!}{\includegraphics[angle=0]{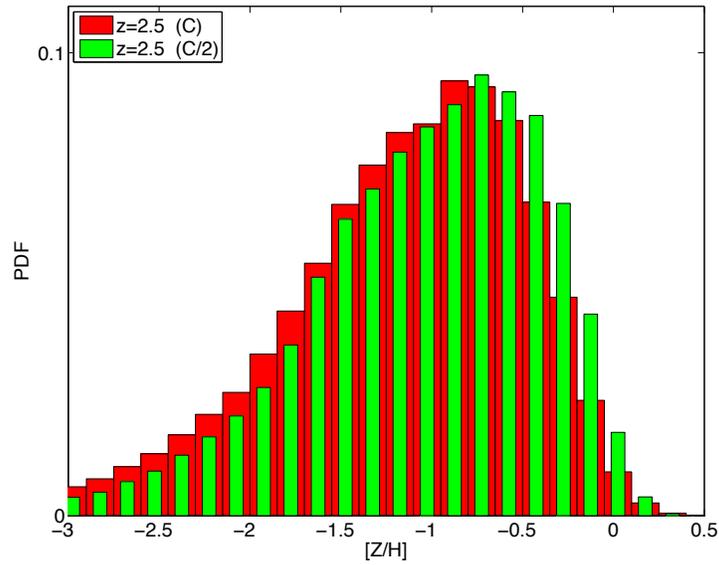}}
\caption{
a comparison of DLA metallicity distributions at $z=2.5$ for ``C" run (red histogram)
and ``C/2" run (green histogram).
}
\label{fig:mtlhisconverge}
\end{figure}

\begin{figure}[h]
\centering
\resizebox{4.5in}{!}{\includegraphics[angle=0]{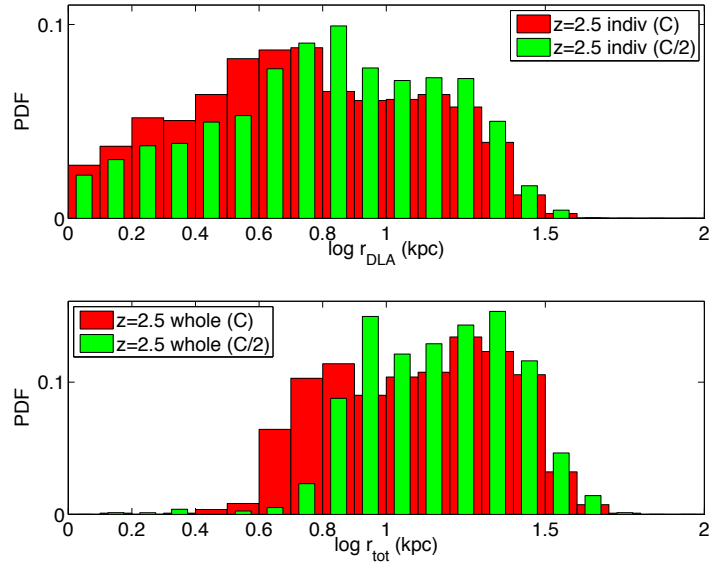}}
\caption{
a comparison of DLA size distributions at $z=2.5$ 
for ``C" run (red histogram)
and ``C/2" run (green histogram).
}
\label{fig:sizehisconverge}
\end{figure}

\begin{figure}[h]
\centering
\resizebox{4.5in}{!}{\includegraphics[angle=0]{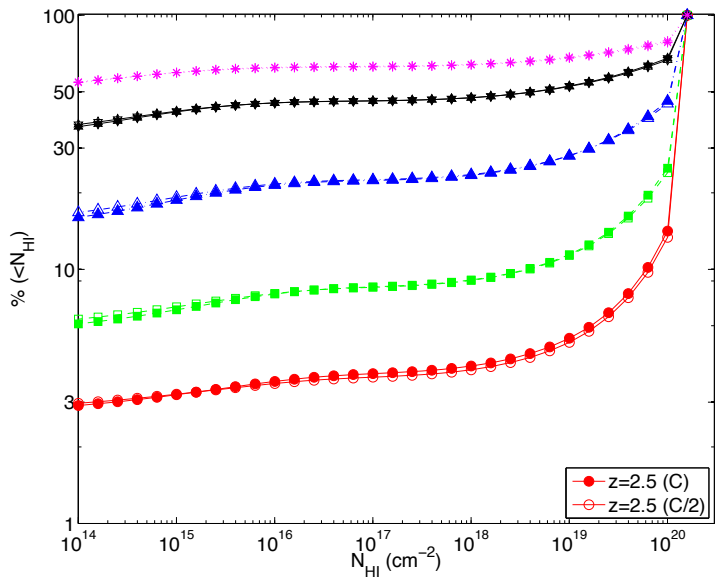}}
\caption{
a comparison of the probability of the second LOS having a HI column lower than the value shown in the x-axis,
while the first LOS is known to have intercepted a DLA at projected separation  
of $(30,20,10,5,3)$kpc (five curves from top to bottom shown in each panel) at $z=2.5$
for ``C" run (solid symbols)
and ``C/2" run (open symbols).
}
\label{fig:PDFcolradconverge}
\end{figure}

In Figure~\ref{fig:colhisconverge} we show a comparison of DLA column density distributions
for the ``C" and ``C/2" run at $z=2.5$.
Recall that ``C/2" run has a spatial resolution lower than ``C" run 
by a factor of 2.
While the agreement is not perfect, it is reasonably good and the difference is at $10-30\%$
across the dynamic range.
Figure~\ref{fig:v90hisconverge} compares velocity width ($v_{90}$ column density distributions
for the ``C" and ``C/2" run at $z=2.5$,
where we see that the level of agreement is comparable to that seen in Figure~\ref{fig:colhisconverge}.
Figure~\ref{fig:mtlhisconverge} shows a comparison of the metallicity distribution
for the ``C" and ``C/2" run at $z=2.5$.
It it seen that the metallicity distribution agrees to within about $0.2$dex and
it appears that a still higer resolution simulation might give a somewhat lower metallicity,
perhaps in still better agreement with observations.
Figure~\ref{fig:sizehisconverge} shows a comparison of the DLA size distribution
for the ``C" and ``C/2" run at $z=2.5$.
Since the spatial resolution more directly affects the size, here we see 
the worst disagreement between the two runs in terms of the shape
of the distribution, but overall still mostly at a level of $\sim 0.05-0.1$dex.
However, a direct comparison of probability for QSO binary sightlines
shown in Figure~\ref{fig:PDFcolradconverge} 
suggests that the overall convergence is quite good.

\end{document}